%% file: main.tex
\documentclass[11pt,a4paper]{article}
\usepackage{placeins}
\usepackage{graphicx}
\usepackage{xcolor}
\usepackage{float}
\usepackage{subfig}
\usepackage{afterpage}
\usepackage{amssymb,amsmath,bm,fixmath}
\usepackage{braket}
\usepackage{multirow,booktabs}
\usepackage[normalem]{ulem}
\usepackage{slashed}
\usepackage[colorlinks=true, linkcolor=black!50!blue, urlcolor=blue, citecolor=blue, anchorcolor=blue]{hyperref}
\usepackage[font=small,labelfont=bf,labelsep=period]{caption}
\usepackage[a4paper,top=3cm,bottom=2.5cm,left=2.5cm,right=2.5cm,bindingoffset=0mm]{geometry}
\usepackage{tabularx}
\newcolumntype{C}[1]{>{\centering\arraybackslash}p{#1}}
\usepackage{scalerel}

\def\vecsign#1{\rule[1.388\LMex]{\dimexpr#1-2.5pt}{.36\LMpt}%
  \kern-6.0\LMpt\mathchar"017E}

\usepackage{stackengine,amsmath}
\stackMath

\newcommand{\be}{\begin{equation}}
\newcommand{\ee}{\end{equation}}
\newcommand{\bea}{\begin{eqnarray}}
\newcommand{\eea}{\end{eqnarray}}
\newcommand{\bi}{\begin{itemize}}
\newcommand{\ei}{\end{itemize}}
\newcommand{\ben}{\begin{enumerate}}
\newcommand{\een}{\end{enumerate}}
\newcommand{\la}{\left\langle}
\newcommand{\ra}{\right\rangle}
\newcommand{\lc}{\left[}
\newcommand{\rc}{\right]}
\newcommand{\lp}{\left(}
\newcommand{\rp}{\right)}

\def\frac#1#2{{{#1}\over {#2}}}
\def\gsim{\gtrsim}

\newcommand{\draft}[1]{}

\newcommand{\sss}{\scriptscriptstyle}
\newcommand{\OO}{\ensuremath{\mathcal{O}}}

\def\beq{\begin{equation}}
\def\eeq{\end{equation}}
\def\({\left(}
\def\){\right)}
\def\[{\left[}
\def\]{\right]}
\let\originalleft\left
\let\originalright\right
\renewcommand{\left}{\mathopen{}\mathclose\bgroup\originalleft}
\renewcommand{\right}{\aftergroup\egroup\originalright}
\numberwithin{equation}{section}
\numberwithin{figure}{section}
\numberwithin{table}{section}
\let\oldsubsection\subsection
\renewcommand\subsection[2][\subsectiontoc]{%
  \def\subsectiontoc{#2}%
  \oldsubsection[#1]{\boldmath #2}%
}
\let\oldsubsubsection\subsubsection
\renewcommand\subsubsection[2][\subsubsectiontoc]{%
  \def\subsubsectiontoc{#2}%
  \oldsubsubsection[#1]{\boldmath #2}%
}

\newcommand{\pwg}{\texttt{POWHEG-box}}

\def \n {$(\checkmark)$}

\newcommand{\lag}{\mathcal{L}}
\newcommand{\op}{\mathcal{O}}

\emergencystretch=\maxdimen
\usepackage{xcolor}
\input{Macros.tex}

\begin{document}
\newgeometry{top=1.5cm,bottom=1.5cm,left=2.5cm,right=2.5cm,bindingoffset=0mm}
\begin{titlepage}
\thispagestyle{empty}
\noindent
\begin{flushright}
Nikhef-2020-039 ,
IPPP/20/71 \\
VBSCAN-PUB-01-21 \\
\end{flushright}
\vspace{0.7cm}
\begin{center}
  {\LARGE \bf\boldmath SMEFT analysis of vector boson scattering \\[0.3cm] and diboson data from the LHC Run II}\vspace{1.4cm}

Jacob~J.~Ethier,$^{1,2}$
Raquel~Gomez-Ambrosio,$^{3,4,\dagger}$
Giacomo~Magni,$^{1,2}$
Juan~Rojo,$^{1,2}$\\[0.1cm]
\vspace{0.7cm}

{\it \small
~$^1$ Department of Physics and Astronomy, Vrije Universiteit Amsterdam,\\ NL-1081 HV Amsterdam, The Netherlands\\[0.1cm]
 ~$^2$  Nikhef Theory Group, Science Park 105, 1098 XG Amsterdam, The Netherlands\\[0.1cm]
 ~$^3$
Dipartimento di Fisica, Universita degli Studi di Milano Bicocca \\
and INFN Sezione di Milano Bicocca, Milan, Italy.
\\[0.1cm]
~$^4$ Institute for Particle Physics Phenomenology
, Durham University, \\ South Road DH1 3LE Durham, UK
\\[0.1cm]
}

\vspace{0.7cm}

{\bf \large Abstract}
\end{center}

We present a systematic interpretation of vector boson scattering (VBS)
and diboson measurements from the LHC in the framework of the dimension-six Standard Model
Effective Field Theory (SMEFT).
We consider all available measurements of VBS fiducial cross-sections and differential
distributions from ATLAS and CMS, in most cases based on the full Run II luminosity,
and use them to constrain 16 independent directions in the dimension-six EFT parameter space.
Compared to the diboson measurements, we find that VBS provides complementary information on several of the operators relevant for the description of the electroweak sector.
We also quantify the ultimate EFT reach of VBS measurements via dedicated projections for the High Luminosity LHC.
Our results motivate the integration of VBS processes
in future global SMEFT interpretations of particle physics data.

\vspace{9cm}
\par\noindent\rule{\textwidth}{0.4pt}
$^\dagger$ {\it \small corresponding author: raquel.gomezambrosio@unimib.it}
\end{titlepage}
\restoregeometry
\tableofcontents

\input{sec-introduction.tex}
\input{sec-theory.tex}
\input{sec-expdata.tex}
\input{sec-results.tex}
\input{sec-hllhc.tex}
\input{sec-summary.tex}

\FloatBarrier
\phantomsection
\addcontentsline{toc}{section}{References}

\bibliographystyle{JHEP}
\bibliography{references}
\end{document}

%% file: Macros.tex
\usepackage[utf8]{inputenc}
\usepackage[T1]{fontenc}
\usepackage[numbers, elide, sort&compress]{natbib}
\usepackage{lmodern, amsmath, amssymb, xcolor, a4wide, graphicx, datetime, rotating, adjustbox, array}
\makeatletter\g@addto@macro\bfseries{\boldmath}\makeatother

\def\equationautorefname~#1\null{Eq.\,(#1)\null}
\newcommand{\fullref}[2]{\hyperref[#2]{#1\,\ref*{#2}}}
\makeatletter
\newcommand{\mg}{\texttt{MG5\_aMC@NLO}}
\newcommand{\rcite}[1]{\hyper@@link[cite]{}{cite.#1}{Ref.\,\cite*{#1}}}
\makeatother
\newcommand*{\eqsref}[2]{\hyperref[#1]{Eqs.\,(\ref*{#1}--}\hyperref[#2]{\ref*{#2})}}
\allowdisplaybreaks[4]
\makeatletter
\newcommand{\parenbar}{\mathpalette\p@renb@r}
\def\p@renb@r#1#2{%
	\vbox{%
		\ifx#1\scriptscriptstyle\dimen@.7em\dimen@ii.20em\else%
		\ifx#1\scriptstyle	\dimen@.8em\dimen@ii.25em\else%
					\dimen@.9em\dimen@ii.35em\fi\fi%
		\offinterlineskip%
		\ialign{%
			\hfill##\hfill\cr
			\vbox{\hrule width\dimen@ii height .35pt}\cr
			\noalign{\vskip-.35ex}%
			\hbox to\dimen@{$\mathchar300\hfil\mathchar301$}\cr
			\noalign{\vskip-.35ex}%
			$#1#2$\cr%
		}%
	}%
}%
\makeatother

\newcommand*{\tmp}[4]{\ensuremath{%
	{#4%
	\ifx\empty#3\empty\ifx\empty#1\empty\else^{#1}\fi\else^{#1(#3)}\fi%
	\ifx\empty#2\empty\else_{#2}\fi}%
}}

\newcommand{\rien}[1]{}
\newcounter{affiliation}

%% file: sec-introduction.tex
\section{Introduction}
\label{sec:introduction}

Since the dawn of the Standard Model (SM), the vector boson scattering (VBS) process has been heralded as a cornerstone to test the high-energy behaviour of the electroweak sector.
Such importance originated in calculations
of scattering amplitudes involving longitudinally
polarised vector bosons which, in the absence of a Higgs boson,
were shown to grow quadratically with energy and eventually violate unitarity bounds~\cite{PhysRevLett.17.616,PhysRevLett.57.2344, PhysRevD.36.1490,PhysRevD.10.1145,PhysRevD.22.200, PASSARINO199031}.
The ability to fully scrutinise the VBS process was therefore one of the
motivations to project the ill-fated Superconducting Super Collider (SSC) with a center of mass energy of $\sqrt{s}=40$ TeV~\cite{Eichten:1984eu}.
If the Higgs boson were not responsible for electroweak symmetry breaking, the SSC might have been able to discover new resonances in the high-energy tail of VBS events.

While we
know now that the Higgs boson, following its discovery in 2012~\cite{Hdiscovery:atlas,Hdiscovery:cms}, unitarises the VBS cross-sections, such processes still provide unique sensitivity to deformations of the SM at high energies,
such as those parametrised by the Standard Model Effective Field Theory
(SMEFT)~\cite{Weinberg:1979sa,Buchmuller:1985jz,Georgi:1994qn}.
VBS therefore provides a fully complementary probe to investigate
the electroweak sector of the SMEFT compared to processes such
as on-shell Higgs production or gauge-boson pair production, both
in terms of covering a different energy regime (up to the TeV scale) and by its contributions from different EFT operator combinations.
A particularly attractive feature of VBS in this context
is the appearance of quartic gauge couplings
(QGCs), which have often led to a theoretical interpretation of VBS data
in terms of anomalous QGCs (aQGCs).

One significant challenge in studying the VBS process
at the LHC is the rather small signal-to-noise ratios
due to its electroweak nature, with backgrounds being
dominated by QCD-induced diboson production.
Fortunately, VBS also benefits from  a characteristic signature that allows for a relatively clean isolation,
defined by two energetic jets in the forward region and a
large rapidity gap between them that contains reduced hadronic activity.\footnote{
This is same  kinematic signature relevant
to identify single Higgs~\cite{Cacciari:2015jma} and Higgs pair~\cite{Bishara:2016kjn} production
in vector boson function (VBF).}
The combination of this characteristic topology together with the improved analysis of the high statistics delivered during Run II of the
LHC ($\mathcal{L}=140$ fb$^{-1}$ at $\sqrt{s}=13$ TeV)
has made possible not only the
identification of VBS events with reasonable statistical significance, but also
the measurement of the associated unfolded cross-sections
and differential distributions in the fiducial region~\citep{WWjjWZjj:cms,WWjj:atlas,WZjj:atlas,ZZjj:atlas,ZZjj:cms137,AZjj:atlas,AZjj:cms,Sirunyan:2020gvn}.
In particular, VBS measurements from ATLAS and CMS
based on the full Run II dataset have recently been presented for
different final states, from $W^{\pm}W^{\pm}jj$ and $ZW^{\pm}jj$~\cite{WWjjWZjj:cms}
to $ZZjj$~\citep{ZZjj:cms137,ZZjj:atlas},
including one analysis targeting polarized $W^{\pm}W^{\pm}$ scattering~\cite{Sirunyan:2020gvn}.

In the past, searches for new physics using  VBS processes have either been based on
unitarisation techniques~\cite{Kilian:2014zja,Perez:2018kav,Corbett:2014ora,Sekulla:2016yku} or
interpreted in terms of
anomalous gauge couplings, where the SM couplings are rescaled by phenomenological
parameters fitted from the data~\cite{2017380,Gounaris:1995ed, Khachatryan:2016vif,ZZjj:cms137}.
However, this approach is only beneficial for bookkeeping
purposes since, among other limitations, it violates gauge invariance.
For this reason, different strategies based on effective field theories have been
advocated~\cite{Hagiwara:1993ck,Eboli:2006wa,Degrande:2012wf,Gritsan:2020pib}  to interpret multi-boson and
VBS measurements.
These EFT-based approaches have numerous advantages over the previous
phenomenological approaches: they respect the fundamental symmetries
of the SM, are systematically improvable in perturbation theory, allow
the correlation of eventual deviations between
different processes, and can accommodate a meaningful quantification of theoretical uncertainties.
We note that, beyond the SMEFT, other effective theory interpretations of VBS data
have been considered such as those based on the Electroweak Chiral Lagrangian~\cite{Delgado:2019ucx, Delgado:2018nnh,Delgado:2017cls,Kozow:2019txg}, where the Higgs boson
is not necessarily part of an SU(2) doublet.

With this motivation, VBS measurements have often been
interpreted in the SMEFT framework to
identify, parametrise, and correlate possible deviations in the  structure of the electroweak gauge couplings
compared to the SM predictions.
However, these studies have so far~\cite{Sirunyan:2019der,Khachatryan:2017jub,Aad:2015uqa,Kalinowski:2018oxd} been mostly
restricted to a selection of dimension-eight
operators~\cite{Eboli:2006wa,Brass:2018hfw}, in particular those that induce aQGCs
without modifying the Triple Gauge Couplings (TGCs).
As emphasized in Ref.~\cite{Gomez-Ambrosio:2018pnl}, it is theoretically inconsistent to derive
bounds on aQGCs from VBS data accounting for dimension-eight operators
while neglecting the dimension-six ones, which also modifying the electroweak interactions that enter the same observables.
The fact that available EFT interpretations of VBS processes ignore the contribution from
dimension-six operators casts doubts on the robustness of the obtained aQGCs bounds.

While several works have investigated the effects of dimension-six operators on
diboson production~\cite{Grojean:2018dqj,Falkowski:2016cxu,Rahaman:2019mnz},
including the impact of QCD corrections to the EFT
cross-sections~\cite{Baglio:2017bfe,Baglio:2020oqu,Baglio:2019uty},
much less attention has been devoted to the corresponding effects
on VBS processes~\cite{Jager:2013iza, Gomez-Ambrosio:2018pnl,Dedes:2020xmo,Gallinaro:2020cte}.
In this work, we present for the first time
a systematic interpretation of VBS fiducial cross-sections
and unfolded differential
distributions from the LHC in the framework of the dimension-6 SMEFT
at linear order, $\mathcal{O}\lp \Lambda^{-2}\rp$, in the effective theory expansion.
Our study is carried out within the {\tt SMEFiT} framework, a toolbox for global EFT
interpretations of experimental data which has been deployed to characterise the top-quark
sector~\cite{Hartland:2019bjb} and is currently being updated to perform a combined EFT analysis of Higgs boson, top-quark, and diboson measurements from LEP and the LHC in Ref.~\cite{smefittophiggs}.

In the present study, we consider all available VBS measurements of fiducial cross-sections and
distributions, in most cases based on the full Run II integrated luminosity.
These are complemented by the most updated QCD-induced diboson production datasets
from ATLAS and CMS~\citep{WW:atlas,WW:cms,WZ:atlas,WZ:cms,ZZ:cms137},
which are interpreted simultaneously within the same
EFT theoretical framework as the VBS measurements.
We demonstrate how the VBS measurements provide complementary
information on several operators relevant for the description the electroweak sector of
the SMEFT,  in particular those modifying the triple and quartic gauge couplings.
In addition, we quantify the impact of the VBS
data by direct fits and by using statistical
metrics such as information geometry and principal component analysis.
We also highlight the consistency between
the constraints separately provided by the VBS and diboson data on the
dimension-six operators considered,
representing a non-trivial stress-test of the gauge sector of the SMEFT.
Overall, our analysis motivates the systematic
inclusion of VBS data in global SMEFT
interpretations~\cite{Berthier:2015gja,deBlas:2016ojx,Englert:2017aqb,Ellis:2018gqa,
Biekotter:2018rhp,Aebischer:2018iyb,Falkowski:2019hvp,deBlas:2019okz,Ellis:2020unq,Dawson:2020oco}.

While we have now the first VBS unfolded measurements
of cross-sections and differential distributions, they are
limited by statistics.
Accessing the full physics potential associated to VBS processes
will only be achieved with the analysis of the complete dataset from
the High Luminosity LHC~\cite{Azzi:2019yne,Cepeda:2019klc}.
In particular, the HL-LHC will provide access to the high energy region of $VV'\to VV'$ scattering and has the potential to disentangle contributions from $V_LV_L'$ polarised scattering~\cite{CMS:2018mbt,CMS:2018zxa,ATLAS:2018tav,ATLAS:2018ocj}.
To quantify this impact, we present projections for the reach in the EFT parameter space of the VBS measurements expected at the HL-LHC, which demonstrate a significant increase in sensitivity compared to current measurements.

The structure of this paper is as follows.
First, we present the theoretical framework of the analysis in Sect.~\ref{sec:eftth},
in particular our definition of the dimension-six operator basis and the flavour assumptions.
In Sect.~\ref{sec:expdata} we describe the VBS and diboson data used as input for our EFT fit, outline the details of the corresponding SM theoretical calculations, and present different measures
of the expected operator sensitivity.
The main results of this work are then presented in Sect.~\ref{sec:results}, where we derive bounds
on the relevant operators and discuss the interplay between the various data sets.
Finally, we study in Sect.~\ref{sec:hllhc} the impact that future measurements of VBS processes at the HL-LHC will have on the EFT parameter space, followed by a summary and indication of possible future
developments in Sect.~\ref{sec:summary}.

%% file: sec-theory.tex
\section{Theoretical framework}
\label{sec:eftth}
\label{sec:Warsawbasis}

In this section we introduce the dimension-six SMEFT operators that will be considered
for the interpretation
of the vector boson scattering and diboson measurements at the LHC.
Restricting ourselves to dimension-six operators, we can express
the SMEFT Lagrangian as,
\begin{equation}
  \label{eq:SMEFTlag}
\lag_{\rm SMEFT}  = \lag_{\rm SM} + \sum_{i=1}^{n_{\rm op}} \frac{c_{i}}{\Lambda^2}  \OO_i^{(6)} \, ,
\end{equation}
where the $\op_i^{(6)}$ represent a complete basis of operators built upon the SM fields with mass dimension equal to six, and $c_i$ are their corresponding Wilson coefficients.
These operators respect the fundamental symmetries of the SM such as gauge and Lorentz invariance.
In Eq.~(\ref{eq:crosssection}), $\Lambda$ indicates the
energy scale that determines the regime of validity of the EFT approximation.
For instance, $\Lambda$  can be interpreted as the typical
mass of the new heavy particles that arise in the
ultraviolet (UV) completion of the SM.
Note that, from a bottom-up phenomenological analysis, only the ratio $c_i/\Lambda^2$ can
be determined, rather than the two parameters separately.

In this work, we will focus on those
operators that modify the interactions of the electroweak gauge bosons.
These will involve the weak gauge field strength tensors
\begin{equation}
     \label{eq:Wstrength}
 	W^{I}_{\mu \nu } = \partial_\mu W^{I}_\nu - \partial_\nu W^I_\mu - g_2 \epsilon^{IJK}  W^J_\mu W^K_\nu \, ,
 \end{equation}
 \begin{equation}
     \label{eq:Bstrength}
 	B_{\mu \nu } = \partial_\mu B_\nu - \partial_\nu B_\mu \, ,
 \end{equation}
as well as the SM covariant derivative, given by
 \begin{equation}
     D_\mu = \partial_\mu +  i g_2 \frac{\sigma^{I}}{2} W^{I}_\mu + i g_{1}  Y_f B_\mu \, ,
     \label{eq:covderiviative}
 \end{equation}
 where $g_1, g_2$ are the weak couplings,  $\sigma^{I}$ are the Pauli matrices (SU(2)$_L$ generators),
 and $Y_f$ is the fermionic hypercharge.
 Here we neglect strong interaction effects, which play a limited role
 in the description of the VBS process, and set to zero the masses of all leptons and quarks except
 for the top quark.
 Some of the
 relevant dimension-six operators for this analysis will also involve the Higgs doublet field,
  defined in the unitary gauge by
\begin{equation}
    \label{eq:HHdoublet}
   \varphi = \frac{1}{\sqrt{2}} \begin{pmatrix} 0 \\ v + h \end{pmatrix} \, ,
\end{equation}
with $v=246$ GeV being the Higgs vacuum expectation value (vev) and $h$ represents
the $m_h=125$ GeV Higgs boson.
Here we will also consider CP-odd operators, which are constructed in terms of the
dual field strength tensors, defined by
\begin{equation}
    \label{eq:dual}
	\widetilde{X}_{\mu\nu } = \frac{1}{2} \epsilon_{\mu \nu \rho \sigma} X^{\rho \sigma} \, ,
\end{equation}
and whose presence leads to CP-violating effects which are potentially observable
in the
electroweak sector~\cite{DasBakshi:2020ejz,Choudhury:1999fz,Biekotter:2020flu,
Azatov:2019xxn,Banerjee:2020vtm}

There exist several bases that span the SMEFT operator space at dimension-six.
In this work we adopt the  Warsaw basis~\cite{Grzadkowski:2010es}, which contains 59 operators for one fermion generation,
and consider only those operators that contain at least one electroweak gauge field.
This means, in particular,
that we neglect the contributions from four-fermion operators as well as from those
that modify the Yukawa interactions and the Higgs self-coupling.

\paragraph{Flavour assumptions.}
In this work, we will assume that the operator structure is the same across the
three fermionic families, the so-called SU(3)$^5$-symmetric model.
In other words, we assume flavour universality of the UV-complete theory.
In practice, this means that all Warsaw basis operators that contain fermion
generation indices will be understood as diagonal and summed over
generations, {\it e.g.},
\begin{equation}
    [c_{\varphi f}]_{ij} ( \varphi^{\dagger}\overleftrightarrow{D}_{\mu}\varphi )(\Bar{f}_i \gamma^\mu f_j) \longrightarrow
    c_{\varphi f} \sum_{i=1}^3 ( \varphi^{\dagger}\overleftrightarrow{D}_{\mu}\varphi )(\Bar{f}_i \gamma^\mu f_i) \, .
\end{equation}
Note that, as a consequence of this SU(3)$^5$ symmetric flavour structure,
when comparing with constraints obtained
in EFT fits based on more general flavour specific operators, such as those that single
out the top quark, the value of our coefficient will be the average of the flavour-dependent
coefficients in that analysis.

\paragraph{Purely bosonic operators.}
To begin, we define the purely bosonic operators that modify the gauge structure of the theory
as compared to the SM.
In Table~\ref{tab:gauge} we list the dimension-six operators constructed
from bosonic fields that
modify the interactions of the electroweak gauge bosons and which are considered in this work.
For each operator, we indicate its definition in terms of the SM
fields and also the notation conventions adopted
both for the operator and for the Wilson coefficient.
Note that, as mentioned above, we consider both CP-even and CP-odd operators.

The only CP-even modifications of the triple
and quartic gauge couplings arise from $\op_{W}$.
In addition, we account for
possible CP-odd contributions to the aTGC and aQGC from the
$\lbrace \op_{\widetilde{W}} ,  \op_{ \varphi \widetilde{W}} , \op_{ \varphi \widetilde{B}} , \op_{ \varphi \widetilde{W}B} \rbrace $ operators.
The remaining operators in this category
modify the Higgs-gauge ($hVV$ and $hhVV$) vertices.
They appear in the processes either by means of
Higgs decays (through the interference of $gg \to h \to 4 \ell / 2 \ell 2\nu$ with diboson production), or through the $t$-channel Higgs exchange  contributions to the VBS cross-sections.
Furthermore, the operators  $\op_{ \varphi WB}$ and $\op_{\varphi D}$ also enter the definitions of the gauge masses and mixing angle in the SMEFT Lagrangian,
and are hence both dependent of our scheme choice.

\input{tables/table-gauge.tex}

\paragraph{Two-fermion operators.}
Another relevant class of dimension-six operators that modify the interactions of the electroweak gauge bosons
are those composed by two fermion fields and two Higgs fields, where the gauge bosons enter via
the covariant derivative.
These operators describe new contact interactions
involving fermions with gauge and Higgs bosons
which are unrelated to the Yukawa couplings.
They generate corrections to the $V\ell \ell$ and $V q \bar{q}$
vertices and can be constrained, among other processes,
from the electroweak precision observables (EWPOs) measured by LEP~\cite{ALEPH:2005ab}.
They also generate contact interactions of the form
$h V f \bar{f}$ which affect specific Higgs boson production and decay processes.
The two-fermion operators that will be considered in this
work are listed in Table~\ref{tab:2fermion}, and consist of seven  CP-even operators containing each
two Higgs doublets, a covariant derivative, and two fermionic fields.
In the definition of these operators, we have introduced
\be
\overleftrightarrow{D}_\mu \equiv ( D_\mu + \overleftarrow{D}_\mu )\, ,
\ee
which is required to ensure that operators with fermionic neutral currents are Hermitian.
All the operators listed in Table~\ref{tab:2fermion} are CP-even.

\input{tables/table-2fermion.tex}

\paragraph{Dipole operators.}
These operators involve the direct interactions between gauge bosons and fermions, rather
than the indirect ones that proceed via the covariant derivative such as the operators listed in Table~\ref{tab:2fermion}.
They have a special Lorentz structure connecting same-helicity fermions. In general, they
do not interfere with the SM, except for a few cases where the light Yukawa couplings are
taken to be nonzero. Since our analysis is restricted to the $\mathcal{O}\lp \Lambda^{-2}\rp$ corrections
to the VBS and diboson cross-sections and we neglect quark masses, we
do not need to consider these operators here.

\paragraph{Parameter shifts and EWPOs.}
Some of the dimension-six SMEFT operators generate a contribution to the relevant electroweak parameters,
\be
 m_Z, m_W, G_F, \sin^2 \theta_W , \alpha_{\rm EW} \, ,
\ee
and depending on which input parameter scheme (IPS) one adopts, the expressions for $\lbrace g_1 , g_2 , v \rbrace $, and hence
for the resulting SM Lagrangian and Feynman rules, will be different.
The operators affecting these electroweak input parameters are closely connected with
the EWPOs
and are thus significantly constrained by the former.
In particular, the  $c_{\varphi l}^{(3)}$ and $c_{l l}$ coefficients modify
the definition of Fermi's constant $G_F$, while $c_{\varphi W B}$ and $ c_{\varphi D}$ enter the $Z$ mass and mixing angle.
They can be well constrained through the measurement of the muon lifetime and of the EW oblique  parameter respectively~\cite{Alonso:2013hga}:  $c_{\varphi W B}$ affects directly the value of the $S$ parameter, also known as $\rho$~\cite{Ross:1975fq},
whereas $c_{\varphi D}$ contributes to the $T$ parameter.

Several BSM and EFT fits of these EWPO have been performed in recent years~\cite{Ciuchini2013,deBlas:2016ojx,Ciuchini:2014dea}, and furthermore various LHC analyses tackle the extraction of the same EWPOs from LHC data~\cite{A.Savin:2018bmz,Aaboud:2017svj,Khachatryan:2016yte,Aad:2015uau,Aaij:2015lka,Chatrchyan:2011ya}, mostly relying on Drell-Yan production
and related processes.
Here we choose not to account for these constraints in our study, and constrain the coefficients
of the operators listed in Tables~\ref{tab:gauge} and~\ref{tab:2fermion} solely from the VBS
and diboson measurements.
In the future, once the VBS measurements are integrated in the global EFT analysis,
one will be able to constrain these electroweak parameter shifts
by including both the LEP's EWPOs, the LHC Drell-Yan data directly~\cite{Farina:2016rws,Franceschini:2017xkh,Alioli:2017nzr,Dawson:2018dxp,Ricci:2020xre,Alioli:2018ljm}, and all other measurements ({\it e.g.} Higgs production) sensitive to them.

\paragraph{Overview of fitted degrees of freedom.}
We summarise in Table~\ref{tab:table-operatorbasis}
the degrees of freedom  considered in the present work,  categorised
into purely bosonic  and two-fermion
operators.
We also indicate the notation that will be used in some of the plots
and tables of the following sections.
We end up with $n_{\rm op}=16$ independent coefficients,
of which 9 are purely bosonic and 7 are two-fermion
operators.
Of the purely bosonic operators, 5 are CP-even and 4 are CP-odd.
Recall that we use symmetric flavour assumptions and thus the operators involving quarks or leptons
are summed over the three SM generations.

\input{tables/table-operatorbasis.tex}


\paragraph{Amplitudes and cross-sections.}
The  dimension-six operators that compose the SMEFT Lagrangian Eq.~(\ref{eq:SMEFTlag})
modify a generic SM cross-section to be,
\begin{equation}
    \label{eq:crosssection}
    \sigma_{\rm SMEFT} = \sigma_{\rm SM} + \sum_i^{n_{\rm op}} \frac{c_i}{\Lambda^2}\sigma^{(\rm eft)}_i + \sum_{i,j}^{n_{\rm op}} \frac{c_i c_j}{\Lambda^4} \widetilde{\sigma}^{(\rm eft)}_{ij} \, ,
\end{equation}
where $\sigma_{\rm SM}$ indicates the SM prediction and the Wilson coefficients are assumed to be real.
The $\mathcal{O}(\Lambda^{-2})$ terms arise from EFT operators interfering with the SM amplitude
and in most cases  correspond to the dominant correction.
For this reason, the cross-sections $\sigma^{(\rm eft)}_i$ are usually denoted
as the SMEFT linear interference terms.

The third term in the RHS of Eq.~(\ref{eq:crosssection}) contains the quadratic contribution
arising from the square of the amplitudes involving dimension-six operators,
and scales as $\mathcal{O}(\Lambda^{-4})$.
These quadratic terms are of the same order of the dimension-eight operators that interfere
with the SM amplitudes and that modify the TGCs and QGCs.
Given that we consider here only dimension-six operators,
 the consistent inclusion
of $\mathcal{O}(\Lambda^{-4})$ corrections to VBS processes is left
for future work and we restrict ourselves to the linear approximation.

We note that linear EFT interference effects due to CP-odd operators remain CP-odd,
while squared CP-odd terms become CP-even and thus are difficult to disentangle
from their CP-even counterparts~\cite{Englert:2019xhk}.
For this reason, it is interesting to study CP-odd operators in processes
for which the linear EFT terms are dominant, such as the high energy bins of differential distributions, or by looking at specific observables such as asymmetries.
Separating the impact of CP-even and CP-odd operators has been studied
mostly in the context of EFT analysis of the Higgs sector~\cite{Dolan:2014upa,Cirigliano:2019vfc,Englert:2019xhk,Bernlochner:2018opw,Biekotter:2020flu}.

The SMEFT is defined to be valid for energies satisfying $E \ll \Lambda$.
A lower bound on the value of $\Lambda$ is given by the highest energy scale of the data included in our fit,
which as discussed in Sect.~\ref{sec:expdata} turns out to be around $E\simeq 3$ TeV.
An upper bound on $\Lambda$ cannot be set from first principles and requires
the observation of a hypothetical heavy resonance.
In the rest of this paper, we will assume for simplicity $\Lambda=1$ TeV,
with the caveat that results for any other
values of $\Lambda$ can be obtained by a trivial re-scaling.

\paragraph{Interplay between VBS and diboson production.}
Gauge boson pair production has been extensively studied as a precision probe
of the electroweak sector of the SM and its various extensions,
first in the context of precision SM electroweak tests at LEP
and more recently in the  EFT framework and accounting for the corresponding
LHC measurements~\cite{Grojean:2018dqj,Falkowski:2016cxu,Rahaman:2019mnz,Baglio:2017bfe,Baglio:2020oqu,Baglio:2019uty}.
Since diboson production is a relatively clean process with large cross-sections~\cite{Campbell:2011bn}, fiducial cross-sections and differential distributions have been measured
with high precision by ATLAS and CMS.

Most of the dimension-six operators listed in
Table~\ref{tab:table-operatorbasis} modify also the theoretical calculation
of diboson cross-sections, and thus it would seem that VBS  data might be
redundant for EFT studies.
While indeed dimension-six EFT effects can be well constrained by diboson
production at the LHC~\cite{Grojean:2018dqj,Azatov:2017kzw}, here we will show
that VBS measurement provide non-trivial, complementary information for many
of these operators.
Furthermore, the role of VBS measurememnts is only bound to increase as more data is accumulated,
in particular at the HL-LHC.

\begin{figure}[t]
  \centering
      \includegraphics[width=1.\textwidth]{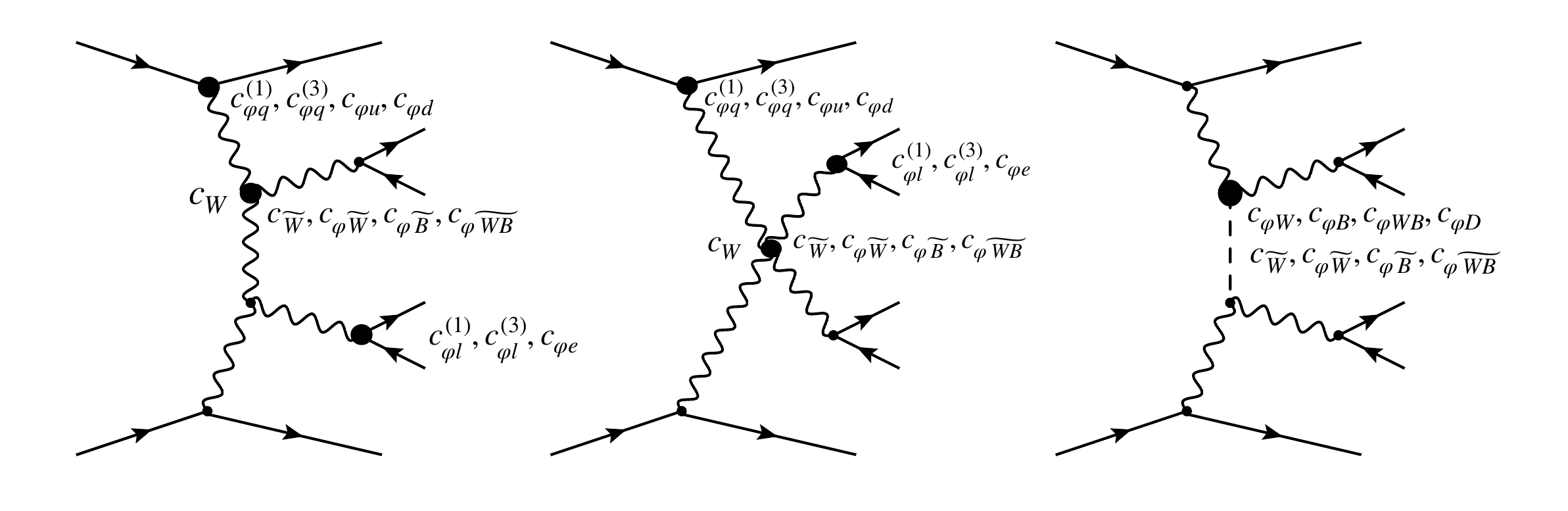}
      \caption{\small EFT corrections modifying the quartic (left panel) and triple
        (middle panel) gauge couplings in vector-boson scattering,
        as well as the  the $t$-channel Higgs exchange contribution
        (right panel) and the $V f \bar{f}$ interaction vertices.
        In this work we consider only
        final states where the gauge bosons decay leptonically.
      }
  \label{fig:feynman_EFT1}
\end{figure}

In VBS, only one CP-even
operator in the Warsaw basis affects directly the triple and quartic gauge couplings,
 with three more operators contributing once CP-odd effects are
allowed.
Beyond these modifications of the TGCs and QGCs, the VBS process is also sensitive
to several other dimension-six operators, given the large amount of vertices and
topologies contributing to the definition of the its final state.
This is illustrated in Fig.~\ref{fig:feynman_EFT1},
where we show representative diagrams for
EFT corrections to quartic and triple
gauge couplings 
as well as the  the $t$-channel Higgs exchange contribution.

\begin{figure}[t]
  \centering
    \subfloat{
     \includegraphics[width=0.44\textwidth]{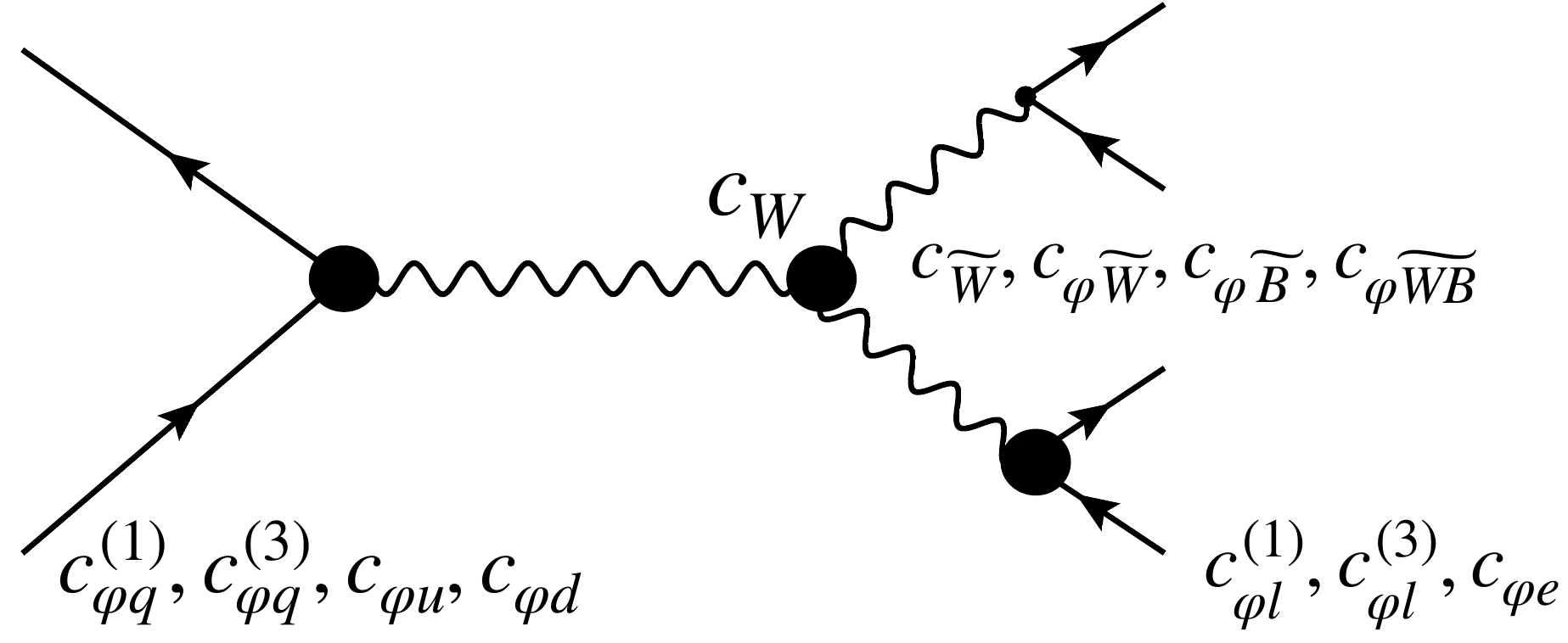}}
     \hspace{0.7cm}
    \subfloat{
     \includegraphics[width=0.44\textwidth]{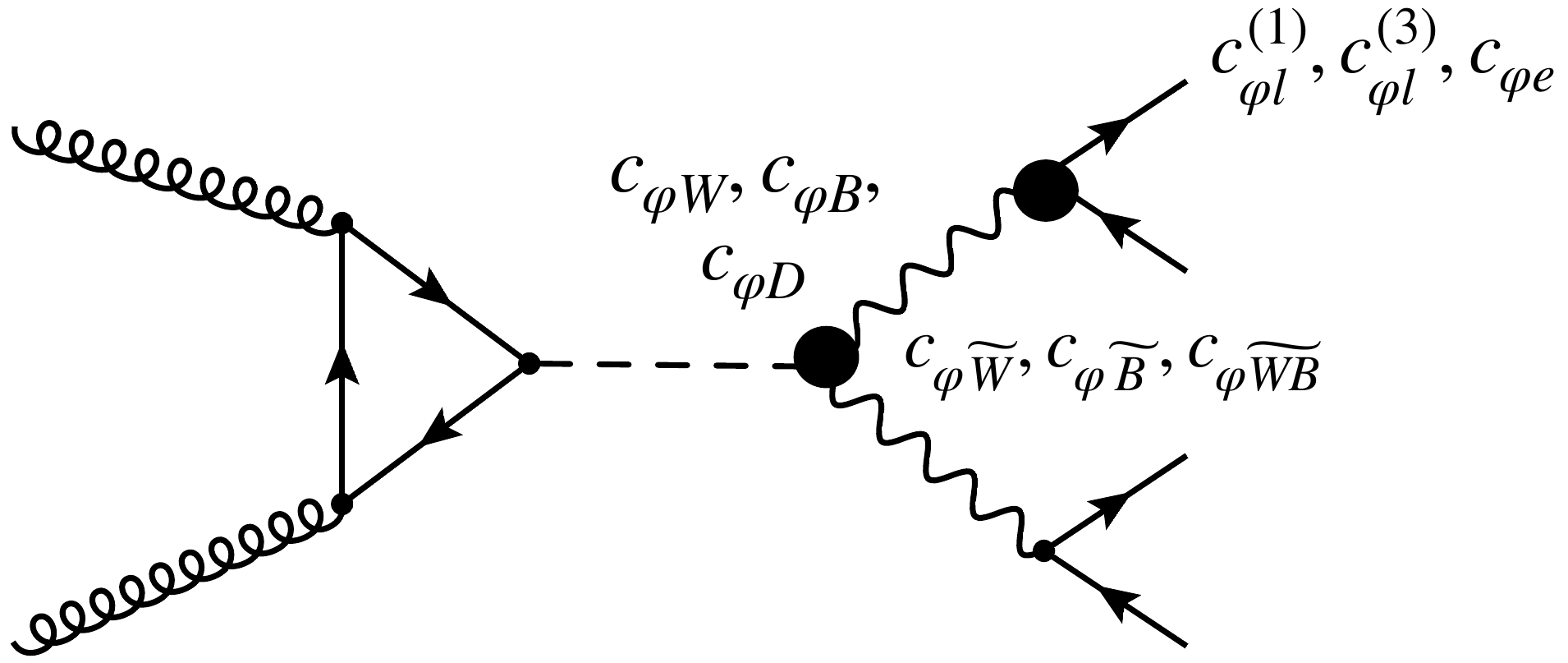}}
    \caption{\small Same as Fig.~\ref{fig:feynman_EFT1} for two representative
      EFT diagrams contributing
      to diboson production: a pure diboson diagram (left)
      and another for which diboson production interferes with the $h\to VV$
      process (right).}
  \label{fig:feynman_EFT2}
\end{figure}

In the case of $WW$ diboson production at LEP, the
process is sensitive to the triple gauge couplings $ZWW$ and $\gamma WW$
at leading order in the EFT expansion,
and thus the corresponding EFT parametrisation will include the modification of the TGC (through $c_W$).
It will also modify the $e \bar{e} Z$ vertex
and the corresponding IPS dependence, which could include $c_{\varphi WB},c_{\varphi D}$ and $c_{\varphi l}^{(3)}$,
and even some contact term of the form $e \bar{e} W W$, generally not interfering with the SM.
Similar considerations apply for diboson production at hadron colliders, although now
a new feature appears, namely the interference with Higgs production in gluon
fusion followed by the $h \to VV$ decay.
This correction induces a non-negligible sensitivity to
the $c_{\varphi B}$ and $c_{\varphi W}$
coefficients in gauge boson pair production at the LHC.
These features are illustrated in Fig.~\ref{fig:feynman_EFT2}.

%% file: tables/table-gauge.tex
\begin{table}[t]
  \begin{center}
    \renewcommand{\arraystretch}{1.75}
    \begin{tabular}{llll}
      \toprule
CP properties    &  Operator $\qquad$ & Coefficient $\qquad\qquad\qquad$ & Definition \\
      \midrule
\multirow{6}{*}{CP-even} &     $\op_W$ & $c_{W}$  & $\epsilon^{IJK}W^{I \nu}_{\mu} W^{J \rho}_{\nu}W^{K \mu}_{\rho}$
      \\ \cmidrule(lr{0.7em}){2-4}
    &  $\op_{\varphi W} $  & $ c_{\varphi W}$  & $ (\varphi^{\dagger} \varphi - \frac{v^2}{2}) W^{I}_{\mu\nu} W^{I \mu \nu}$\\ \cmidrule(lr{0.7em}){2-4}
    &  $\op_{\varphi B}$    &$c_{\varphi B}$ & $ (\varphi^{\dagger} \varphi - \frac{v^2}{2}) B_{\mu\nu} B^{\mu \nu}$
      \\ \cmidrule(lr{0.7em}){2-4}
    &  $ \op_{\varphi WB}$  & $c_{\varphi WB}$ &
      $ \lp \varphi^{\dagger} \sigma_{I} \varphi\rp W^I_{\mu\nu} B^{\mu \nu}$
      \\ \cmidrule(lr{0.7em}){2-4}
   &   $\op_{\varphi D} $   &	$c_{\varphi D}$ &
      $ (\varphi^{\dagger} D^{\mu} \varphi)^*(\varphi^{\dagger} D_{\mu} \varphi)$ \\
      \midrule
\multirow{5}{*}{CP-odd}    &  $\op_{\widetilde{W}}$ & $c_{\widetilde{W}}$ &
      $\epsilon^{IJK}\widetilde{W}^{I \nu}_{\mu} W^{J \rho}_{\nu}W^{K \mu}_{\rho}$ \\ \cmidrule(lr{0.7em}){2-4}
     & $\op_{\varphi\widetilde{ W}} $  & $ c_{\varphi \widetilde{W}}$  & $
      (\varphi^{\dagger} \varphi) \widetilde{W}^{I}_{\mu\nu} W^{I \mu \nu}$\\ \cmidrule(lr{0.7em}){2-4}
     & $ \op_{\varphi \widetilde{ W}B}$  & $c_{\varphi \widetilde{W}B}$ &
      $ \lp \varphi^{\dagger} \sigma_{I} \varphi\rp \widetilde{W}^I_{\mu\nu} B^{\mu \nu}$
      \\ \cmidrule(lr{0.7em}){2-4}
    &  $\op_{\varphi \widetilde{B}}$    &$c_{\varphi \widetilde{B}}$ &
      $ (\varphi^{\dagger} \varphi) \widetilde{B}_{\mu\nu} B^{\mu \nu}$      \\
      \bottomrule
    \end{tabular}
    \caption{Dimension-six purely bosonic operators that
      modify the interactions of the electroweak gauge bosons.
      For each operator, we indicate its definition in terms of the SM
      fields,
      and the notational conventions that will be used
      both for the operator and for the Wilson coefficient.
      \label{tab:gauge}}
  \end{center}
\end{table}

%% file: tables/table-2fermion.tex
\begin{table}[t]
  \begin{center}
    \renewcommand{\arraystretch}{1.75}
    \begin{tabular}{lll}
      \toprule
  Operator $\qquad$ & Coefficient $\qquad\qquad\qquad$ & Definition \\
  \midrule
 $\op_{\varphi l}^{(1)}$ &    	$c_{\varphi l}^{(1)}$ & $ \sum\limits_{\sss j}i ( \varphi^{\dagger}\overleftrightarrow{D}_{\mu}\varphi )(\Bar{l}_j \gamma^\mu l_j)$ \\
  $\op_{\varphi l}^{(3)}$  &
  $c_{\varphi l}^{(3)}$ &
  $\sum\limits_{\sss j} i ( \varphi^{\dagger}\overleftrightarrow{D}^I_{\mu}\varphi )(\Bar{l}_j \sigma^I \gamma^\mu l_j)$ \\
  $\op_{\varphi q}^{(1)}$&		$c_{\varphi q}^{(1)}$ &
  $ \sum\limits_{\sss j}i ( \varphi^{\dagger}\overleftrightarrow{D}_{\mu}\varphi )(\Bar{q}_j \gamma^\mu q_j)$ \\
  $\op_{\varphi q}^{(3)}$ &	$c_{\varphi q}^{(3)}$	 &
  $\sum\limits_{\sss j} i ( \varphi^{\dagger}\overleftrightarrow{D}^I_{\mu}\varphi )(\Bar{q}_j \sigma^i \gamma^\mu q_j)$ \\
  \hline
  $\op_{\varphi e}$&
  $c_{\varphi e}$ & $ \sum\limits_{\sss j}i ( \varphi^{\dagger}\overleftrightarrow{D}_{\mu}\varphi )(\Bar{e}_j \gamma^\mu e_j)$ \\
  $\op_{\varphi u}$ &$c_{\varphi u}$	 &
  $ \sum\limits_{\sss j}i ( \varphi^{\dagger}\overleftrightarrow{D}_{\mu}\varphi )(\Bar{u}_j \gamma^\mu u_j)$ \\
  $\op_{\varphi d}$ &$c_{\varphi d}$	 &
  $\sum\limits_{\sss j} i ( \varphi^{\dagger}\overleftrightarrow{D}_{\mu}\varphi )(\Bar{d}_j \gamma^\mu d_j)$ \\
      \bottomrule
    \end{tabular}
    \caption{Dimension-six  operators that
      modify the interactions of the electroweak gauge bosons
      and that are composed by two fermion fields, two Higgs fields, and one covariant derivative.
      The sum over the  index $j$ runs over the three SM generations,
      which are treated symmetrically in this study.
      \label{tab:2fermion}}
  \end{center}
\end{table}

%% file: tables/table-operatorbasis.tex

\begin{table}[t]
  \begin{center}
    \renewcommand{\arraystretch}{1.80}
        \begin{tabular}{ccc}
          \toprule
  Class  &   $\qquad\qquad$ DoF$\qquad\qquad$   & $\qquad \qquad$ Notation $\qquad\qquad$ \\
  \midrule
       \multirow{1}{*}{Purely bosonic}      &  $c_{\varphi B}$, $c_{\varphi W}$, $c_{\varphi W B}$
                                            &   {\tt cpB}, {\tt cpW}, {\tt cpWB} \\
   \multirow{1}{*}{(CP-even)}     &    $c_{\varphi D}$, $c_{WWW}$
       &    {\tt cpD}, {\tt   cWWW} \\

  \midrule
       \multirow{1}{*}{Purely bosonic}      &  $c_{\widetilde{W}}$, $c_{\varphi \widetilde{W}}$
                                            &   {\tt   cWWWtil}, {\tt cpWtil} \\
   \multirow{1}{*}{(CP-odd)}     &    $c_{\varphi \widetilde{B}}$, $c_{\varphi \widetilde{W} B}$
       &     {\tt   cpBtil}, {\tt cpWBtil} \\
  \midrule
  \multirow{1}{*}{two-fermion}      & $c_{\varphi l}^{(1)}$, $c_{\varphi l}^{(3)}$,
  	$c_{\varphi q}^{(1)}$  & {\tt cpl}, {\tt c3pl}, {\tt cpq }\\
            \multirow{1}{*}{(+ bosonic fields)}&    $c_{\varphi q}^{(3)}$, $c_{\varphi u}$, $c_{\varphi d}$, $c_{\varphi e}$ &  {\tt c3pq}, {\tt cpu}, {\tt cpd}, {\tt cpe}  \\
            \bottomrule
  \end{tabular}
        \caption{Overview of the degrees of freedom considered in this analysis,
          separated into purely bosonic (upper) and two-fermion (lower part)
          operators.
          We also indicate the notation that will be used in some of the plots
          and tables of the following sections.
          See Tables~\ref{tab:gauge} and~\ref{tab:2fermion} for their
          definitions in terms of the SM fields.
\label{tab:table-operatorbasis}}
\end{center}
\end{table}

%% file: sec-expdata.tex
\section{Experimental data and theoretical calculations}
\label{sec:expdata}

In this section we describe the experimental data sets that will be used
in the present analysis as well as the corresponding theoretical predictions
both in the SM and at the EFT level.
We also quantify the sensitivity that each of the VBS and diboson data have on the coefficients
associated to the dimension-six operators introduced in Sect.~\ref{sec:eftth}.

\subsection{Vector boson scattering}
\label{sec:VBSproduction}

\begin{figure}[t]
  \centering
    \subfloat{
     \includegraphics[width=0.22\textwidth]{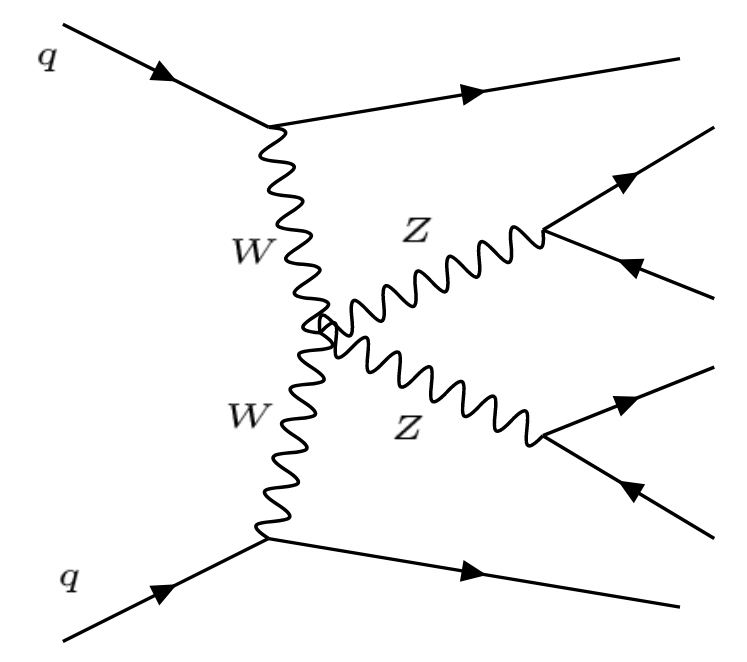}}
     \hspace{0.8cm}
     \subfloat{
     \includegraphics[width=0.22\textwidth]{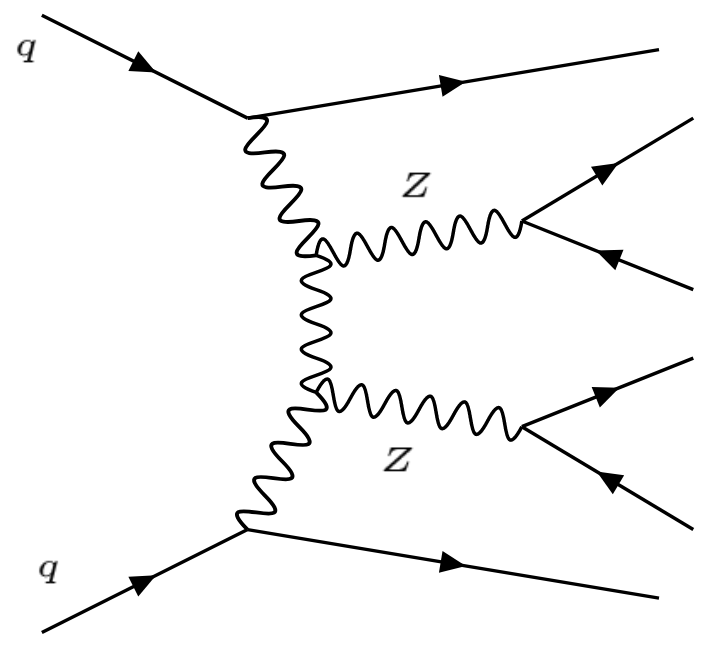}}
     \hspace{0.8cm}
     \subfloat{
     \includegraphics[width=0.22\textwidth]{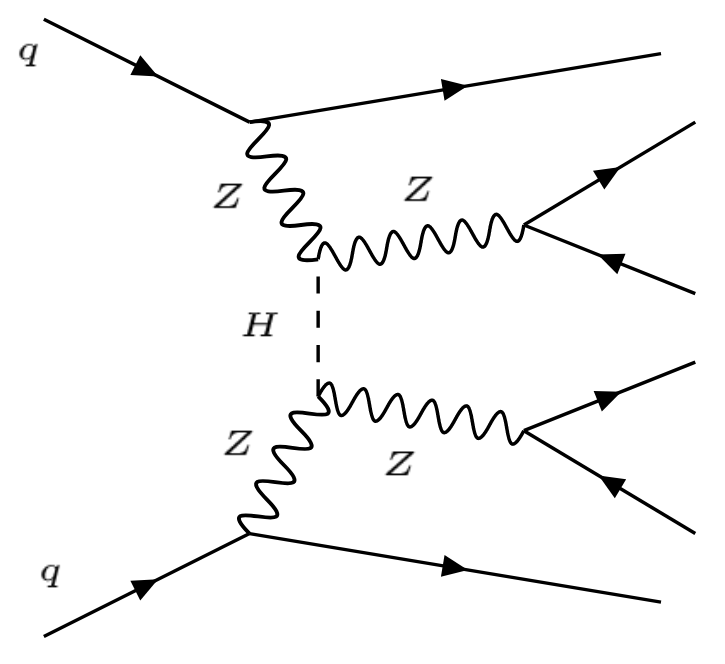}}

     \subfloat{
     \includegraphics[width=0.25\textwidth]{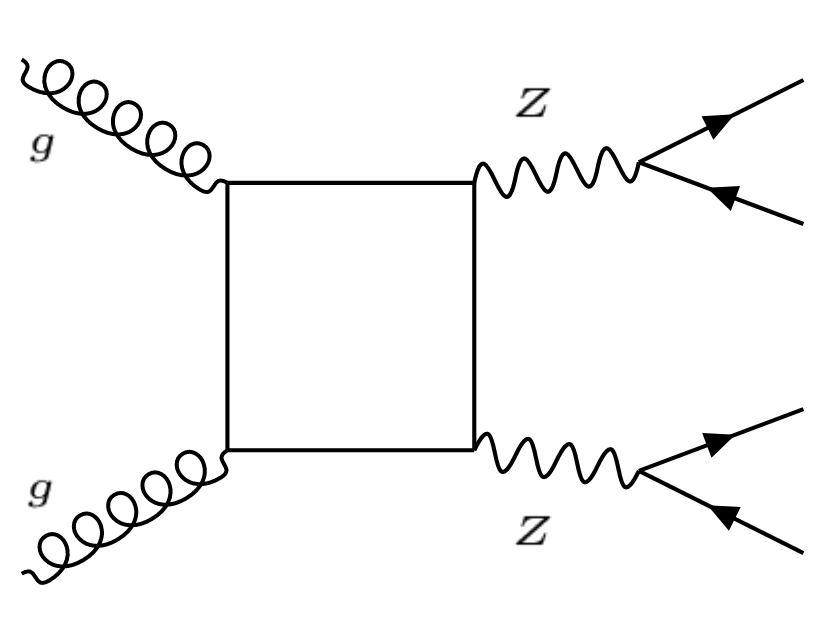}}
     \hspace{1.2cm}
     \subfloat{
     \includegraphics[width=0.2\textwidth]{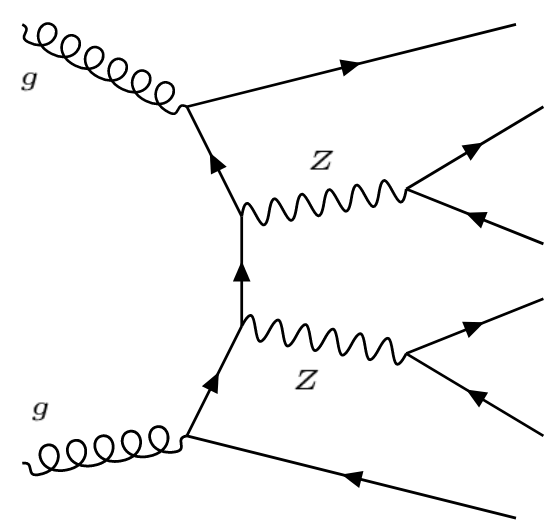}
     }
  \caption{\small Representative Feynman diagrams for vector boson scattering in the $ZZjj$ final state (top row)
    and its main background, the QCD-induced diboson production (bottom row).
  }
  \label{fig:feynman_ewVBS_ZZ}
\end{figure}

At hadron colliders, vector boson scattering occurs when two vector bosons
are radiated off incoming quark lines and scatter into another pair
of vector bosons, $VV'\to VV'$.
The latter  decay either leptonically or hadronically,
and thus the VBS amplitude will be proportional to $\alpha_{\rm EW}^6$.
Fig.~\ref{fig:feynman_ewVBS_ZZ} displays representative Feynman diagrams
associated to vector boson scattering at the LHC for the $ZZjj$ channel.
The sensitivity to quartic gauge couplings is a unique feature of this process,
and in particular the longitudinally polarised scattering amplitude $V_LV'_L\to V_LV_L'$ provides
a direct probe of the high-energy behaviour of the theory.
We emphasize again that QGCs represent only a fraction of the VBS events, and thus a complete
description of the process requires accounting for EFT effects in all possible topologies,
as discussed in Sect.~\ref{sec:eftth}.

The characteristic VBS topology is defined by two energetic jets with moderate transverse momenta,
$p_T\sim M_V/2$, which therefore are produced relatively close to the beam pipe and
appear predominantly in the forward region of the detectors.
The specific final-state signature that we will focus on in this work
is thus composed by four leptons (either charged or neutral) and two jets
in the forward region exhibiting a large invariant mass $m_{jj}$
and wide rapidity separation $\Delta y_{jj}$.
Furthermore, being a purely electroweak process, there is no color flow between the two incoming quark lines.
This implies that the central rapidity region between the two tagging jets will have a reduced amount of hadronic activity, known as the ``rapidity gap''.

As highlighted by the bottom diagrams of  Fig.~\ref{fig:feynman_ewVBS_ZZ},
the vector boson scattering process is affected
by large backgrounds from QCD-induced
diboson production processes with similar topology, with amplitudes proportional instead
to $\alpha_{\rm EW}^4 \alpha_s^2$.
The interference terms between the diboson and VBS processes are usually small and therefore
will be neglected in this analysis.
Beyond diboson production, other sources of background to VBS include  $t+V$, $t\bar{t}$, $V$+jets and QCD multijet production and are generally small.
While the diboson inclusive cross-section is much larger than the VBS one, provided the statistics are large enough,
one can efficiently disentangle the two processes by focusing
on the large $m_{jj}$ and $\Delta y_{jj}$ region (or related kinematic variables)
where the VBS processes dominates.

\paragraph{Theoretical simulations.}
In order to evaluate the expected cross-sections and differential
distribution for the VBS (and the diboson) processes, we use
two Monte Carlo generators, \mg~\cite{Alwall:2014hca} and \pwg~\cite{Nason:2004rx, Frixione:2007vw, Alioli:2010xd}, to generate NLO QCD matrix elements.
QCD corrections represent up to a $\mathcal{O}(100\%)$ effect
for diboson processes~\cite{Grazzini:2017ckn, WZ:atlas, WZ:cms},
while in VBS, a purely electroweak process,
they amount to a few percent~\cite{Denner:2019tmn,Bozzi:2007ur,Biedermann:2017bss,Chiesa:2019ulk,Denner:2020bcz,Denner:2020zit,Denner:2012dz}, although they can modify the shape of distributions.
Here we adopt the NNPDF3.1NNLO no-top PDF set~\cite{Ball:2017nwa}.

The fixed-order NLO events  are then showered with
\texttt{Pythia8}~\cite{Hoche:2014rga, Sjostrand:2007gs, Sjostrand:2006za}.
Accounting for parton shower effects is especially relevant for the modelling of additional soft QCD radiation in diboson production. It is also convenient to facilitate the matching between the  theoretical predictions with the  experimental
analyses.
However, since we restrict ourselves to fully leptonic final states,
both hadronisation, underlying event, and multiple parton interactions are switched off in
the \texttt{Pythia8} simulation.
The showered events are further processed with {\tt Rivet}~\cite{Bierlich:2019rhm},
a crucial step to reproduce the experimental selection requirements
and acceptance cuts, given that only a subset of these can be implemented at the generation level. Moreover, this allows us to compare directly with the datasets published in {\tt HEPData}~\cite{Maguire:2017ypu}.

Bottom quarks are always included in the initial state ($n_f=5$ scheme) and sometimes also
in the definition of the final state jets, following the prescription in the
associated experimental analysis.

The signal to background ratio in VBS is generally small, and for this reason
most VBS differential results are only available as a sum of EW- and QCD-induced processes, which can only be disentangled at the level of fiducial cross-sections.
To account for this, in the simulation of VBS processes
we generate MC events corresponding to both
the EW-induced contributions (signal) and the QCD-induced contributions (background),
with EFT corrections included only in the former\footnote{Note that EFT corrections to some of these backgrounds
  are already being constrained by the diboson production measurements included here.}.

The evaluation of the linear EFT cross-sections, $\sigma^{(\rm eft)}_i$ in  Eq.~(\ref{eq:crosssection}),
 is carried out with \mg $\:$ interfaced with
{\tt SMEFTsim}~\cite{Brivio:2017btx}, in its $\lbrace m_W , m_Z , G_F \rbrace$ IPS implementation.
Specifically, we compute the linear EFT cross-sections at LO in the SMEFT, and then calculate an NLO/LO $K$-factor assuming that the QCD corrections to the SM cross-sections factorise such that they can be assumed to be the same in the EFT.
Nevertheless, we found the impact of this assumption to be rather small at the level of our fit results.
In future work, it would be advisable to use exact NLO QCD calculations for the EFT
cross-sections, such as the ones presented in~\cite{Baglio:2017bfe,Baglio:2020oqu,Baglio:2019uty}, or by using for example {\tt SMEFT@NLO}~\cite{Degrande:2020evl}.

In Table~\ref{tab:calc_details} we summarize the settings of the SM and EFT theoretical calculations
used to evaluate the LHC VBS and diboson cross-sections included in the fit.
The perturbative accuracy and the codes used
to produce the corresponding predictions for both the SM and the EFT contributions are also given.

\input{tables/calc_details}

\begin{figure}[t]
    \centering
    \includegraphics[width=0.6\textwidth]{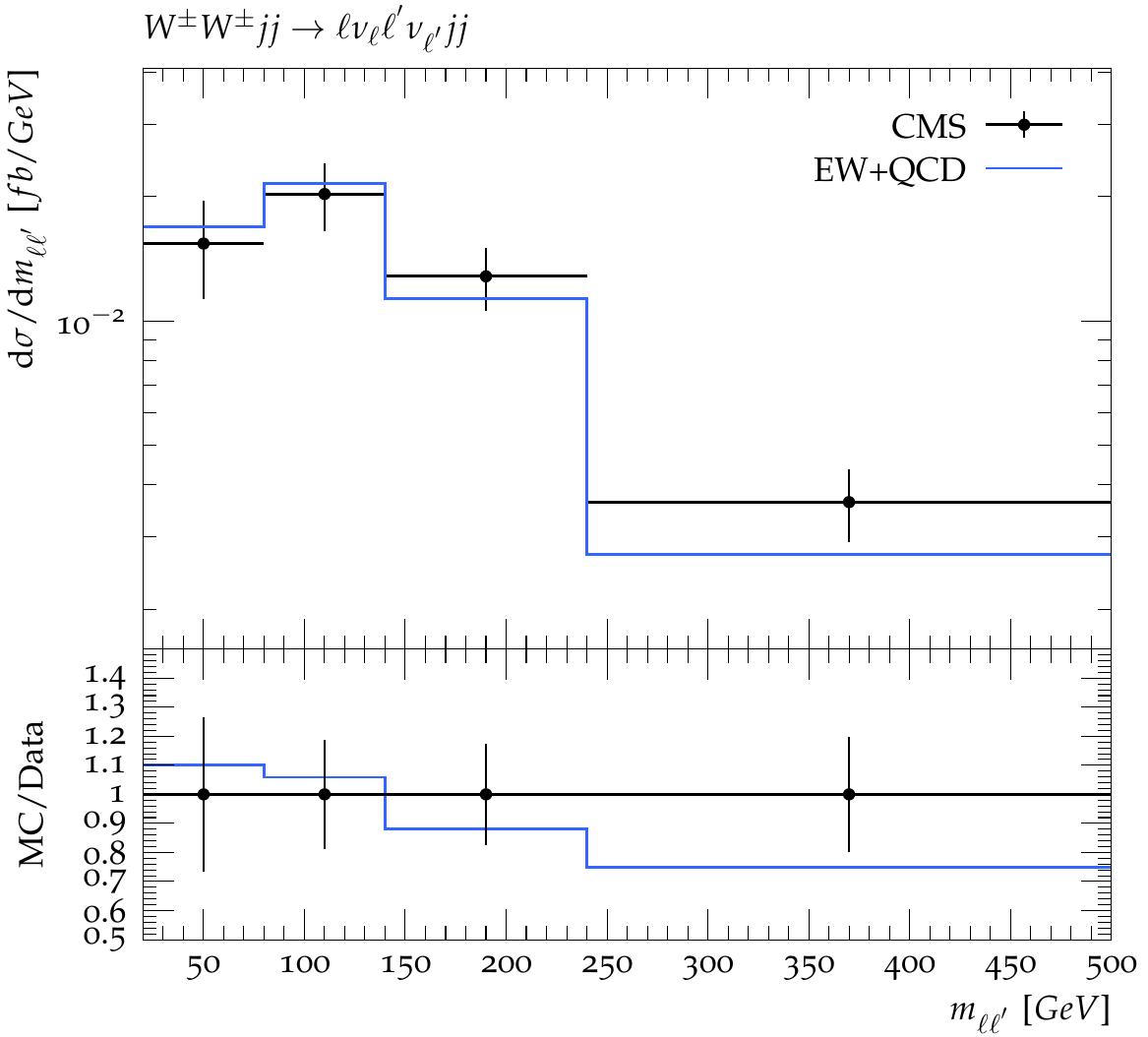}
    \caption{\small The dilepton invariant mass distribution $m_{ll'}$ from the CMS
      measurements of same-sign $W^{\pm}W^{\pm}jj$ production~\cite{WWjjWZjj:cms},
      compared with the corresponding diboson (QCD-induced) plus the VBS (EW-induced) theoretical predictions.
      The  error bar on the CMS data points indicates the total experimental uncertainty.
      In the lower panel, we display the ratio of theory over data.
    }
    \label{fig:WWjjmll}
\end{figure}

\paragraph{Same sign $\mathbold{W^{\pm}W^{\pm}jj}$ production.}
In this category we consider two data sets, one from ATLAS~\cite{WWjj:atlas,WWjj:atlas:hepdata} based on $\mathcal{L}=36$ fb$^{-1}$
and another from CMS based on the full Run II luminosity~\cite{WWjjWZjj:cms, WWjjWZjj:cms:hepdata}, $\mathcal{L}=137$ fb$^{-1}$.
Theoretical predictions are evaluated using \mg~ and then showered with {\tt Pythia8}.
Only the fiducial cross section measurement from ATLAS is used in the fit,  since no differential distributions
are available.
Concerning the CMS measurement, the input to the fit is the differential
distribution in the mass of the charged lepton pair $m_{ll'}$,
which includes the sum of VBS (EW-induced) and diboson (QCD-induced) contributions.
In addition, we include the VBS-only fiducial cross section measurement.
To avoid double counting, we remove one bin of the aforementioned distribution.
Fig.~\ref{fig:WWjjmll} displays the CMS $m_{ll'}$ measurement together with
the corresponding EW+QCD-induced theoretical predictions, finding
good agreement.
These theoretical predictions also agree with
those presented in the original CMS publication~\cite{WWjjWZjj:cms}.

\paragraph{$\mathbold{W^{\pm}Zjj}$ production.}
In this category we include the $m_{T}^{WZ}$
differential distribution from the ATLAS measurement~\cite{WZjj:atlas,WZjj:atlas:hepdata}
based on $\mathcal{L}=36$ fb$^{-1}$,
which consists again on the sum of VBS signal and diboson background.
For this dataset
the full bin-by-bin correlation matrix is available and is accounted for in the fit.
We also include the (signal plus background) differential
distribution in the dijet invariant mass from CMS, $d\sigma / d m_{jj}$,
based on the full Run II dataset luminosity of
$\mathcal{L}=137$ fb$^{-1}$~\cite{WWjjWZjj:cms,WWjjWZjj:cms:hepdata}.
Again, we include the  EW-only fiducial cross-section from CMS in addition to the differential distribution,
and remove a bin from the latter to avoid double counting.
Theoretical predictions for this process are
evaluated at NLO with \pwg $\:$
for the EW component and at LO for the QCD diboson background with \mg.

\begin{figure}[h]
    \centering
    \subfloat{
    \includegraphics[width=0.45\textwidth]{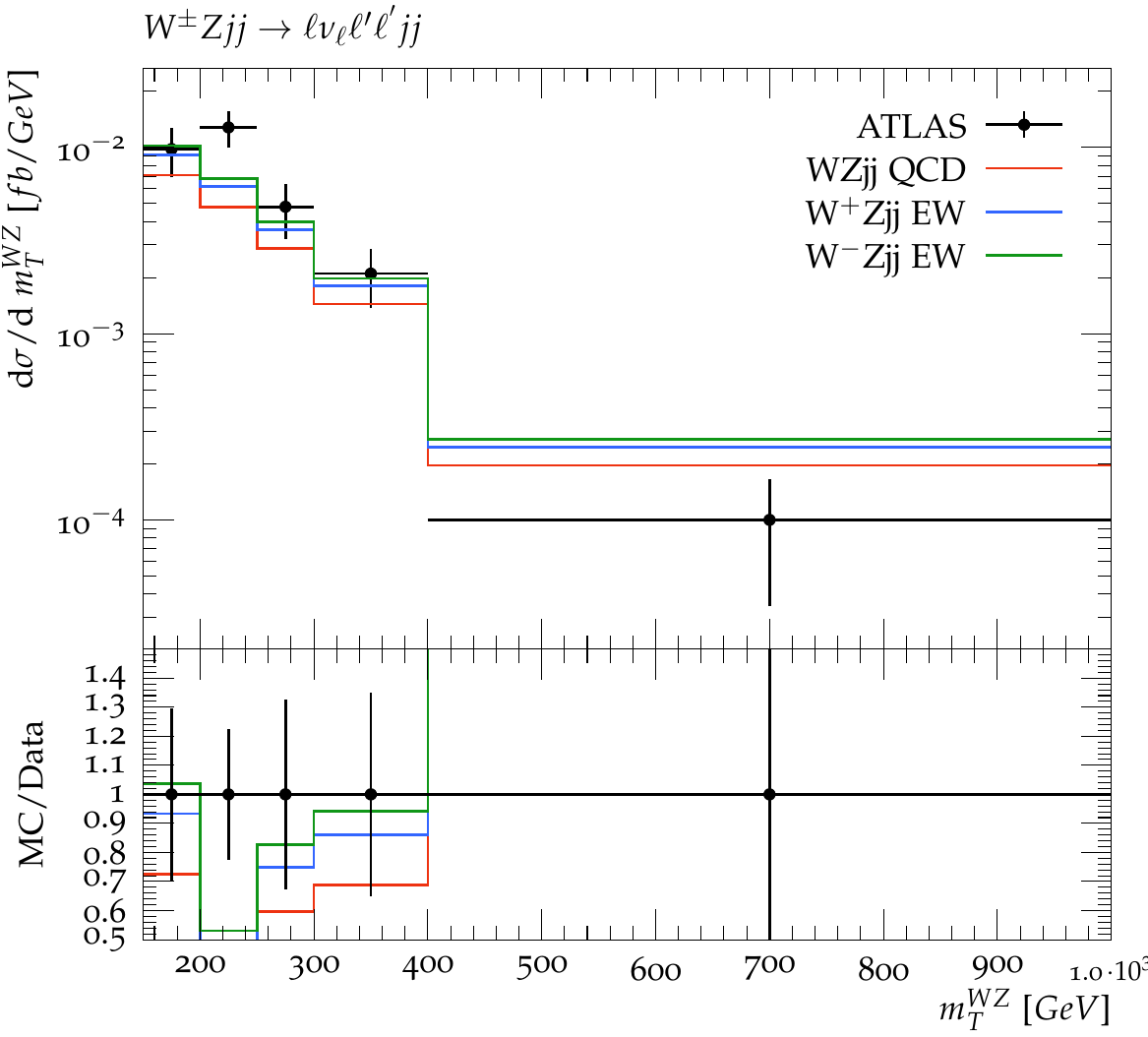}
    }\quad
    \subfloat{
    \includegraphics[width=0.45\textwidth]{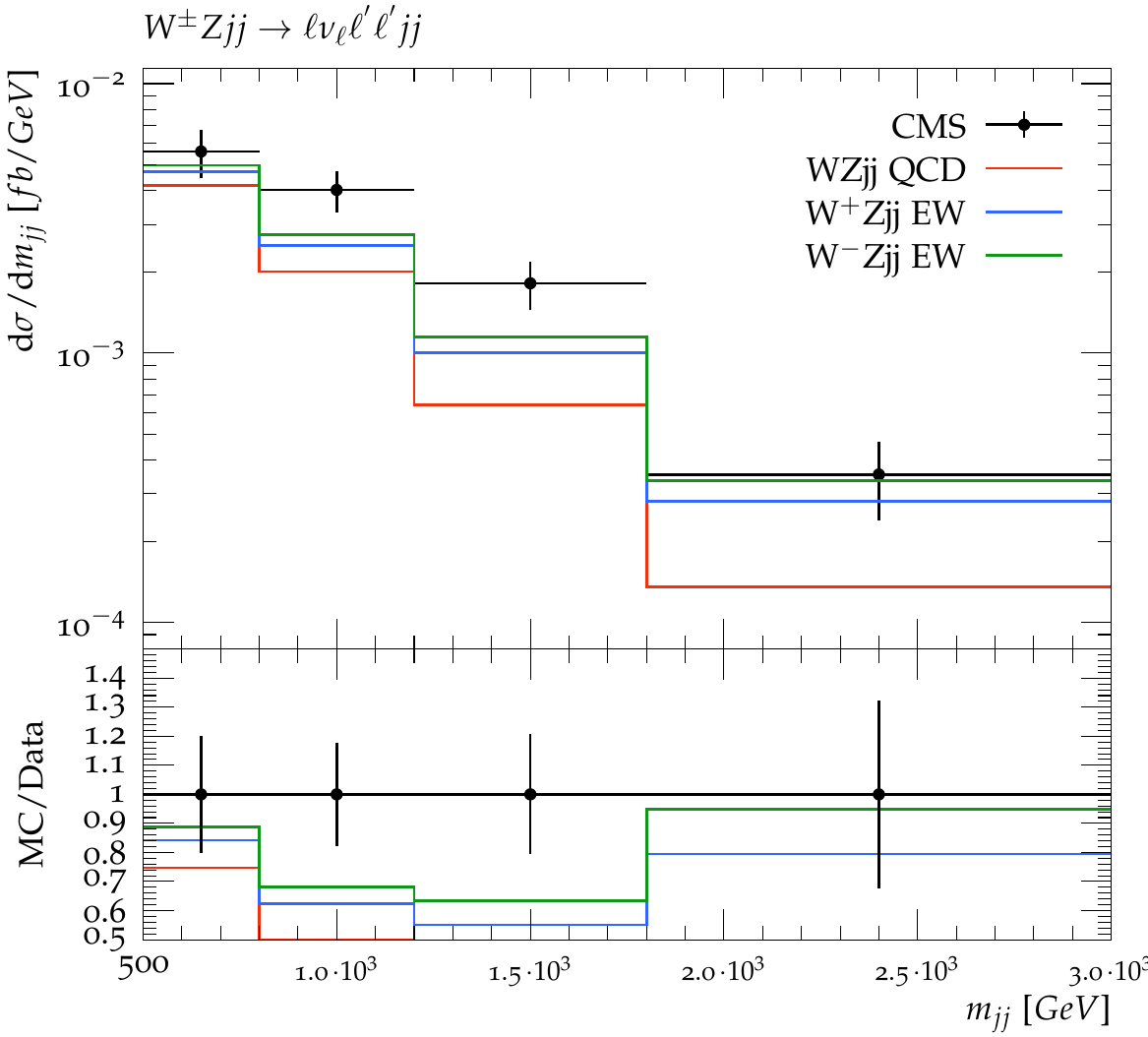}
    }
    \caption{\small  The $W^{\pm}Zjj$
      production measurements from ATLAS~\cite{WZjj:atlas} (left)
      and  CMS~\cite{WWjjWZjj:cms} (right).
      In both cases the EW-induced induced contributions, which are being added to the QCD-induced
      ones, are  separated into $W^{+}Zjj$ and $W^{-}Zjj$.}
    \label{fig:WZjj}
\end{figure}

Fig.~\ref{fig:WZjj} displays a comparison between our theoretical predictions and the $W^{\pm}Zjj$
production measurements from ATLAS~\cite{WZjj:atlas} (absolute $m^{WZ}_T$ distribution)
and from CMS~\cite{WWjjWZjj:cms} (absolute  $m_{jj}$ distribution).
For completeness, the EW-induced induced contributions, which are being added to the QCD ones,
have been separated into $W^{+}Zjj$ and $W^{-}Zjj$.
In the case of the CMS $m_{jj}$ measurement,
there is good agreement between data and theory,
and one can observe how the VBS contribution clearly
dominates over the QCD-induced processes at large dijet
invariant masses $m_{jj}$.
For the ATLAS measurement, we observe some tension on the second bin in $m_{T}^{WZ}$ where the theory undershoots
the data, a behaviour that was also observed in the original analysis~\cite{WZjj:atlas}.
Both the ATLAS and CMS $W^{\pm}Zjj$ measurements
benefit from sensitivity to the high-energy region, covering kinematics of
up to $m_{T}^{WZ}\simeq 1$ TeV for ATLAS and $m_{jj}=3$ TeV for CMS, which
highlights their potential
for constraining EFT operators that modify the VBS process.


\paragraph{$\mathbold{ZZjj}$ production.}
Here we consider two recently released measurements from ATLAS~\cite{ZZjj:atlas,ZZjj:atlas:hepdata} and
CMS~\cite{ZZjj:cms137,ZZjj:cms137:hepdata} based on the full Run II luminosity of $\mathcal{L} \approx 140$ fb$^{-1}$.
The ATLAS analysis represents their first VBS measurement in the $ZZjj$ final state, while the CMS
one updates a previous study of the same final state~\cite{ZZjj:cms}.
In the ATLAS case,  we include the fiducial VBS cross section, which accounts for both EW- and QCD-induced
contributions, while from CMS we include the EW-induced fiducial cross section together
with the detector-level differential distribution in $m_{ZZ}$ for the sum of the
EW and QCD-induced contributions.
Since the latter is not unfolded, it requires some modelling of detector effects.
For this reason, our baseline dataset used in the fit will include only unfolded measurements,
with the detector-level ones used as an additional cross-check\footnote{Moreover, we only include bins
  of the detector-level
  distribution containing more than 30 events to ensure the validity of the
  Gaussian approximation.}.

\begin{figure}[t]
  \centering
  \includegraphics[width=0.6\textwidth]{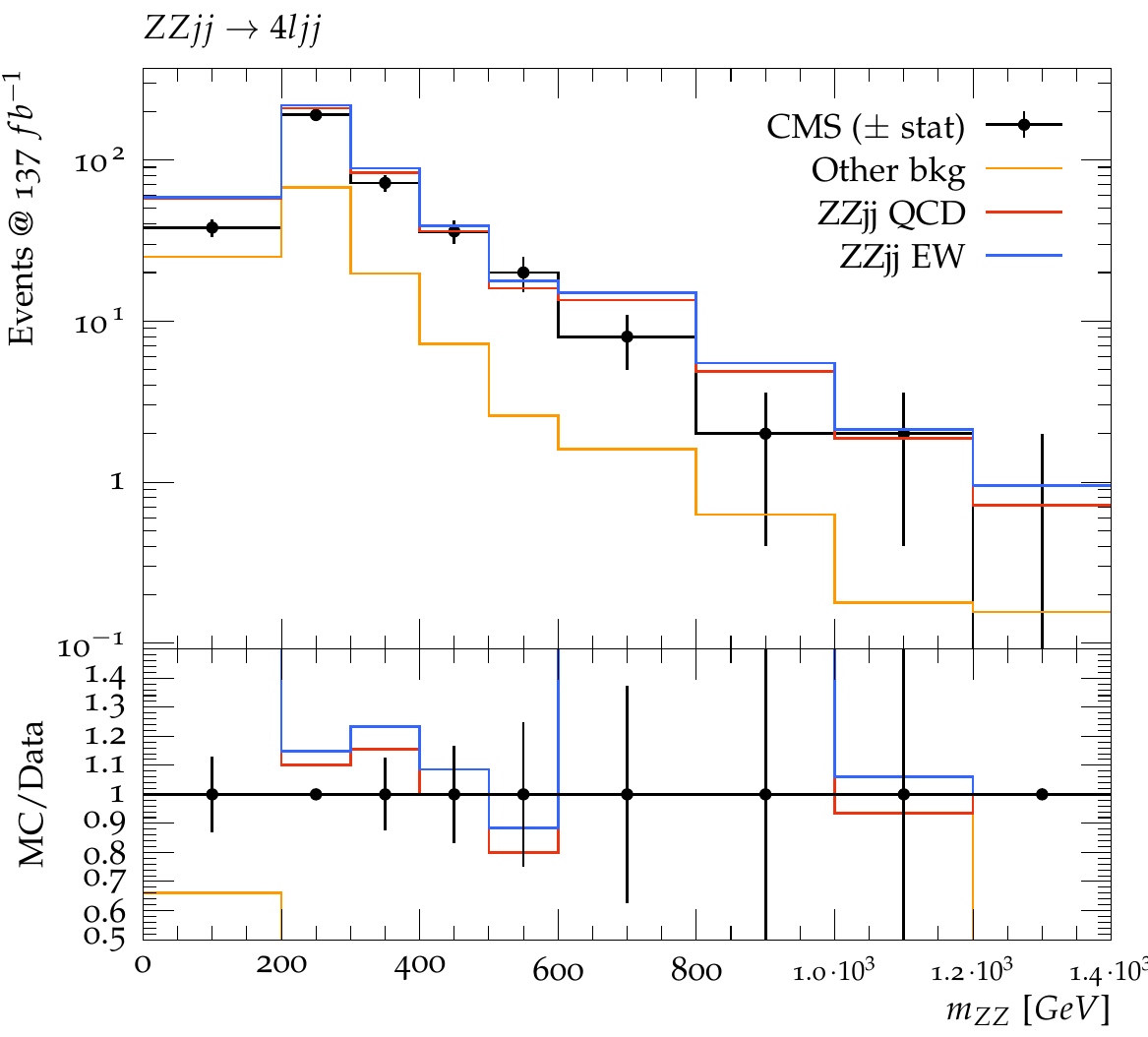}
  \caption{\small The CMS detector-level
    VBS measurement in the $ZZjj$ final state based on the full Run II luminosity~\cite{ZZjj:cms137}.
  Here we compare the number of events
  per $m_{ZZ}$ bin between the theoretical predictions and the experimental data.
  The error bars display only the statistical uncertainty.
  }
  \label{fig:ZZjj_mzz}
\end{figure}

The theoretical calculation for the $ZZjj$ process for the signal (EW-induced) events is simulated
at NLO using \pwg $\:$ \cite{Jager:2013iza} and at LO with \mg~for the QCD-induced background.
As discussed in Sect.~\ref{sec:EFTsensitivity}, the $ZZjj$ final state exhibits a large sensitivity to the EFT operators considered
in this work, but their practical impact in the fit is moderate
due to the large experimental uncertainties.
In Fig.~\ref{fig:ZZjj_mzz} we compare the number of events
per $m_{ZZ}$ bin between the theoretical predictions and the detector-level experimental data
from CMS in the $ZZjj$ final state based on the full Run II luminosity.
In this comparison,  our simulations account for the QCD- and EW-induced $ZZjj$ contributions,
while the other sources of  background are taken from the original publication~\cite{ZZjj:cms137}.
Note that the error band on the data points includes only the statistical uncertainty, which is dominant.
The overall detector selection efficiency is modelled here by comparing the theory prediction for the fiducial cross-section with the expected yields in the folded distribution.
In general, we observe a fair agreement between the theory simulations and the experimental data
once the experimental uncertainties are accounted for.

\begin{figure}[t]
  \centering
  \includegraphics[width=0.6\textwidth]{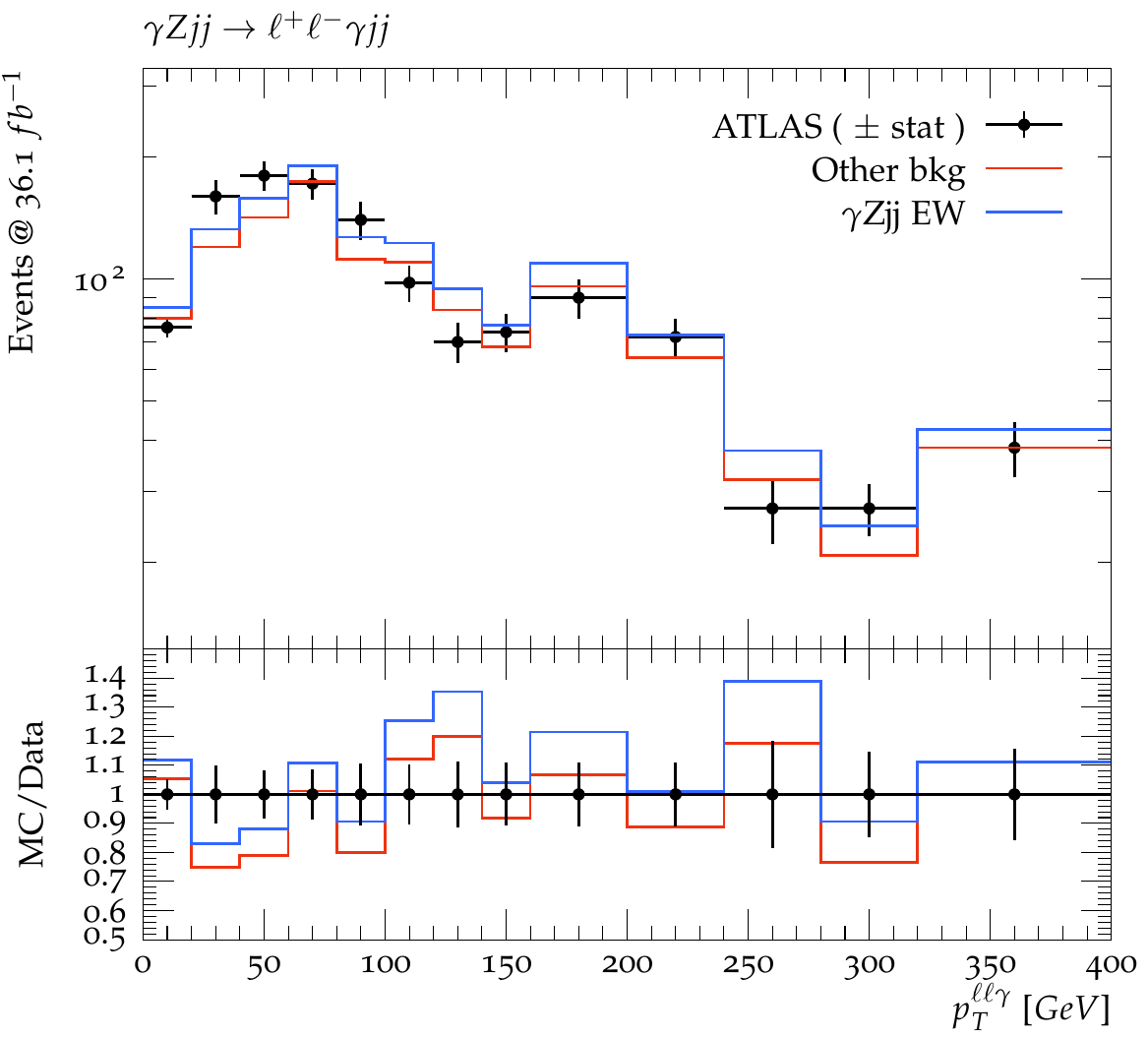}
  \caption{\small Comparison between data and theory predictions
    for the ATLAS  measurement of VBS in the $\gamma Zjj$ final state~\cite{AZjj:atlas}.
  Here we compare the number of events
  per $p_{T}^{\ell \ell \gamma}$ bin between our  predictions and the  experimental data.
  }
  \label{fig:AZjj_ptlla}
\end{figure}

\paragraph{$\mathbold{\gamma Zjj}$ production.}
Finally, we consider the rare VBS final state composed by a photon $\gamma$ and a $Z$ boson
which subsequently decays leptonically.
In this case, we have available two fiducial cross-section measurements for the electroweak
production of a $Z\gamma$ pair in association with two jets from ATLAS~\cite{AZjj:atlas} and
CMS~\cite{AZjj:cms,AZjj:cms:hepdata} based on the 2016 dataset with $\mathcal{L}\simeq 36$~fb$^{-1}$.
As for the $ZZjj$ final state, we will consider here
one detector-level  distribution from ATLAS as a consistency check.
Our theoretical predictions for this channel are evaluated at LO with \mg~ and are found to be
in good agreement with the data.
This channel is interesting for our study both because of its sensitivity to neutral Higgs couplings
as well as its ability to break degenerate solutions in the EFT parameter space.
Moreover, we found that ATLAS and CMS have taken very different approaches to the definition of the phase space, which is already useful at the level of the cross-section and would mean an increased EFT sensitivity if unfolded distributions were also available.
In Fig.~\ref{fig:AZjj_ptlla} we report the reconstructed differential distribution.
Our theoretical simulation includes only the EW signal,
while other sources of background (QCD-induced $\gamma Z$, $Z+{\rm jets}$, and
$t\bar{t}\gamma$) are taken from~\cite{AZjj:atlas}.
For this process, EW-induced VBS
contributes only to $\sim 10 \%$ of the total events, thus the impact of this distribution to the EFT fit is expected to be moderate.

\paragraph{Overview of VBS measurements.}
A summary of the VBS datasets to be considered in our EFT
interpretation is collected in Table~\ref{tab:datasettable_VBS}.
For each dataset, we indicate the final state, the selection
criteria ({\it e.g.} EW-only versus EW+QCD contributions),
the experimental observable, the number of data points
$n_{\rm dat}$ and integrated luminosity $\mathcal{L}$,
as well as the dataset label and the original reference.
In the data labelled with $^{(*)}$, one
bin from the differential distribution
has been traded by the associated fiducial cross section to avoid double counting.
In those cases, the latter corresponds to the EW-only component and thus exhibits
increased sensitivity to the EFT operators, and
 $n_{\rm dat}$ indicates the actual number of fitted data points.
In this overview we separate the unfolded from the folded, detector-level data, since only the former
will be part of the baseline dataset.
Overall, we end up with $n_{\rm dat}=18$  unfolded VBS cross-sections and $n_{\rm dat}=15$ bins
for the detector-level distributions, giving a total of $n_{\rm dat}=33$ fitted data points.
As will be shown in Sect.~\ref{sec:results}, the addition of the detector-level
distributions has a significant impact in a VBS-only EFT fit, but only a marginal effect
in the joint VBS+diboson analysis.

\input{tables/table-dataset-VBS.tex}

\subsection{Diboson production}
\label{sec:VVproduction}

In this work, gauge boson pair production is defined as the
process whereby, at leading order, two vector bosons
are produced on shell and then decay.
This implies that the tree-level scattering amplitude will be proportional to $\alpha_{\rm EW}^4$.
Higher-order QCD corrections will lead to additional hard radiation and thus
the QCD-induced $VV'jj$ final state becomes a background to the VBS processes.
This final state scales as $\alpha_{\rm EW}^4\alpha_s^2$, and therefore
in general will dominate over the EW-induced diagrams except in regions of the phase space
where the VBS topology is enhanced.

\begin{figure}[t]
  \centering
  \subfloat{
  \includegraphics[width=0.18\textwidth]{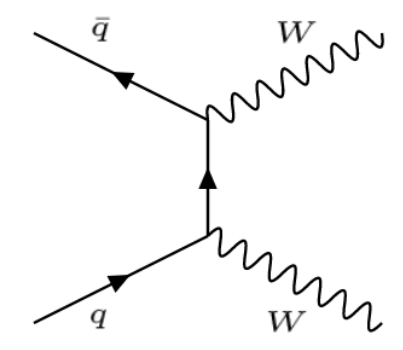}
  }
  \hspace{0.8cm}
  \subfloat{
  \includegraphics[width=0.16\textwidth]{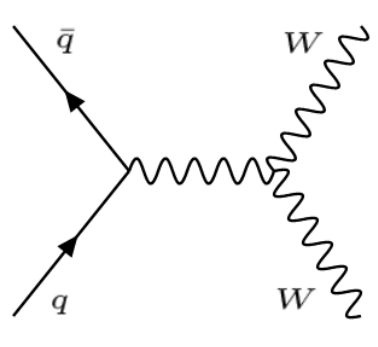}
  }
  \hspace{0.8cm}
  \subfloat{
  \includegraphics[width=0.20\textwidth]{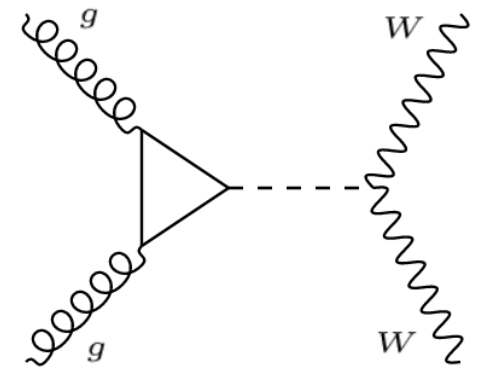}
  }
  \hspace{0.8cm}
  \subfloat{
  \includegraphics[width=0.16\textwidth]{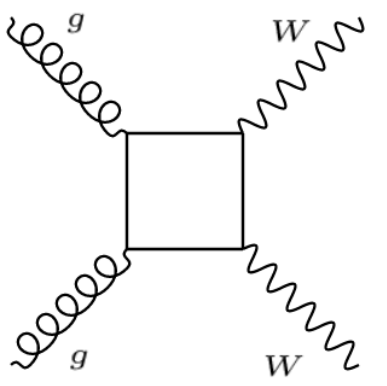}
  }
  \caption{\small Representative Feynman diagrams for opposite-sign $W^\pm W^\mp$ diboson production,
    where
    the first two diagrams correspond to leading order processes
    while other two to gluon-initiated loop-induced contributions.}
  \label{fig:VVfeynman}
\end{figure}

Fig.~\ref{fig:VVfeynman} displays representative Feynman diagrams for opposite sign $W^\pm W^\mp$ production,
a typical example of a diboson process.
One can observe how diboson production is sensitive to the TGCs at the Born level and that
the QGCs do not enter the theoretical description of this process.
The gluon-gluon-initiated contributions are usually quite suppressed in VBS-like analysis, since their
topology does not have the characteristic forward tagging jets.
In this work, we will focus on the diboson production data with leptonic final states,
in correspondence with the VBS case.

The standard experimental selection cuts for diboson processes are $p_T$ cuts
in the leading and subleading charged leptons, leptonic
rapidities being restricted to the central region,
and in the presence of $W$ bosons, a cut on the missing transverse energy, $E_{T}^{\rm miss} \gsim 30 $ GeV.
Furthermore, additional cuts on the transverse masses of the reconstructed leptons around $m_W$ and $m_Z$
are required to minimise the contribution from Higgs $s$-channel production.
The resulting fiducial cross-sections are relatively large,
and already at $\mathcal{L}\simeq 36$~fb$^{-1}$ they
become limited by systematic uncertainties.
These large cross-sections explain why unfolded differential cross-sections
for different kinematic variables have been available for some time already.

\paragraph{Opposite-sign $\mathbold{W^\pm W^\mp}$ production.}
This channel has been measured by ATLAS based on the $\mathcal{L}=36$ fb$^{-1}$~\cite{WW:atlas,WW:atlas:hepdata} data in the
$e \mu$ final state.
Several differential distributions are available with their corresponding bin-by-bin correlation matrices.
From CMS, we include their recent measurement~\cite{WW:cms,WW:cms:hepdata} based on the same luminosity,
where events containing two oppositely charged leptons (electrons or muons) are selected.
In our EFT analysis, we will include the same differential distribution, $m_{\mu e}$, from
both ATLAS and CMS consisting of $n_{\rm dat}=13$ data points in each case.
While the ATLAS distribution is provided as an absolute distribution, the CMS
is normalised to the fiducial cross-section.
Since the EFT total cross-section is different to the SM one, we revert this
normalisation to maximise our EFT sensitivity.

\begin{figure}[t]
  \centering
  \subfloat{
    \includegraphics[width=0.46\textwidth]{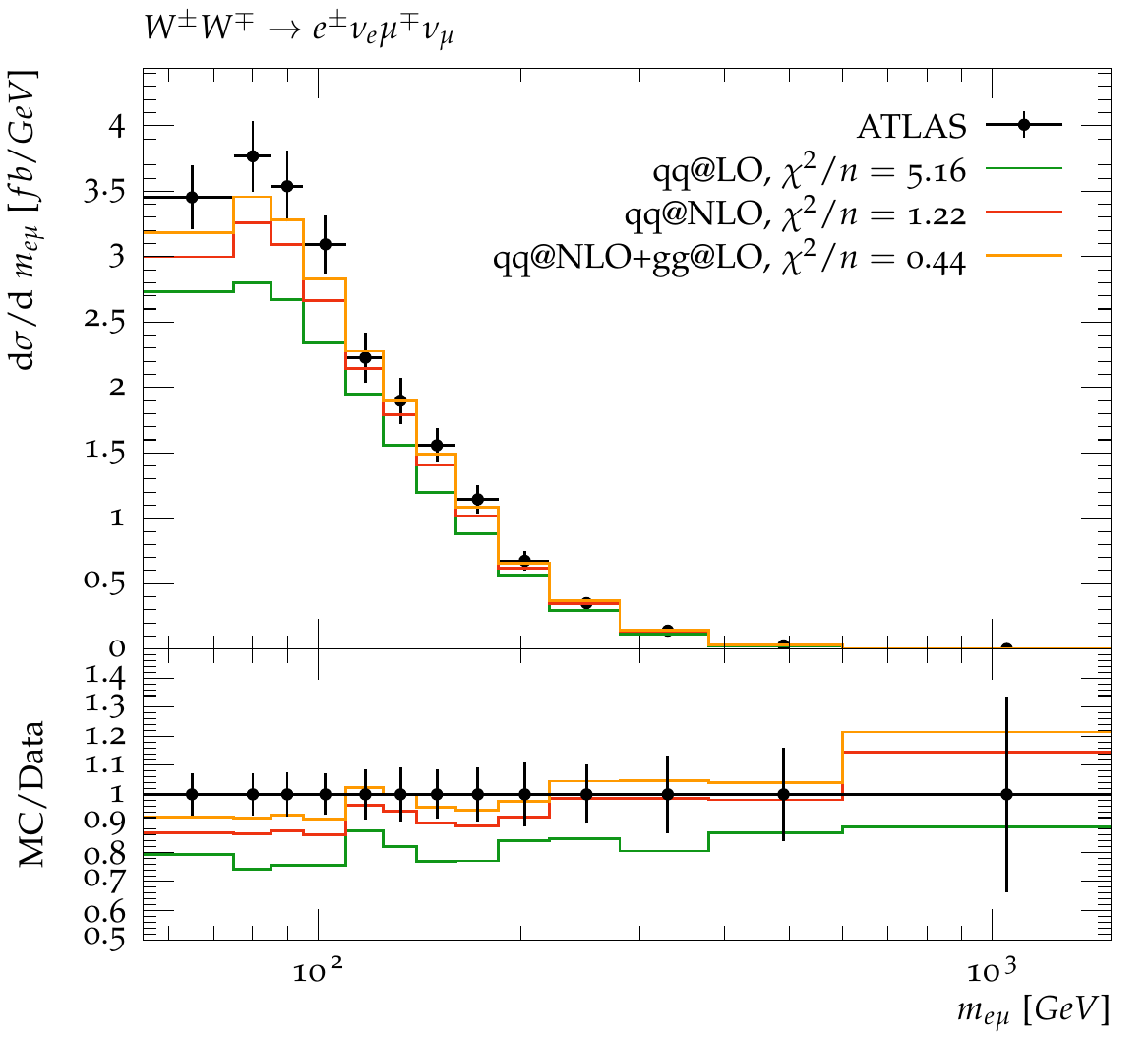}
  }\quad
  \subfloat{
    \includegraphics[width=0.46\textwidth]{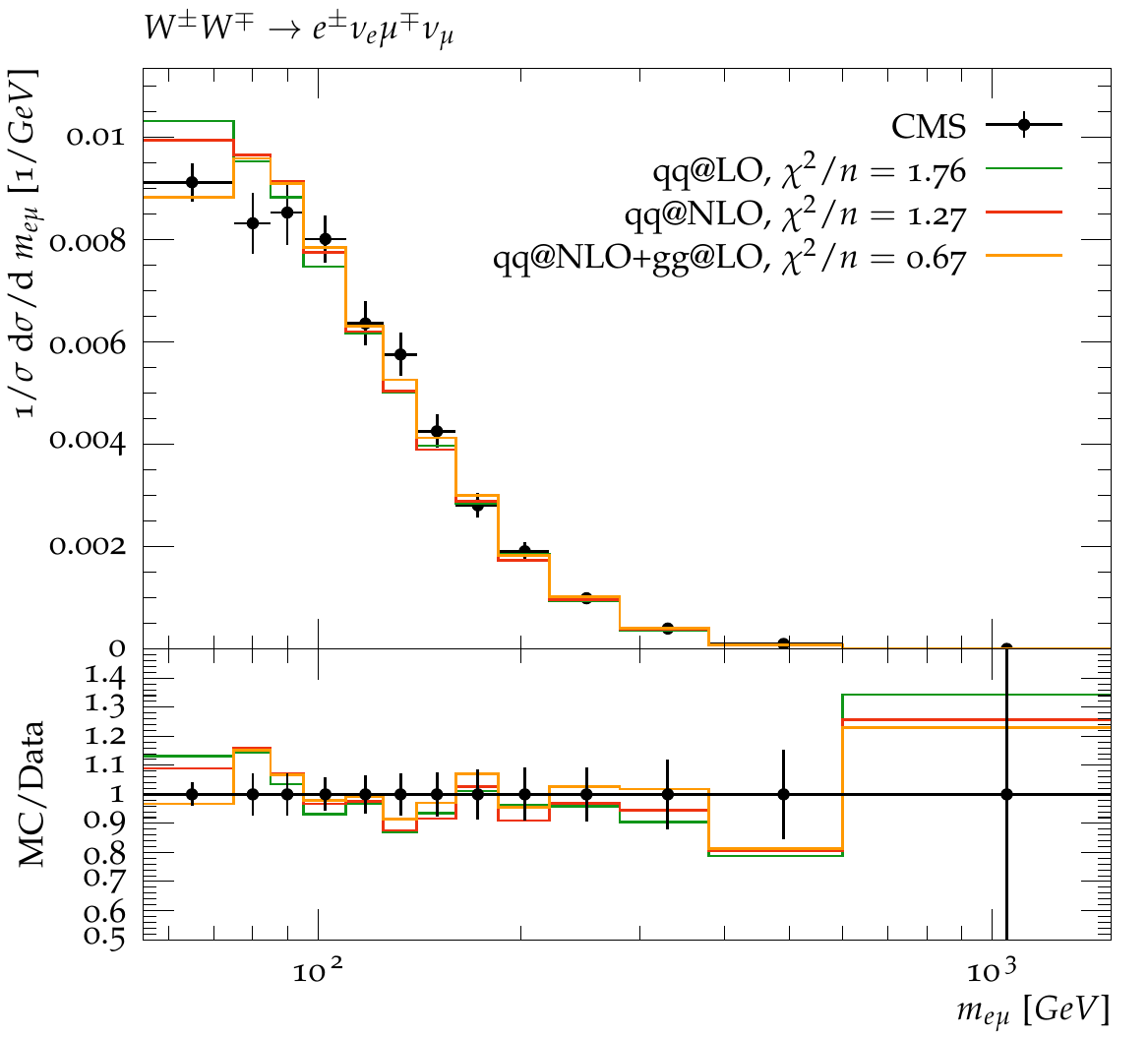}
  }
  \caption{\small The $m_{e\mu}$ differential
    distributions in opposite-sign $W^\pm W^\mp$ diboson production at $\sqrt{s}=13$ TeV from ATLAS (left)
    and CMS (right panel).
    The legend indicates the values of the $\chi^2$ per data point associated to
    different theoretical predictions: $q\bar{q}$-initiated at LO and NLO, and the latter plus
    $gg$-initiated
    at LO.  \label{fig:WW_atlas}
  }
\end{figure}

Fig.~\ref{fig:WW_atlas} displays a comparison between our theory predictions
and the experimental data.
The measurement extends up to values of the dilepton invariant mass of $m_{e\mu}\simeq 1.5$ TeV.
Here one can observe that the inclusion of higher-order QCD and gluon-initiated contributions
is essential to achieve a good agreement with experimental data, which turns out to be similarly
good for the two data sets.
Furthermore, the effect of NLO QCD corrections is seen to be smaller for the normalised distribution
than the absolute one, indicating that the NLO $K$-factor depends only mildly on the value
of the invariant mass $m_{e\mu}$.

\begin{figure}[t]
    \centering
    \subfloat{
    \includegraphics[width=0.45\textwidth]{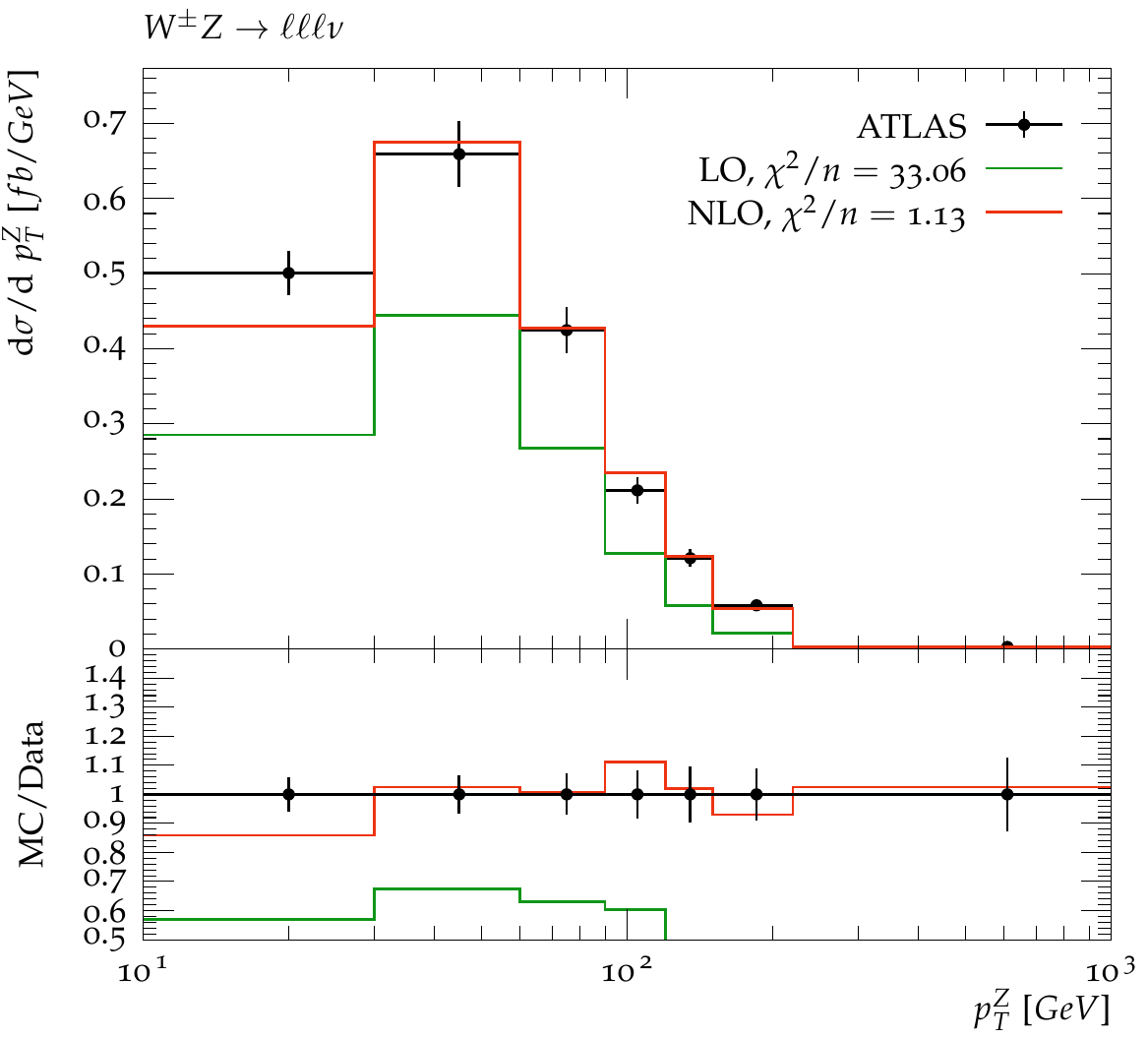}
    }\quad
    \subfloat {
    \includegraphics[width=0.45\textwidth]{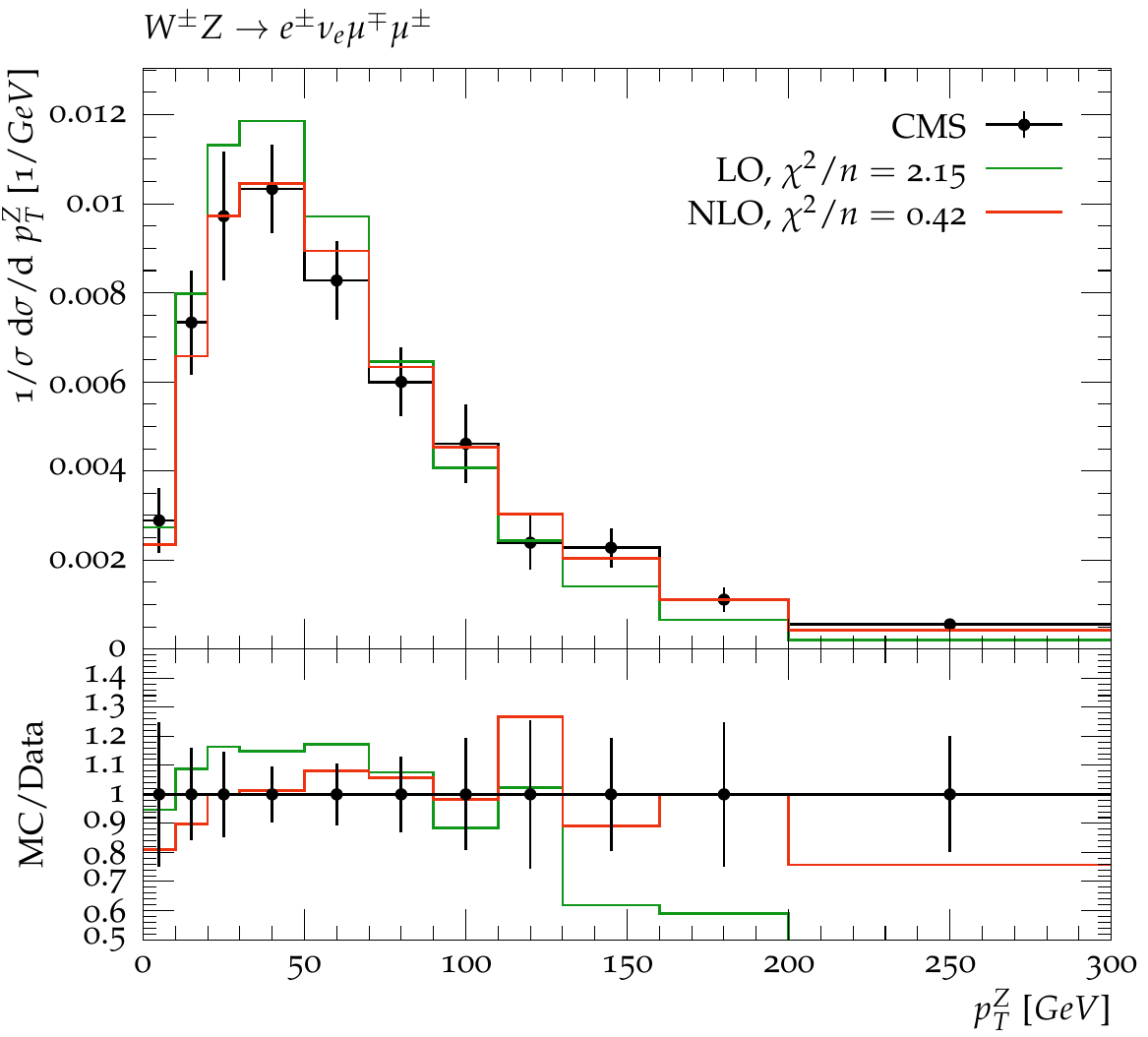}
    }
    \caption{\small The $Z$ boson transverse momentum distribution,
      $p_T^Z$, as measured in $W^\pm Z$ production from ATLAS~\cite{WZ:atlas} and CMS~\cite{WZ:cms} at 13 TeV
      based on $\mathcal{L}=36$ fb$^{-1}$.
      Note that while ATLAS provides an absolute distribution, the CMS one is instead normalised.
      }
    \label{fig:WZ_SM}
\end{figure}

\paragraph{$\mathbold{W^\pm Z}$ production.}
In this channel, we  consider the ATLAS~\cite{WZ:atlas,WZ:atlas:hepdata} and CMS~\cite{WZ:cms}
measurements at 13 TeV based on $\mathcal{L}=36$ fb$^{-1}$. In particular we chose the $e \mu \mu$ final state as a benchmark, although other combinations are available.
The ATLAS and CMS $p_T^Z$ distributions contain $n_{\rm dat}=7$ and 11 data points
and their kinematic reach is $p_T^Z\sim 1$ TeV and 300 GeV, respectively.
For the ATLAS measurement, the information on the
bin-by-bin correlated systematic uncertainties is made available
and therefore are included.
Moreover, we note that an EFT interpretation in terms of a subset of dimension-six operators has been presented in the CMS analysis of Ref.~\cite{WZ:cms}.
We display in Fig.~\ref{fig:WZ_SM} the comparison to our theoretical predictions at LO and at NLO. The latter in particular provides an excellent description to the experimental data.
Here the effects of the NLO QCD corrections are reduced
in the normalised distributions as was the case in $W^\pm W^\mp$ production.
Finally, as we will show in Sect.~\ref{sec:results}, this channel provides the strongest bounds on the TGC/QGC operator $\mathcal{O}_W$.

\begin{figure}[t]
  \centering
  \includegraphics[width=0.6\textwidth]{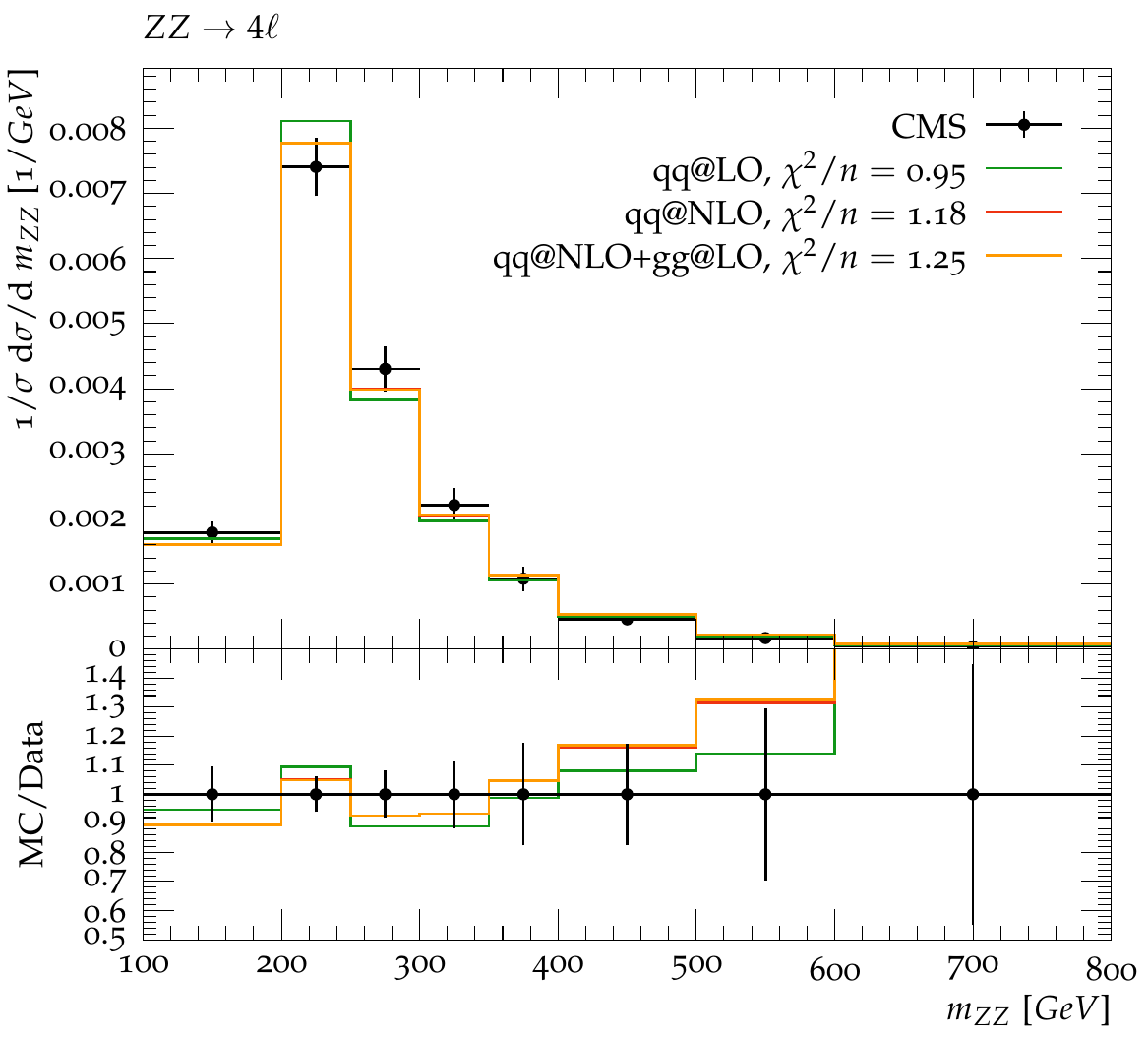}
  \caption{\small Normalized $d \sigma / d m_{ZZ}$ distribution in the fiducial phase space
    from the CMS measurement based on  $\mathcal{L}=137$ fb$^{-1}$.
  }
  \label{fig:ZZ_cms}
\end{figure}

\paragraph{$\mathbold{ZZ}$ production.}
For this channel, we use the recent CMS measurements based on $\mathcal{L}=137$~fb$^{-1}$
corresponding to the four-lepton
final state~\cite{ZZ:cms137}, which supersedes a previous publication based on 36 fb$^{-1}$~\cite{ZZ:cms,ZZ:cms:hepdata}.
For the theoretical predictions, the $qq \rightarrow ZZ$ and $gg \rightarrow ZZ$ contributions are simulated with \pwg $\:$ at NLO and with \mg $\:$ at LO, respectively.
Fig.~\ref{fig:ZZ_cms} displays
the normalized $d \sigma / d m_{ZZ}$ distribution in the fiducial phase space
from this CMS  $ZZ\to 4\ell$ measurement, which contains $n_{\rm dat}=8$ data points.
We find that the agreement with the normalised distribution at LO
is good, and that the contribution from the gluon-dominated diagrams is quite small.
The most updated ATLAS analysis related to the $ZZ$ final state is the measurement of
the four-lepton invariant mass spectrum at 13 TeV based on $\mathcal{L}=36$ fb$^{-1}$~\cite{Aaboud:2019lxo},
which receives contributions also from single-$Z$ and from Higgs production (via $h\to ZZ^*$ decays) and therefore is not considered further here.

\paragraph{Overview of diboson measurements.}
The diboson  measurements that will be considered
in this analysis are summarised in Table~\ref{tab:datasettable_VV}.
In total we have $n_{\rm dat}=52$  diboson  cross-sections
from the $W^\pm W^\mp$, $W^\pm Z$, and $ZZ$ channels, three times more data points than
the corresponding VBS unfolded cross-sections.
In Sect.~\ref{sec:results} we will compare the impact in the EFT parameter space between these
two families of measurements.

\input{tables/table-dataset-VV.tex}

\subsection{Sensitivity on the dimension-six EFT operators }
\label{sec:EFTsensitivity}

Quantifying the sensitivity of each VBS and diboson data set to the various dimension-six
EFT operators is an important step towards understanding the fit results.
It is also relevant to understand if
there are flat directions in our fit basis, and identify which data sets will provide the dominant constraints in the parameter space.
In the following, we summarise the dependence of each process to the EFT
operators considered and determine their relative sensitivity by means of the Fisher information.
We also apply a principal component analysis (PCA) to identify the hierarchy of directions
in the parameter space and assess the possible presence of flat
directions.

\paragraph{General discussion.}
In Table~\ref{tab:sensitivitytable} we list the contributions of the dimension-six EFT operators that constitute our fitting basis to the various VBS and diboson processes.
Overall the complementarity between the diboson and VBS can be seen, with VBS providing direct access to the $\mathcal{O}_{\varphi B}$ and $\mathcal{O}_{\varphi W}$
operators (and their CP-odd counterparts) which are essentially
unconstrained from diboson-only data.

\input{tables/sensitivity_table}

The operators $\mathcal{O}_W$  and $\mathcal{O}_{\widetilde{W}}$   modify both the TGCs and
the QGCs, and thus are not relevant
for the description of the  diboson production in the $ZZ$ channel.
The operators $\OO_{\varphi D}$ and $\OO_{\varphi WB}$ contribute to all the diboson and VBS
channels, since they lead to modifications of the SM parameters as discussed in Sect.~\ref{sec:eftth}.
Given that $\OO_{\varphi B}$ modifies only couplings involving the Higgs boson and/or $Z$ and
$\gamma$, it  will be unconstrained from the $WW$ and $WZ$ diboson channels
as well as from the  $WWjj$ and $WZjj$ processes.
The operator $\OO_{\varphi W}$ induces additional modifications
compared to $\OO_{\varphi B}$, contributing to diboson processes by means of the $hZZ$ and $hWW$ vertices.

The two-fermion interaction vertices
$\gamma \bar{\psi} \psi$ and $Z\bar{\psi} \psi$ are modified by some of the two-fermion
operators, specifically by  $\OO_{\varphi l}^{(1)}$, $\OO_{\varphi e}$, $\OO_{\varphi l}^{(3)}$, $\OO_{\varphi q}^{(3)}$, $\OO_{\varphi q}^{(1)}$, $\OO_{\varphi d}$, and
$\OO_{\varphi u}$, while the $W\bar{\psi} \psi$ vertex will be
affected by $\OO_{\varphi l}^{(3)}$ and $\OO_{\varphi q}^{(3)}$.
Furthermore, the $pp \to VV \to 4\ell$ and $pp \to VVjj \to 4\ell jj$ processes provide
sensitivity to two-fermion operators of the form $\varphi D \psi^2$ in all channels except for the ones with two $WW$ bosons.
Moreover, since the experimental phase space selection in the diboson production is designed to be orthogonal to the Higgs production, we expect that the $WW$ and $ZZ$ channel will be less sensitive to these
operators compared to VBS.
This justifies why the contributions from $\OO_{\varphi B}$ and $\OO_{\varphi W}$ (and their
corresponding CP-odd counterparts) are negligible in this channel.

\paragraph{The Fisher information matrix.}
While certainly informative, Table~\ref{tab:sensitivitytable} does not allow one to compare
the sensitivity brought in by different data sets on a given EFT degree of freedom.
In particular, we would like to quantify the relative impact that the diboson and VBS observables have for each coefficient.
To achieve this, it is convenient to resort to the Fisher information matrix~\cite{Ellis:2018gqa,Brehmer:2017lrt} which,
when restricted to linear contributions only, is given by
\be
\label{eq:fisherinformation2}
I_{ij} = \sum_{m=1}^{n_{\rm dat}} \frac{\sigma^{\rm (eft)}_{m,i}\sigma^{\rm (eft)}_{m,j}}{\delta_{{\rm exp},m}^2} \, ,\quad
i,j=1,\ldots,n_{\rm op} \, ,
\ee
where the EFT coefficients are defined in Eq.~(\ref{eq:crosssection}) and where $\delta_{{\rm exp},m}$
stands for the total experimental error associated to the $m$-th data point.
In Eq.~(\ref{eq:fisherinformation2}), the sum extends over all the data points that belong to a given
data set or family of processes.
While the absolute values of the entries of the Fisher matrix $I_{ij}$ are not physically
meaningful (since the overall normalisation of the EFT operators is arbitrary), the ratios
of the diagonal entries $I_{ii}$ for the $i$-th degree of freedom between two different groups
of process is well-defined, since there the operator normalizations cancel out.

\begin{figure}[htbp]
  \centering
    \includegraphics[width=0.9\textwidth]{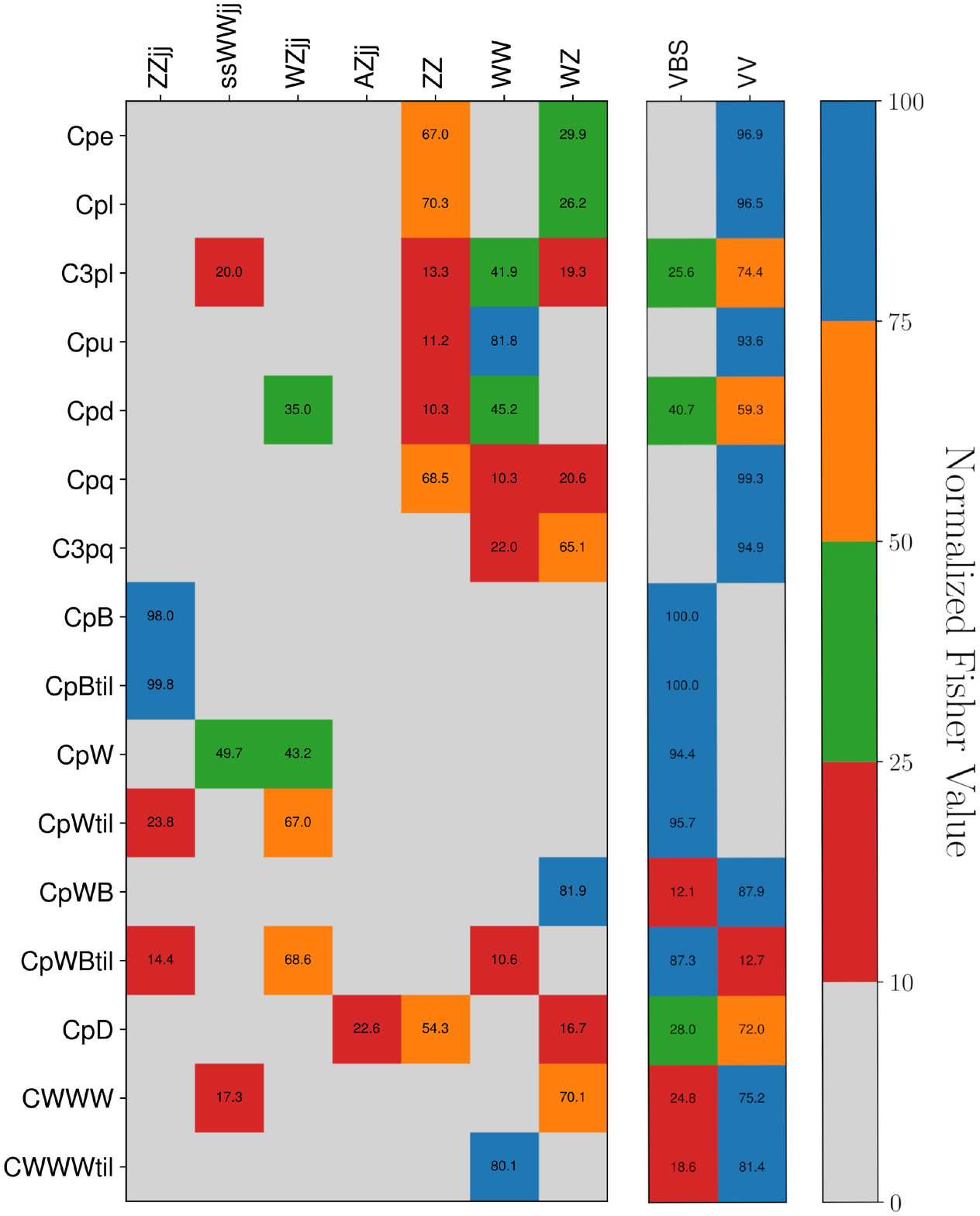}
    \caption{\small The diagonal entries of the Fisher information matrix, $I_{ii}$,
      evaluated for each of the coefficients  that form our fitting basis.
      We display results separately for each channel (left)
      and  when clustering all VBS and diboson datasets together (right panel).
      For those entries greater than 10\%, we also indicate the numerical
value in the heat map.
     \label{fig:FisherMatrix}
    }
\end{figure}

The diagonal entries of the Fisher information matrix
evaluated for each of the degrees of freedom that form our basis
are displayed in Fig.~\ref{fig:FisherMatrix}.
Its entries have been normalised such that the sum
over the elements of a given row adds up to 100.
We show results both for the individual groups of processes
as well as the comparison between the overall impact of the VBS and the diboson datasets.
For those entries greater than 10\%, we also indicate its numerical
value in the heat map.

One can observe from Fig.~\ref{fig:FisherMatrix}
that the VBS data provide the dominant sensitivity
for several of the operators considered in this analysis,
in particular for three of the CP-odd ones.
In general, we find that VBS process can provide complementary information
on the EFT parameter space compared to the diboson data.
Specifically, one finds that VBS measurements provide the dominant sensitivity (more than 50\%
of the Fisher information) for $c_{\varphi B}$ and  $c_{\varphi W}$ (and their CP-odd versions)
as well as for $c_{\varphi \widetilde{W}B}$.
Moreover, they
provide a competitive sensitivity (defined as more than 20\%) for $c_{\varphi l}^{(3)}$,
$c_{\varphi d}$, $c_{\varphi D}$ and for the triple gauge operator $c_{W}$.
The latter result illustrates how VBS measurements, while still providing
less information that diboson measurements to constrain modifications of the TGCs,
do indeed provide useful information.
In the case of the triple gauge operator $c_{W}$, we also note that the $WZ$ diboson final state
dominates the sensitivity, with the contribution from the $WW$ one being negligible.
In terms of identifying which VBS final states lead to higher relative sensitivities, we observe
that $ZZjj$ provides most of the information for $c_{\varphi B}$ and
$c_{\varphi \widetilde{B}}$, $W^\pm W^\mp jj$ dominates for $c_{\varphi W}$, and $WZjj$ leads in constraining the CP-odd operators
$c_{\varphi \widetilde{W}}$ and $c_{\varphi \widetilde{W}B}$.

\begin{figure}[t]
  \centering
  \includegraphics[width=0.49\textwidth]{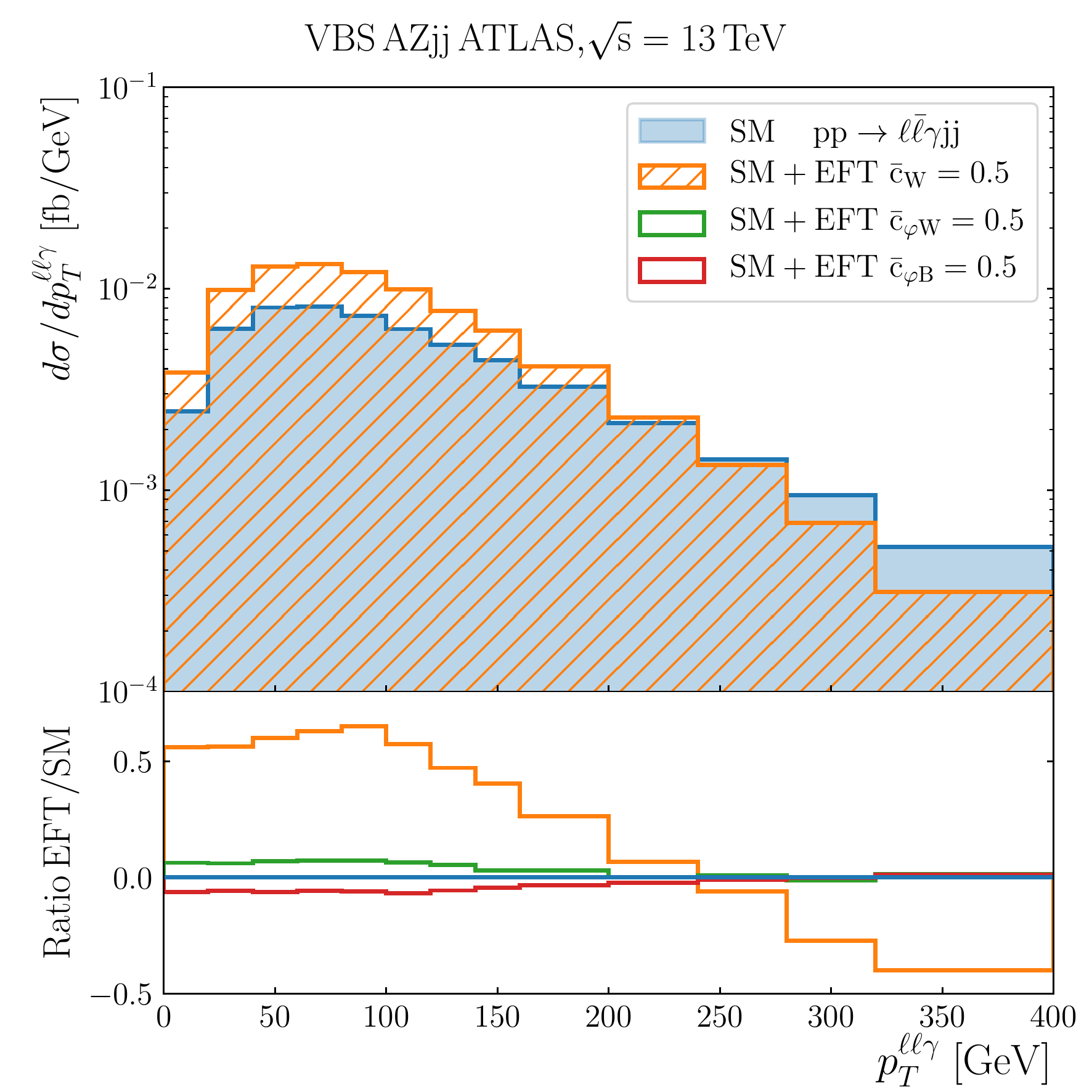}
  \includegraphics[width=0.49\textwidth]{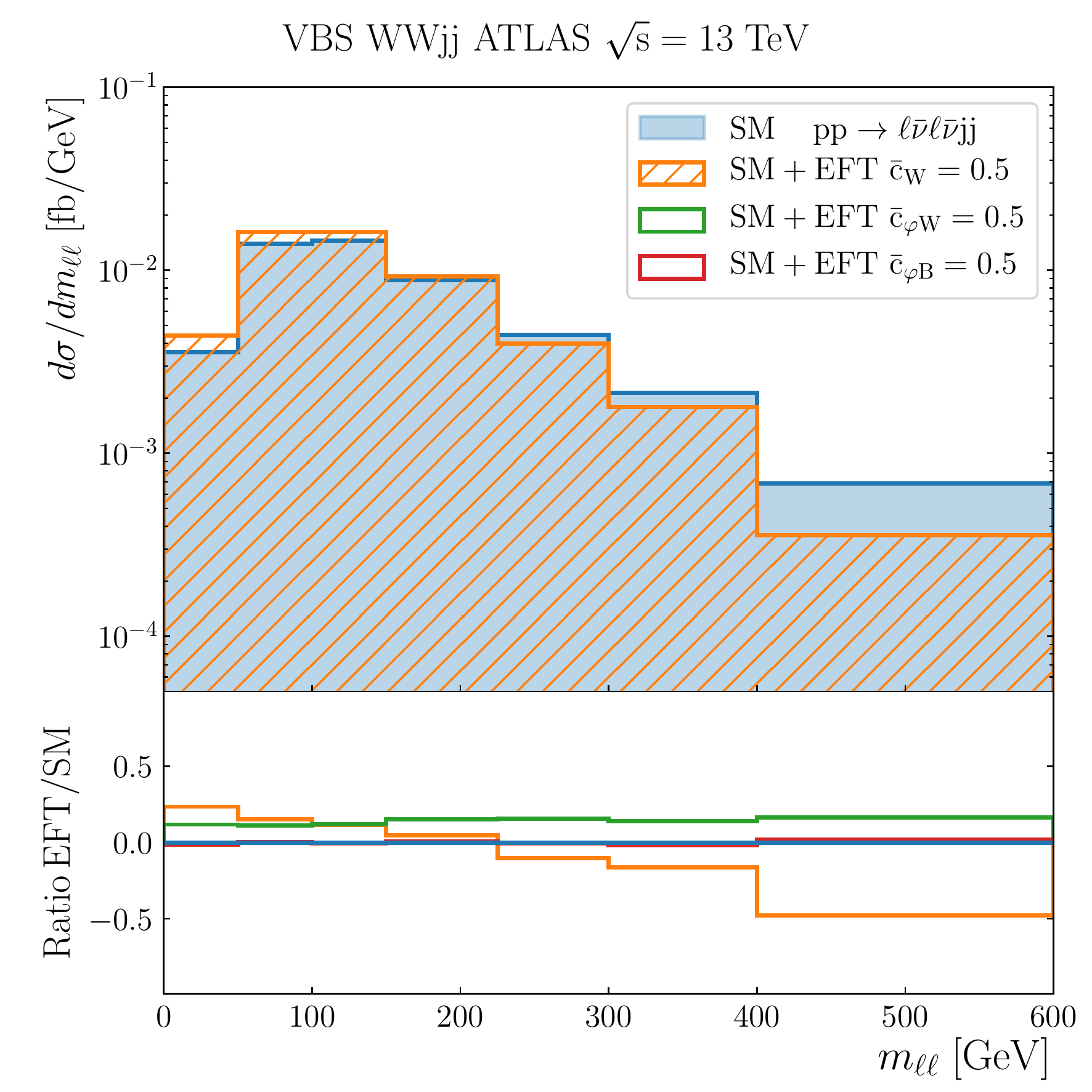}
  \caption{\small Theoretical predictions for the VBS signal (EW-induced component only)
      for different final states at $\sqrt{s}=13$ TeV.
      We show the dilepton $p_{T_{\gamma \ell \ell }}$ distributions
      for the $\gamma Zjj$ (left) and  $m_{\ell \ell }$ for $W^\pm W^\pm jj$ (right) final states based on the selection
      cuts of the corresponding ATLAS measurements.
      In each case, we compare the SM predictions with three EFT benchmark points in terms of the dimensionless quantities $\bar{c} = c v^2 / \Lambda^2  $. Either $\bar{c}_{W}$, $\bar{c}_{\varphi W}$, or $\bar{c}_{\varphi B}$
      are set to $0.5$ and the other coefficients to zero.
      In the upper panels, only the EFT prediction with $\bar{c}_{W}=0.5$ are shown to improve readability.
   \label{fig:sensitivity_benchmarks_ATLAS}
  }
\end{figure}

\begin{figure}[H]
  \centering
    \includegraphics[width=0.49\textwidth]{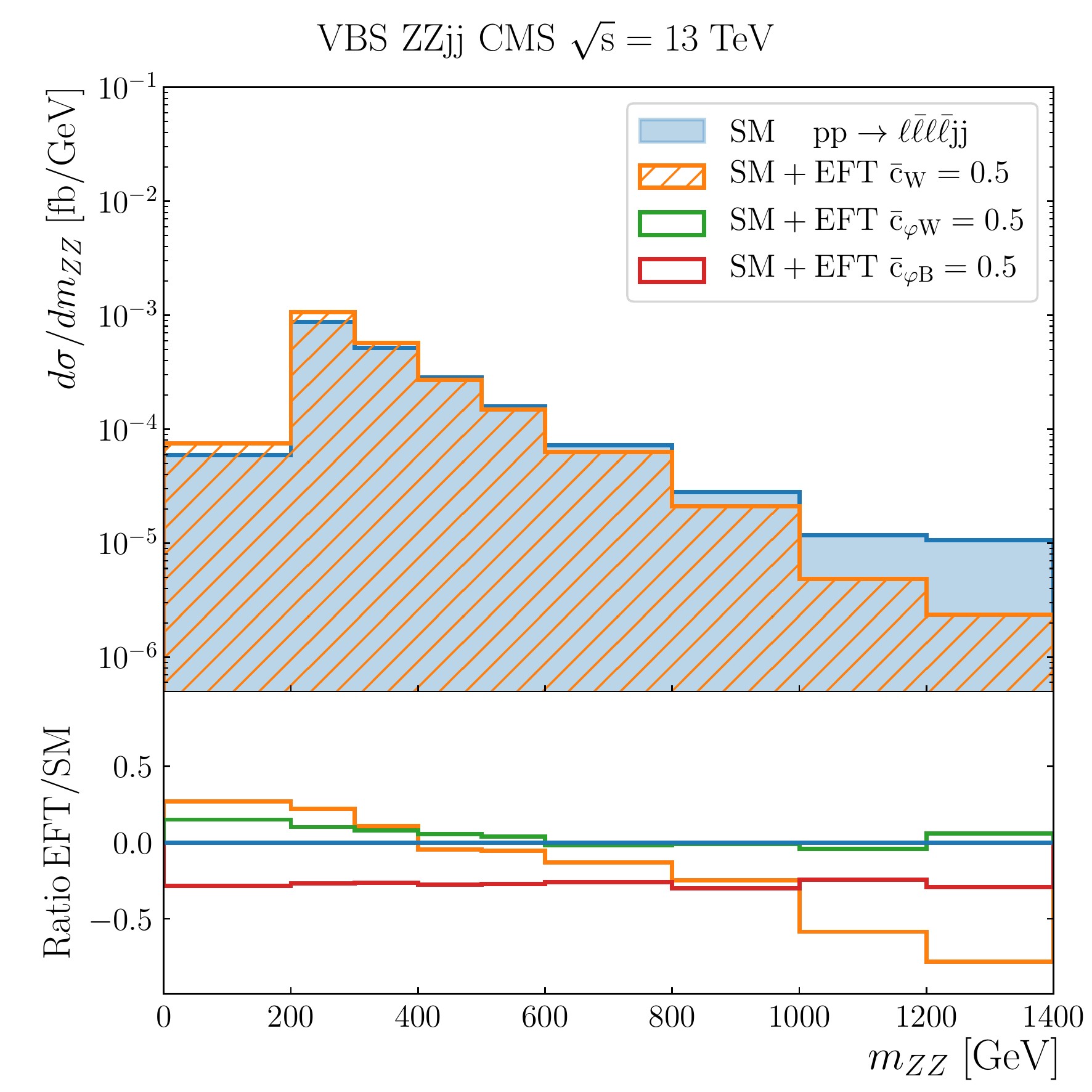}
    \includegraphics[width=0.49\textwidth]{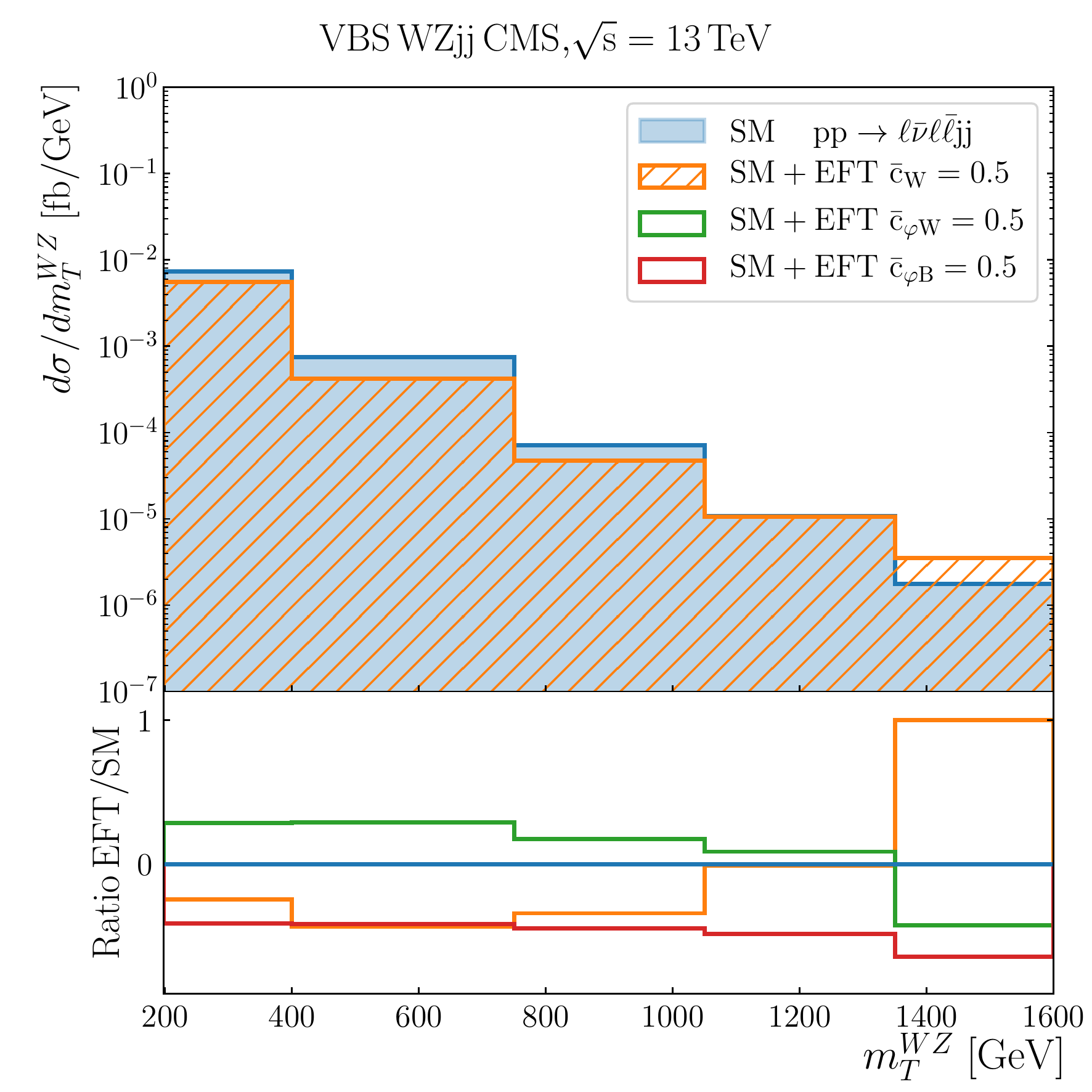}
    \caption{\small The $m_{ZZ}$ and $m_T^{WZ}$ distributions in the $ZZjj$ and $W^\pm Zjj$ final states,
       based on the same selection
      cuts as the corresponding CMS measurements.
   \label{fig:sensitivity_benchmarks_CMS}
  }
\end{figure}

\paragraph{EFT benchmark points.}
Another strategy to quantify the sensitivity to the different Wilson
coefficients is to compare the size of the SM and EFT
cross-sections for representative benchmark points in the parameter space.
Here we present only representative results for these comparisons, since
compatible information is found for the complete
set of final states and EFT operators.
In Figs.~\ref{fig:sensitivity_benchmarks_ATLAS} we display
the theoretical predictions for the VBS signal (EW-induced component only) at $\sqrt{s}=13$ TeV.
We show the differential distributions
for the $\gamma Zjj$  and $W^\pm W^\pm jj$  final states based on the selection cuts of the corresponding ATLAS reference measurements.
In each case, we compare the SM predictions with three EFT benchmark points,
in which either of the dimensionless quantities $c_{W} v^2 / \Lambda^2 $, $c_{\varphi W} v^2 / \Lambda^2$, or $c_{\varphi B} v^2 / \Lambda^2$
are set to 0.5, and the rest are set to zero.
In the upper panels, only the EFT prediction for  $c_{W} v^2 / \Lambda^2 $ is shown, to improve readability.
We also display in Fig.~\ref{fig:sensitivity_benchmarks_CMS}
the corresponding comparisons
for the $m_{ZZ}$ and $m_T^{WZ}$ distributions in the $ZZjj$ and $W^\pm Zjj$ final states
based on the same selection cuts as in the associated CMS measurements.

From the comparisons in
Figs.~\ref{fig:sensitivity_benchmarks_ATLAS} and~\ref{fig:sensitivity_benchmarks_CMS},
one can observe a distinct variation in the EFT sensitivity across the specific final state and differential distribution being considered.
In the case of the $\gamma Zjj$ and $W^\pm W^\pm jj$ final states,
there is good sensitivity to $c_{W}$ but rather less for $c_{\varphi W}$ and $c_{\varphi B}$ assuming the same value for each coefficient.
Interestingly, the sensitivity to $c_{W}$ can arise both from the low energy region as well as from
the high energy tail of the distributions.
The situation concerning $c_{W}$ is similar for the $ZZjj$ and $W^\pm Z jj$ final states, with the difference
being that now one becomes also sensitive to  $c_{\varphi B}$, which suppresses the cross-section compared to the SM expectation in a manner more or less independent from the kinematics.
In the case of the $c_{\varphi W}$ coefficient, the only distribution with comparable sensitivity to the
other benchmark points is $m_T^{WZ}$ in the $W^\pm Z jj$ final state.

\paragraph{Principal component analysis (PCA).}
Lastly, we use PCA in this section to identify the combinations of
Wilson coefficients which exhibit the largest and the smallest variabilities
and determine the possible presence of flat directions.
While PCA is primarily used as a dimensionality reduction tool by removing principal components with the lowest variance, here we use its core steps based on singular value decomposition (SVD) only for diagnosis purposes, and the EFT fitting basis remains the same as that defined in Sect.~\ref{sec:eftth}.
More specifically, we utilize PCA to identify the possible presence of flat directions, assess whether there is a large gap in the variability
between the principal components, and to determine the matching between the physical fitting basis and the principal components.

The starting point of the principal component analysis
is the matrix $K$ of dimensions $n_{\rm dat}\times~n_{\rm op}$ and
(dimensionless) components $K_{mi}=\sigma^{(\rm eft)}_{m,i}/\delta_{{\rm exp},m}$,
where $\delta_{{\rm exp},m}$ is the same total experimental error that appears
in the evaluation of the Fisher information matrix.
Using singular value decomposition (SVD) we can write $K = U W V^\dagger$,
where $U~(V)$ is a $n_{\rm dat} \times n_{\rm dat}$~($n_{\rm op} \times n_{\rm op}$) unitary matrix and $W$ is an $n_{\rm dat}\times n_{\rm op}$ diagonal
matrix with semi-positive real entries, called the singular values, which are ordered by decreasing
magnitude.
The larger a singular value, the higher the variability of the associated principal component.
The elements $V$ contain the (normalised) principal components associated
to each of the singular values, which can be expressed as
a superposition of the original coefficients,
\be
\label{eq:PCdef}
{\rm PC}_k = \sum_{i=1}^{n_{\rm op}} a_{ki}c_i \, , \quad k=1,\ldots,n_{\rm op} \, ,\qquad \lp~ \sum_{i=1}^{n_{\rm op}} a_{ki}^2=1\,~\forall k \rp
\ee
where the larger the value of the coefficient $a_{kl}$, the larger the relative weight
of the associated Wilson in this specific principal component.

\begin{figure}[htbp]
  \centering
  \includegraphics[width=0.9\textwidth]{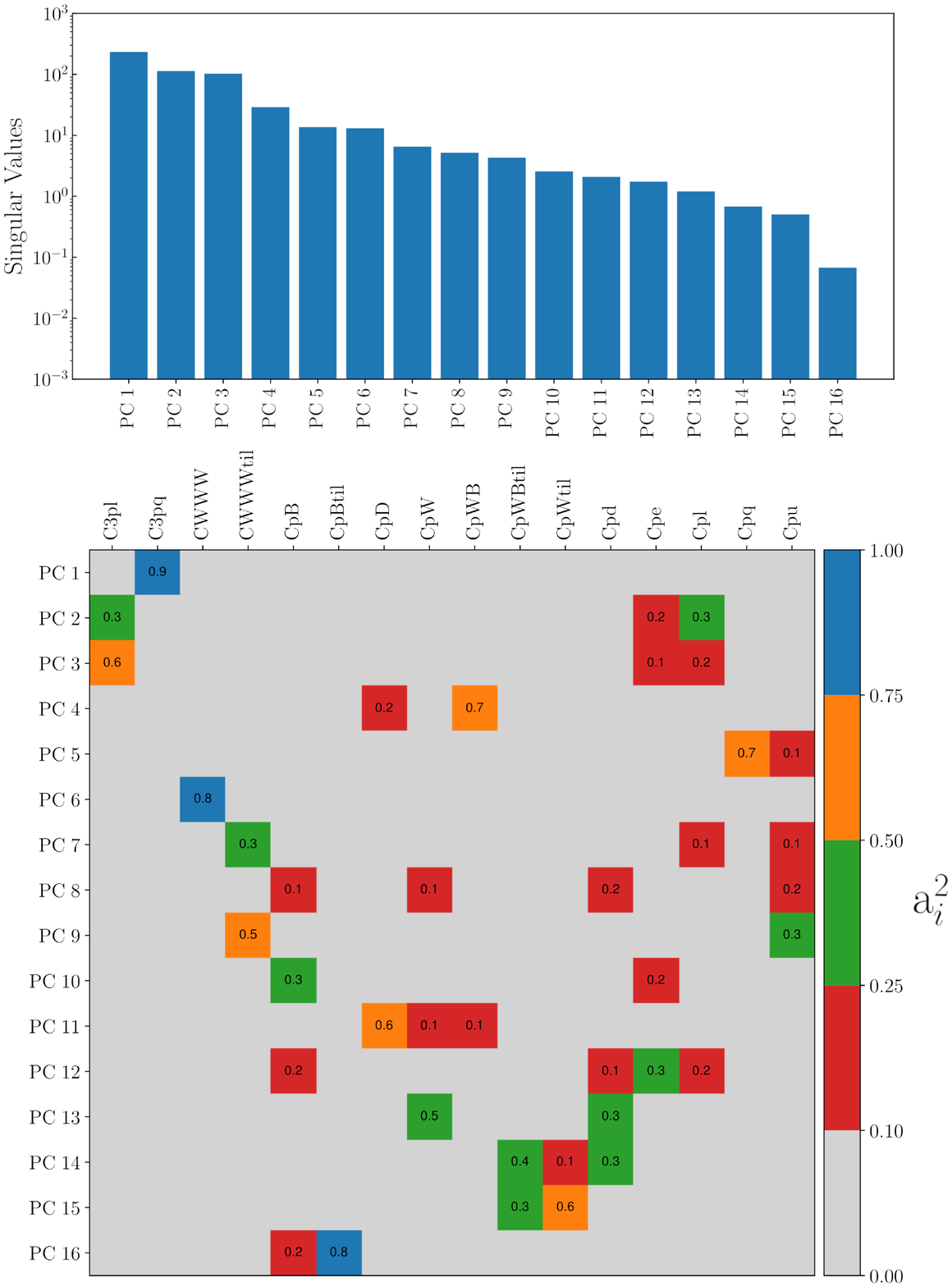}
  \caption{\small \small Results of the principal component
    analysis applied to the baseline VBS+diboson dataset.
    The upper panel shows the distribution of singular values,
    while the lower one displays the squared values for the coefficients $a_{ki}$
    of the principal components in Eq.~(\ref{eq:PCdef})
  }
    \label{fig:PCA}
    \end{figure}

The upper panel of Fig.~\ref{fig:PCA} displays the distribution of singular values
for the $n_{\rm op}=16$ principal
components associated to the  fitting basis
described in Sect.~\ref{sec:Warsawbasis} with the
baseline VBS+diboson dataset.
This analysis confirms that there are no flat directions in our parameter space, which would
appear as a principal component with a  vanishing singular value.
Furthermore, we do not observe large hierarchies in the distribution
of singular values, indicating that the physical dimensionality of our problem coincides with that of the adopted fitting basis.

The lower panel of Fig.~\ref{fig:PCA} displays a heat map indicating
the values of the (squared) coefficients $a_{ki}^2$ that relate the original
fitting basis to the principal components via the rotation in Eq.~(\ref{eq:PCdef}), and whose
associated eigenvalues are displayed in the upper panel.
For entries with $a^2_{ki}\ge 0.1$, we also indicate the numerical value
in the corresponding entry.
The principal component associated with the highest singular value can be attributed to the two-fermion
coefficient $c_{\varphi q}^{(3)}$,
which therefore is expected to be well constrained from the fit (as anticipated in Sect.~\ref{sec:eftth}).
Other principal components which coincide with the coefficients
of our fitting basis are $c_{W}$ and $c_{\varphi \widetilde{B}}$.
In general, the majority of principal components involve a superposition of several basis coefficients $c_i$, for example in PC$_k$ with $k=2,7,8$ or 10, none of the squared coefficients $a_{ki}^2$ is larger than 0.3.

%% file: tables/calc_details.tex
\begin{table}[t]
    \centering
    \footnotesize
    \renewcommand\arraystretch{1.8}
    \begin{tabular}{c|c|c|c|c}
   Process    & SM       & Code & EFT &   Code    \\
    \toprule
    $W^{\pm} W^{\mp}$   & NLO (qq), LO (gg) & \pwg~\citep{Baglio:2018bkm}, \mg  & LO + SM $K$-fact & {\tt SMEFTsim} \\
    $W^{\pm} Z$         &  NLO              & \pwg~\citep{Baglio:2019uty}      & LO + SM $K$-fact &  {\tt SMEFTsim}  \\
    $Z Z$               & NLO (qq), LO (gg) & \pwg~\citep{Nason:2013ydw}, \mg   & LO + SM $K$-fact &  {\tt SMEFTsim}  \\
    \midrule
    $W^{\pm}W^{\pm} jj $&  LO               & \mg                                 & LO + SM $K$-fact &  {\tt SMEFTsim}  \\
    $W^{\pm} Z jj$      &  NLO              & \pwg~\citep{Jager:2018cyo}       & LO + SM $K$-fact &  {\tt SMEFTsim}  \\   
    $Z Z jj$            &  NLO              & \pwg~\citep{Jager:2013iza}        & LO + SM $K$-fact &  {\tt SMEFTsim}  \\
    $\gamma Z jj$       &  LO              & \mg                                 & LO + SM $K$-fact &  {\tt SMEFTsim}  \\
     \bottomrule
    \end{tabular}
    \caption{\small The settings of the theoretical calculations used for the description of the LHC cross-sections included in the present analysis.
      We indicate, for both the SM and the EFT contributions, the perturbative accuracy and the codes used to produce the corresponding predictions.
      All the simulations are first generated at fixed-order and then matched to a parton shower using {\tt Pythia8}.}
    \label{tab:calc_details}
\end{table}

%% file: tables/table-dataset-VBS.tex
\begin{table}[t]
\begin{center}
\scriptsize
\renewcommand{\arraystretch}{1.8}
\begin{tabular}{c|c|c|c|c|c|c}
    Final state & Selection & Observable & $n_{\rm dat}$ & $\mathcal{L}~({\rm fb}^{-1})$ & Label & Ref.
    \\
\toprule
    \multirow{3}*{$W^{\pm}W^{\pm}jj$}
    &  EW-only           & $\sigma_{\rm fid}$
    & 1     & 36.1  & \texttt{ATLAS\_WWjj\_fid} & \citep{WWjj:atlas}
    \\
    \cline{2-7}
    & EW-only    & $\sigma_{\rm fid}$
    & \multirow{2}*{4}  & \multirow{2}*{137}  & \texttt{CMS\_WWjj\_fid}   & \multirow{2}*{\citep{WWjjWZjj:cms}}
    \\
    & EW+QCD & d$\sigma$/d$m_{ll}$     $^{(*)}$
    &                    &                    &  \texttt{CMS\_WWjj\_mll}  &
    \\
\midrule
    \multirow{3}*{$ZW^{\pm}jj$}
    & EW+QCD            & d$\sigma$/d$ m_{T_{WZ}}$
    & 5     & 36.1  & \texttt{ATLAS\_WZjj\_mwz} & \citep{WZjj:atlas}
    \\
    \cline{2-7}
    & EW-only    & $\sigma_{\rm fid}$
    & \multirow{2}*{4}  & \multirow{2}*{137}  & \texttt{CMS\_WZjj\_fid}   & \multirow{2}*{\citep{WWjjWZjj:cms}}
    \\
    & EW+QCD  & d$\sigma$/d$m_{jj}$    $^{(*)}$
    &                    &                    & \texttt{CMS\_WZjj\_mjj}   &
    \\
   \midrule
    \multirow{2}*{$ZZjj$}
    & EW+QCD  & $\sigma_{\rm fid}$
    & 1     & 139   &  \texttt{ATLAS\_ZZjj\_fid} &  \citep{ZZjj:atlas}
    \\
    \cline{2-7}
    & EW-only       &$\sigma_{\rm fid}$
    & \multirow{1}*{1}  & \multirow{1}*{139}  & \texttt{CMS\_ZZjj\_fid}   &  \multirow{1}*{\citep{ZZjj:cms137}} 
    \\
     \midrule
    \multirow{2}*{$\gamma Zjj$}
    & EW-only  & $\sigma_{\rm fid}$
&  \multirow{1}*{1}     & \multirow{1}*{36.1}  &  \texttt{ATLAS\_AZjj\_fid} &   \multirow{1}*{\citep{AZjj:atlas}}
    \\
    \cline{2-7}
    & EW-only  &$\sigma_{\rm fid}$
    & 1  & 35.9  & \texttt{CMS\_AZjj\_fid}   &  \citep{AZjj:cms}
    \\
    \midrule
    {\bf VBS total (unfolded)}  &   &  &   {\bf 18}  &   &   \\
    \midrule
    \midrule
    $ZZjj$      &   EW+QCD+Bkg    & Events/$m_{ZZ}$   & \multirow{1}*{4}  & \multirow{1}*{139}  &  \texttt{CMS\_ZZjj\_mzz}  &  \multirow{1}*{\citep{ZZjj:cms137}} 
    \\
    \midrule
    \multirow{1}{*}{$\gamma Zjj$}    &   EW+QCD+Bkg & Events/$p_{T_{\ell \ell \gamma}}$ & \multirow{1}*{11}
    & \multirow{1}*{36.1}  &  \texttt{ATLAS\_AZjj\_ptlla} &   \multirow{1}*{\citep{AZjj:atlas}} \\
        \midrule
    {\bf VBS total (detector-level)}  &   &  &   {\bf 15}  &   &   \\
 \bottomrule
\end{tabular}
\caption{Overview of the VBS measurements considered
  in this EFT analysis.
  We indicate the final state, the selection
  criteria,  the experimental observable, the number of data points
  $n_{\rm dat}$ and integrated luminosity $\mathcal{L}$. In the datasets labelled with $^{(*)}$, one
  bin from the differential distribution
  has been traded for the fiducial cross section.
  We separate the unfolded (baseline) from the
  detector-level (used for cross-checks) datasets.
}
\label{tab:datasettable_VBS}
\end{center}
\end{table}

%% file: tables/table-dataset-VV.tex
\begin{table}[t]
\begin{center}
\footnotesize
\renewcommand{\arraystretch}{1.8}
\begin{tabular}{c|c|c|c|c|c|c}
    Final state & Selection & Observable & $n_{\rm dat}$ & $\mathcal{L}~({\rm fb}^{-1})$ & Label & Ref.
    \\
\toprule
    \multirow{2}*{$W^{\pm}W^{\mp}$}  & \multirow{2}*{VV}
    & d$\sigma$/d$m_{e\mu}$              & 13    & 36.1 & \texttt{ATLAS\_WW\_memu}  & \citep{WW:atlas}
    \\
    \cline{3-7} &
    & d$\sigma$/d$m_{e\mu}$              & 13    & 35.9  & \texttt{CMS\_WW\_memu} & \citep{WW:cms}
    \\
    \midrule
    \multirow{2}*{$W^{\pm}Z$} & \multirow{2}*{VV}
    & d$\sigma$/d$p_{T_Z}$               & 7     & 36.1 & \texttt{ATLAS\_WZ\_ptz} & \citep{WZ:atlas}
    \\
    \cline{3-7}  &
    & d$\sigma$/d$p_{T_Z}$               & 11    & 35.9  & \texttt{CMS\_WZ\_ptz}   &   \citep{WZ:cms}
    \\
    \midrule
    $ZZ$        & VV
    & d$\sigma$/d$m_{ZZ}$                &  8    & 137  & \texttt{CMS\_ZZ\_mzz}   &   \citep{ZZ:cms137}\\
    \midrule
    \midrule
    {\bf Total diboson}  &   &  & {\bf 52}  &   &    \\
\bottomrule
\end{tabular}
\caption{\small Overview of
  the diboson measurements considered in this work.
}
\label{tab:datasettable_VV}
\end{center}
\end{table}

%% file: tables/sensitivity_table.tex
\begin{table}[t]
    \begin{center}
    \footnotesize
    \renewcommand\arraystretch{1.30}
    \begin{tabular}{c|c|c||c|c|c||c|c|c|c}
      \toprule
      \multirow{2}{*}{Class} & \multirow{2}{*}{Operator} & \multirow{2}{*}{CP-odd?}
      & \multicolumn{3}{c||}{ Diboson production} &\multicolumn{4}{c}{Vector boson scattering} \\
    & &
	& $W^{\pm}W^{\mp}$
	& $WZ$
	& $ZZ$
	& $W^\pm W^\pm jj$
	& $WZ jj$
    & $ZZ jj$
    & $\gamma Z jj$ \\
    \toprule
\multirow{9}{*}{Bosonic} & $\mathcal{O}_W$   &    & \checkmark   & \checkmark &    & \checkmark& \checkmark& \checkmark& \checkmark\\
	& $\OO_{\widetilde{W}}$  & yes& \checkmark& \checkmark&    & \checkmark& \checkmark& \checkmark & \checkmark\\
	&$\OO_{\varphi D}$ 	          &    & \checkmark& \checkmark& \checkmark& \checkmark& \checkmark& \checkmark & \checkmark\\
	& $\OO_{\varphi W}$ 			   &    & \n &    & \n & \checkmark& \checkmark& \checkmark & \checkmark\\
	&$\OO_{\varphi \widetilde{W}}$  & yes& \n &    & \n & \checkmark& \checkmark& \checkmark& \checkmark\\
	&$\OO_{\varphi B}$              &    &    &    & \n &    &    & \checkmark& \checkmark\\
	&$\OO_{\varphi \widetilde{B}}$  & yes&    &    & \n &    &    & \checkmark& \checkmark\\
	&$\OO_{\varphi WB}$ 			   &    & \n & \checkmark& \checkmark& \checkmark& \checkmark& \checkmark& \checkmark\\
	&$\OO_{\varphi \widetilde{W}B}$ & yes & \n & \checkmark& \checkmark& \checkmark& \checkmark& \checkmark& \checkmark\\
	\midrule \multirow{7}{*}{2-fermion}
	&$\OO^{(1)}_{\varphi \ell}$   &    &    & \checkmark& \checkmark&    & \checkmark& \checkmark& \checkmark\\
	&$\OO^{(3)}_{\varphi \ell}$   &    & \checkmark& \checkmark& \checkmark& \checkmark& \checkmark& \checkmark& \checkmark\\
	&$\OO_{\varphi e}$             &    &    & \checkmark& \checkmark&    & \checkmark& \checkmark& \checkmark\\
	&$\OO^{(1)}_{\varphi q}$      &    & \checkmark& \checkmark& \checkmark& \checkmark& \checkmark& \checkmark& \checkmark\\
	&$\OO^{(3)}_{\varphi q}$      &    & \checkmark& \checkmark& \checkmark& \checkmark& \checkmark& \checkmark& \checkmark\\
	&$\OO_{\varphi u}$            &    & \checkmark& \checkmark& \checkmark& \checkmark& \checkmark& \checkmark& \checkmark\\
	&$\OO_{\varphi d}$            &    & \checkmark& \checkmark& \checkmark& \checkmark& \checkmark& \checkmark& \checkmark\\
\bottomrule
    \end{tabular}
    \caption{\small We indicate which of the $n_{\rm op}=16$ dimension-six
      EFT operators considered in this analysis contributes to which of the experimental datasets.
      Furthermore, with \n~we denote those datasets where the corresponding operator sensitivity arises via the gluon fusion contributions, $gg  \rightarrow h \rightarrow VV$, which is known to be a very small effect.
      \label{tab:sensitivitytable}
    }
    \end{center}
\end{table}

%% file: sec-results.tex
\section{Results and discussion}
\label{sec:results}

In this section, we present the main results of this work, namely the dimension-six EFT interpretation
of the VBS and diboson  datasets from the LHC Run II.
We first briefly summarise the fitting strategy
adopted in this analysis and then
present the fit quality by comparing the best-fit results
with the corresponding experimental measurements.
We then present the fit results for the baseline dataset, determine
the 95\% CL intervals
for the $n_{\rm op}=16$ operators considered, and study the dependence of
our results with respect to variations of the input data, in particular
with fits based only on VBS measurements.

\subsection{Fitting strategy}

The EFT analyses carried out in this work are based on the {\tt SMEFiT} global
fitting framework
presented in~\cite{smefittophiggs,Hartland:2019bjb}.
Two options to constrain the EFT parameter are available in this framework: the Monte Carlo replica fit method (MCfit) and Nested Sampling (NS) via {\tt MultiNest}~\cite{Feroz:2013hea}. In this work we adopt the latter technique.
The end result of {\tt SMEFiT} is a representation of the probability density in the space of Wilson coefficients spanned by $N_{\rm spl}$ samples, $\{ c_i^{(k)}\}$, which allows the evaluation of
statistical estimators such as mean values and standard deviations, {\it e.g.},
\be
\label{eq:MCaverage}
\la c_i\ra = \frac{1}{N_{\rm rep}}\sum_{k=1}^{N_{\rm rep}}  c_i^{(k)} \, , \quad i=1,\ldots,n_{\rm op} \, ,
\ee
\be
\delta c_i = \lp \frac{1}{N_{\rm rep}-1}\sum_{k=1}^{N_{\rm rep}} \lp c_i^{(k)} -\la c_i\ra  \rp^2\rp^{1/2} \, , \quad i=1,\ldots,n_{\rm op} \, ,
\ee
and likewise for other estimators such as the correlation coefficients.
Since the present analysis is carried out at the linear level in the EFT expansion,
and there are no flat directions for the baseline dataset
(see Sect.~\ref{sec:EFTsensitivity}), the probability distributions
associated to the coefficients $\boldsymbol{c}$ are expected to be Gaussian.
For this reason, it is not necessary to go beyond the first two moments
of the posterior distributions in $\boldsymbol{c}$.

The overall fit
quality is assessed by means of the $\chi^2$ figure of merit, defined as
\begin{equation}
  \chi^2\lp {\boldsymbol c} \rp \equiv \frac{1}{n_{\rm dat}}\sum_{i,j=1}^{n_{\rm dat}}\lp
  \sigma^{(\rm th)}_i\lp {\boldsymbol c} \rp
  -\sigma^{(\rm exp)}_i\rp ({\rm cov}^{-1})_{ij}
\lp
  \sigma^{(\rm th)}_j\lp {\boldsymbol c}\rp
  -\sigma^{(\rm exp)}_j\rp
 \label{eq:chi2definition2}
    \; ,
\end{equation}
where $\sigma_i^{\rm (exp)}$ corresponds to the
central experimental data point and $\sigma^{(\rm th)}_i\lp {\boldsymbol c}\rp$ is the associated theoretical
prediction, Eq.~(\ref{eq:crosssection}), for the $i-$th cross-section.
The covariance matrix, ${\rm cov}$, is constructed from all available sources of uncorrelated and correlated experimental uncertainties, with the `$t_0$' definition~\cite{Ball:2009qv} used for the fit and the standard experimental
covariance used to quote the resulting $\chi^2$ values.
Whenever appropriate, we also add to the covariance matrix estimates of theoretical uncertainties coming from the input proton PDFs, as well as the MC theory calculations.
The post-fit $\chi^2$ values are then evaluated using the best-fit estimate (mean) of the Wilson coefficients, Eq.~(\ref{eq:MCaverage}), computed from the resulting MC samples obtained by NS.

\subsection{Fit quality and comparison with data} \label{sec:fitquality}

\begin{figure}[t]
    \centering
     \includegraphics[width=\textwidth]{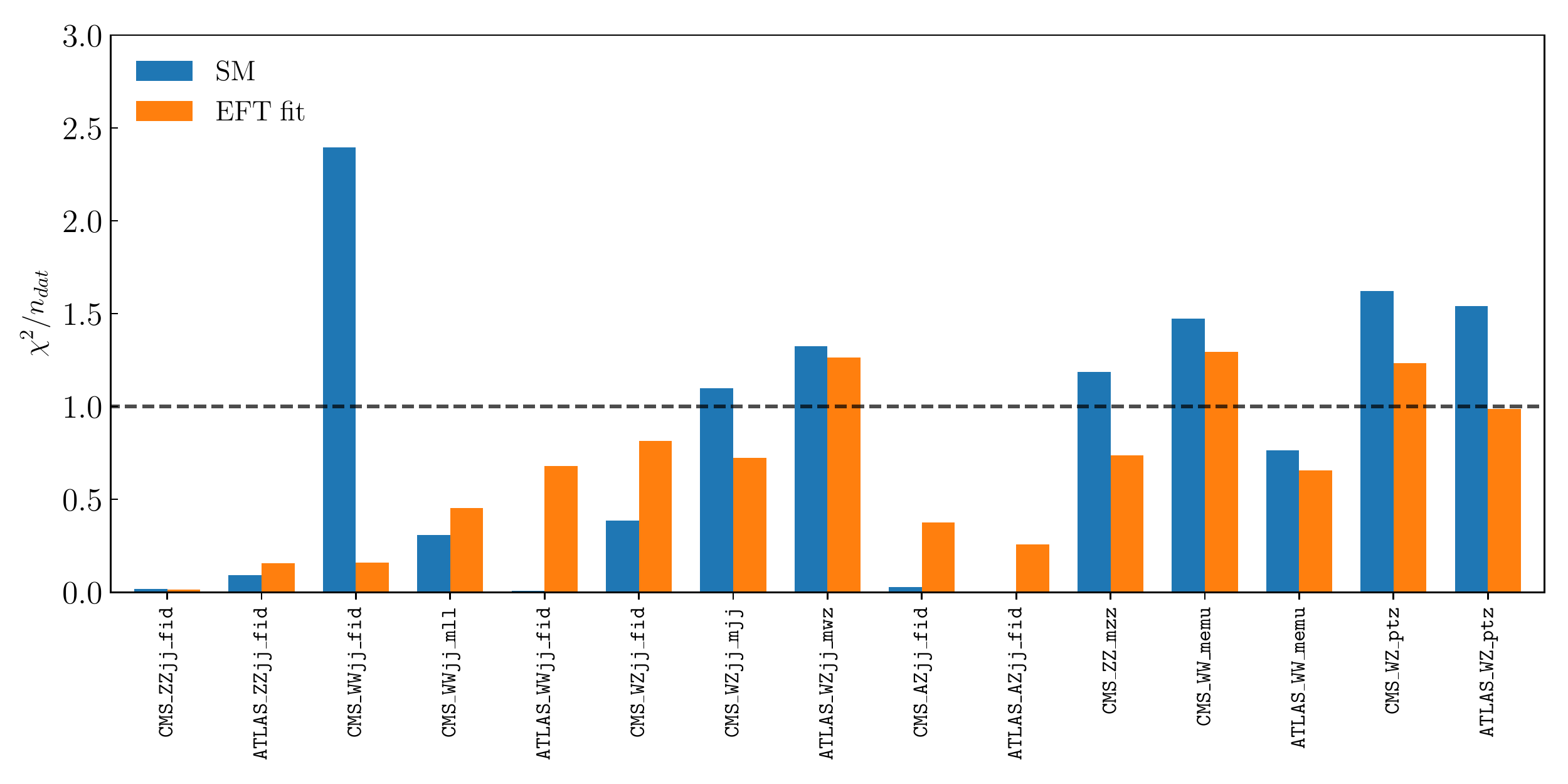}
     \caption{ \small Graphical representation of the $\chi^2$ values
     reported in  Table~\ref{tab:chivals}. }
    \label{fig:chi2_values}
\end{figure}

In Table~\ref{tab:chivals} we display the values of the $\chi^2/n_{\rm dat}$, Eq.~(\ref{eq:chi2definition2}),
for each of the data sets contained in our baseline fit, as well as the total values associated to the diboson
and VBS categories.
We also indicate the $\chi^2$ values corresponding to the Standard Model
predictions (pre-fit) together with the values obtained once the EFT corrections are accounted for
(post-fit).
Note that our baseline dataset does not contain any detector-level
folded distributions.
The graphical representation of these $\chi^2$ values is also displayed in Fig.~\ref{fig:chi2_values}.

\input{tables/chi2_table}

From Table~\ref{tab:chivals} one can observe that for the diboson data,
a $\chi^2$ of around
one per data point is obtained.
Moreover, the total $\chi^2/n_{\rm dat}=1.17$ found at the level
of SM calculations is reduced to 0.97 once EFT effects are included in the fit.
Concerning the VBS dataset, there is a higher spread in the $\chi^2/n_{\rm dat}$ values, which is explained by the fact that each data set is composed of either a single or a few cross-section measurements.
Taking into account the 18 independent cross-section measurements that we includein the fit, the SM value of $\chi^2/n_{\rm dat}=0.83$ is reduced to 0.75 at the post-fit level.
Overall, the combination of the diboson and VBS measurements adds up to $n_{\rm dat}=70$ data points
for which a pre-fit value of $\chi^2/n_{\rm dat}=1.08$ based on the SM predictions is reduced
to 0.92 after the EFT fit.

\begin{figure}[t]
  \centering
  \includegraphics[width=0.49\textwidth]{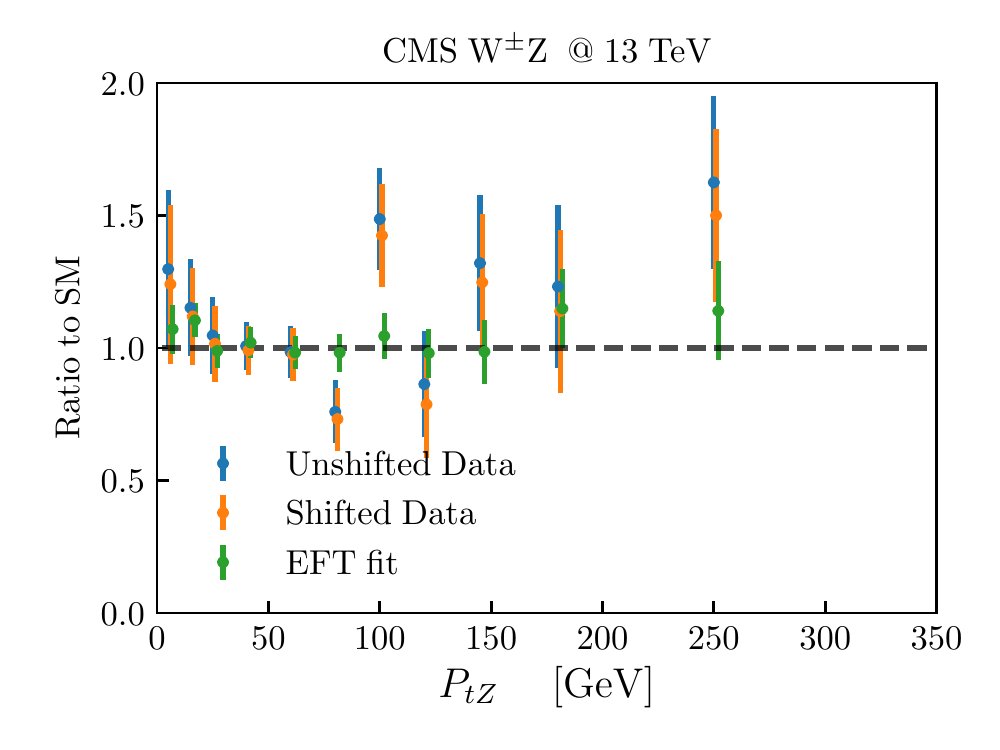}
  \includegraphics[width=0.49\textwidth]{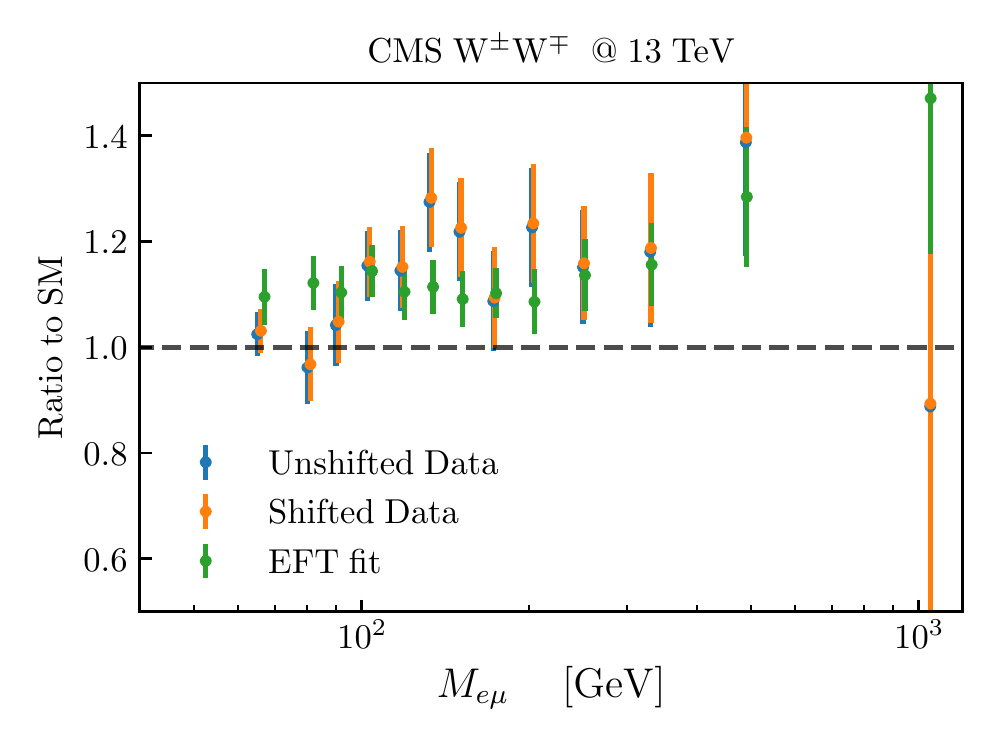}
  \includegraphics[width=0.49\textwidth]{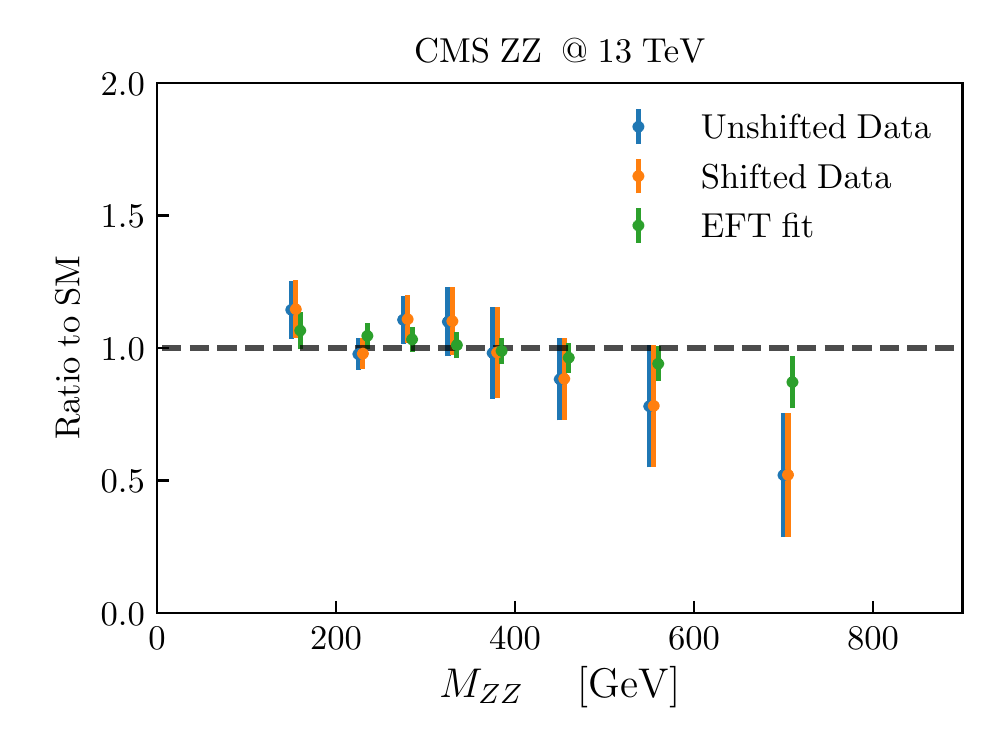}
  \includegraphics[width=0.49\textwidth]{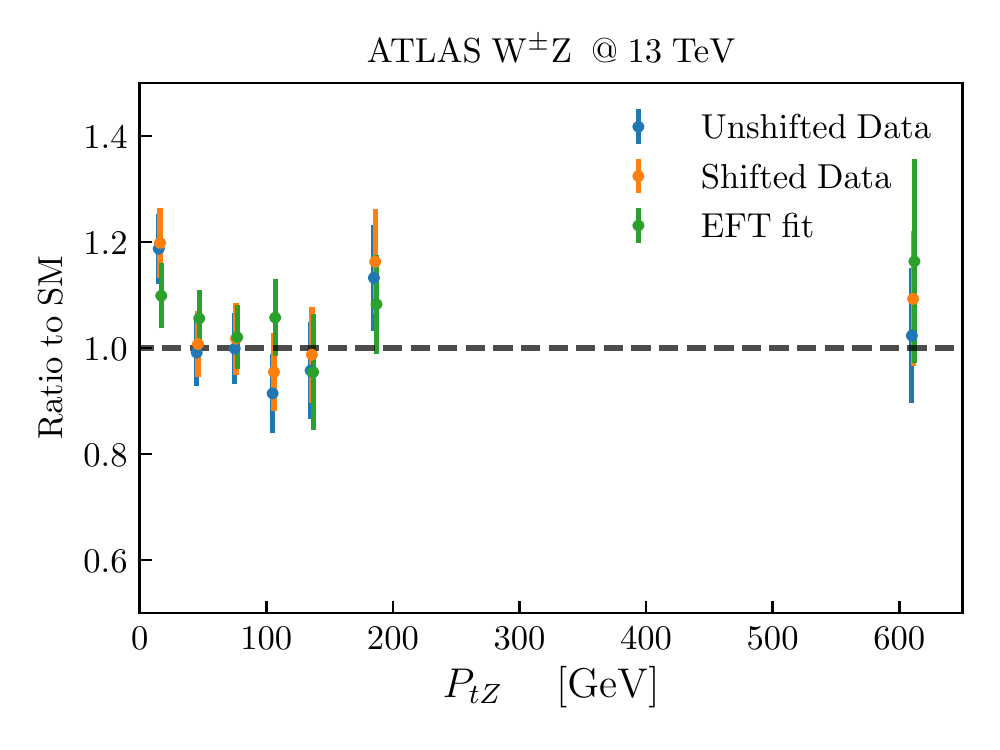}
  \includegraphics[width=0.49\textwidth]{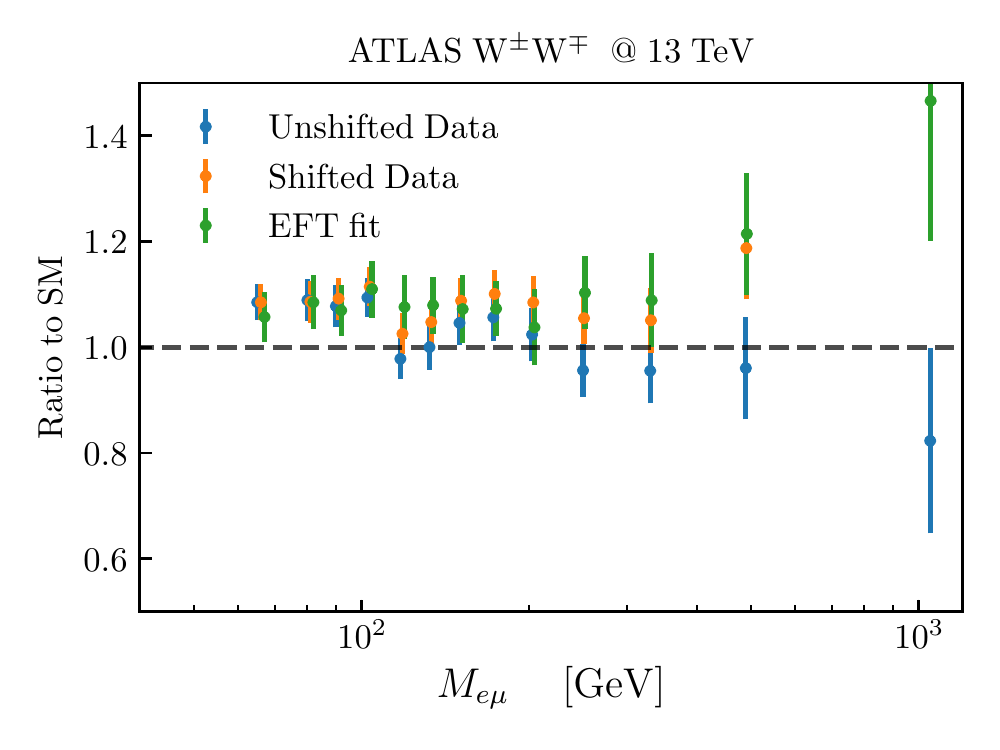}
  \caption{\small Comparison between experimental data and best-fit EFT theory
      predictions
      for the LHC diboson distributions considered in the present analysis.
      Both the data and the EFT fit results are normalised to the central value of the SM prediction.
      The band in the EFT prediction indicates the post-fit 95\% CL uncertainty. }
  \label{fig:DvTdiboson}
\end{figure}

\begin{figure}[t]
  \centering
  \includegraphics[width=0.49\textwidth]{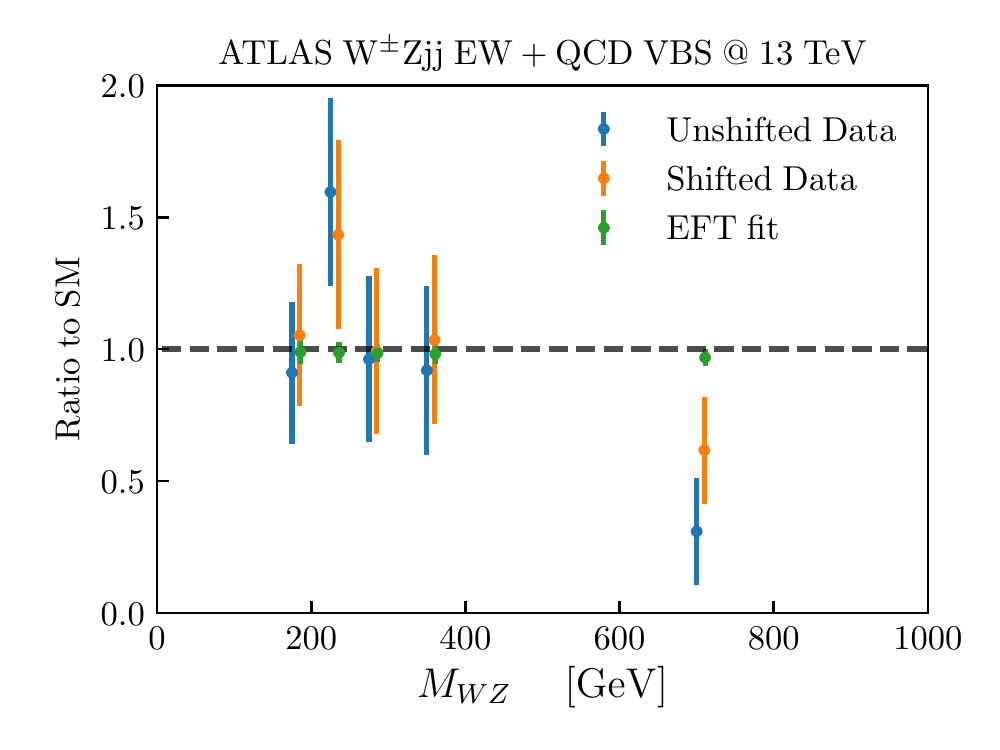}
  \includegraphics[width=0.49\textwidth]{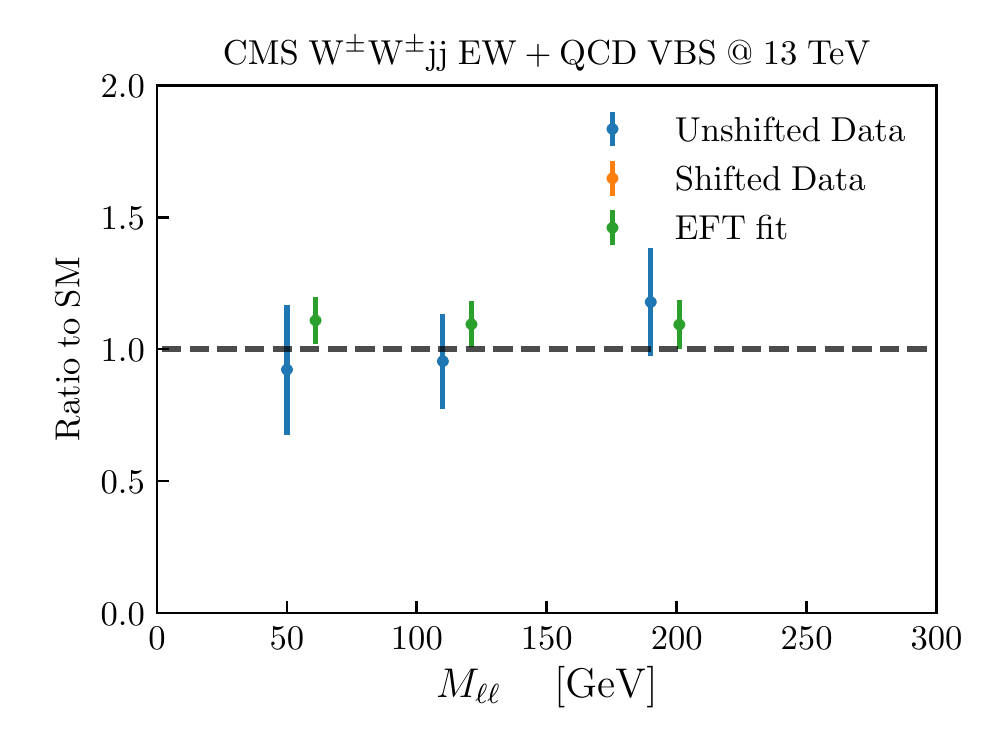}
  \includegraphics[width=0.49\textwidth]{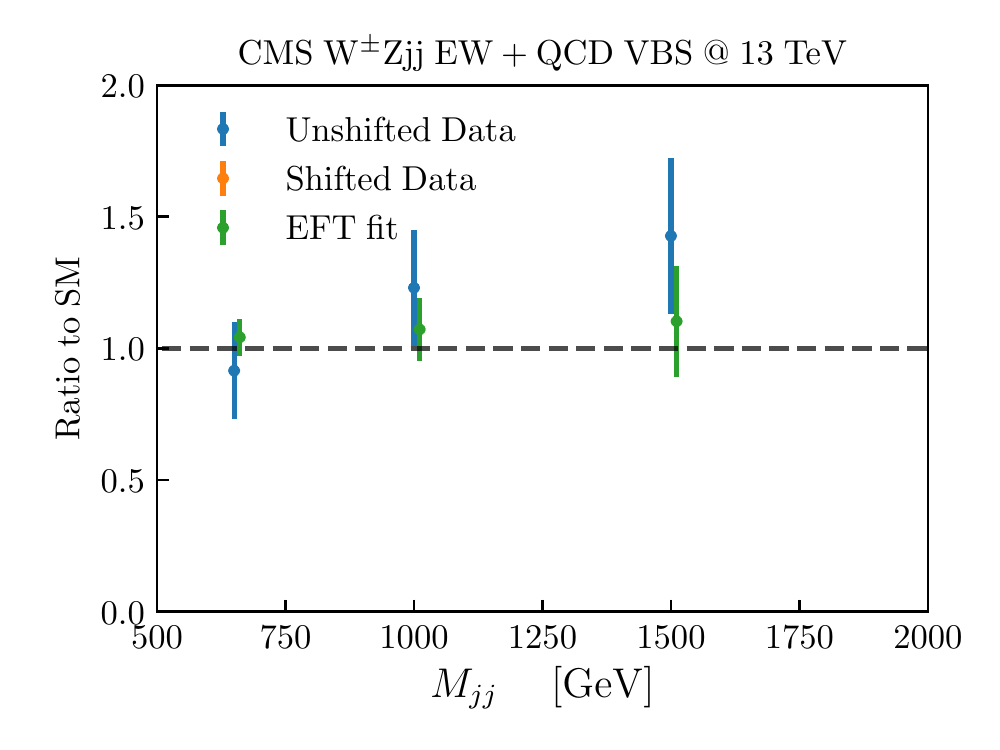}
  \includegraphics[width=0.60\textwidth]{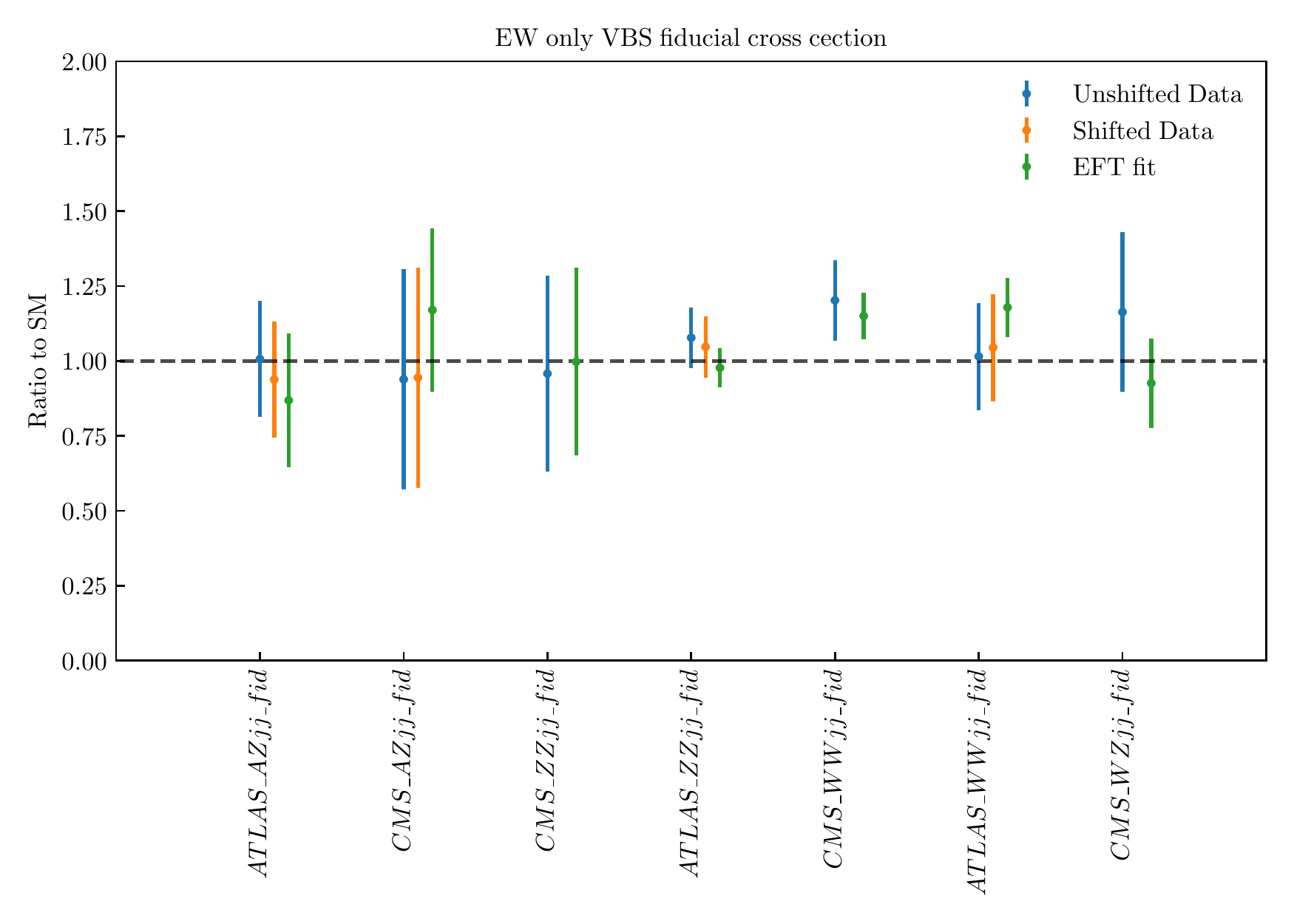}
  \caption{\small  Comparison for the VBS measurements,
    both for the  unfolded differential distributions (top panel),
  and for the EW-only fiducial cross-sections (bottom panel).}
  \label{fig:DvTvbs}
\end{figure}

Fig.~\ref{fig:DvTdiboson} displays a
comparison between experimental data and best-fit EFT theory
predictions
for the LHC diboson distributions considered in the present analysis.
We show the results for the $W^\pm Z$, $W^\pm W^\mp$ and $ZZ$ final states from CMS
in the upper panels and the corresponding  $W^\pm Z$ and $W^\pm W^\mp$
distributions from ATLAS in the lower panels.
Both the data and the EFT fit results are normalised to the central value of the SM prediction.
The experimental data is presented as both unshifted in central values
(where the error band represents the total error) and with
the best-fit systematic shifts having been subtracted (so that the error band
contains only the statistical component).
The band in the EFT prediction indicates the post-fit 95\% CL uncertainty.
For the datasets in which the information on correlated systematics
is not available, only the unshifted data is shown.
In Fig.~\ref{fig:DvTvbs}, we show a similar comparison as that of Fig.~\ref{fig:DvTdiboson} but
now for the VBS measurements.
In all cases a fair agreement is observed between experimental data and SM and EFT
theory predictions, consistent with the $\chi^2$ values reported
in Table~\ref{tab:chivals}.

\subsection{Constraints on the EFT parameter space}
\label{sec:fitresult}

\begin{figure}[t]
    \centering
     \includegraphics[width=\textwidth]{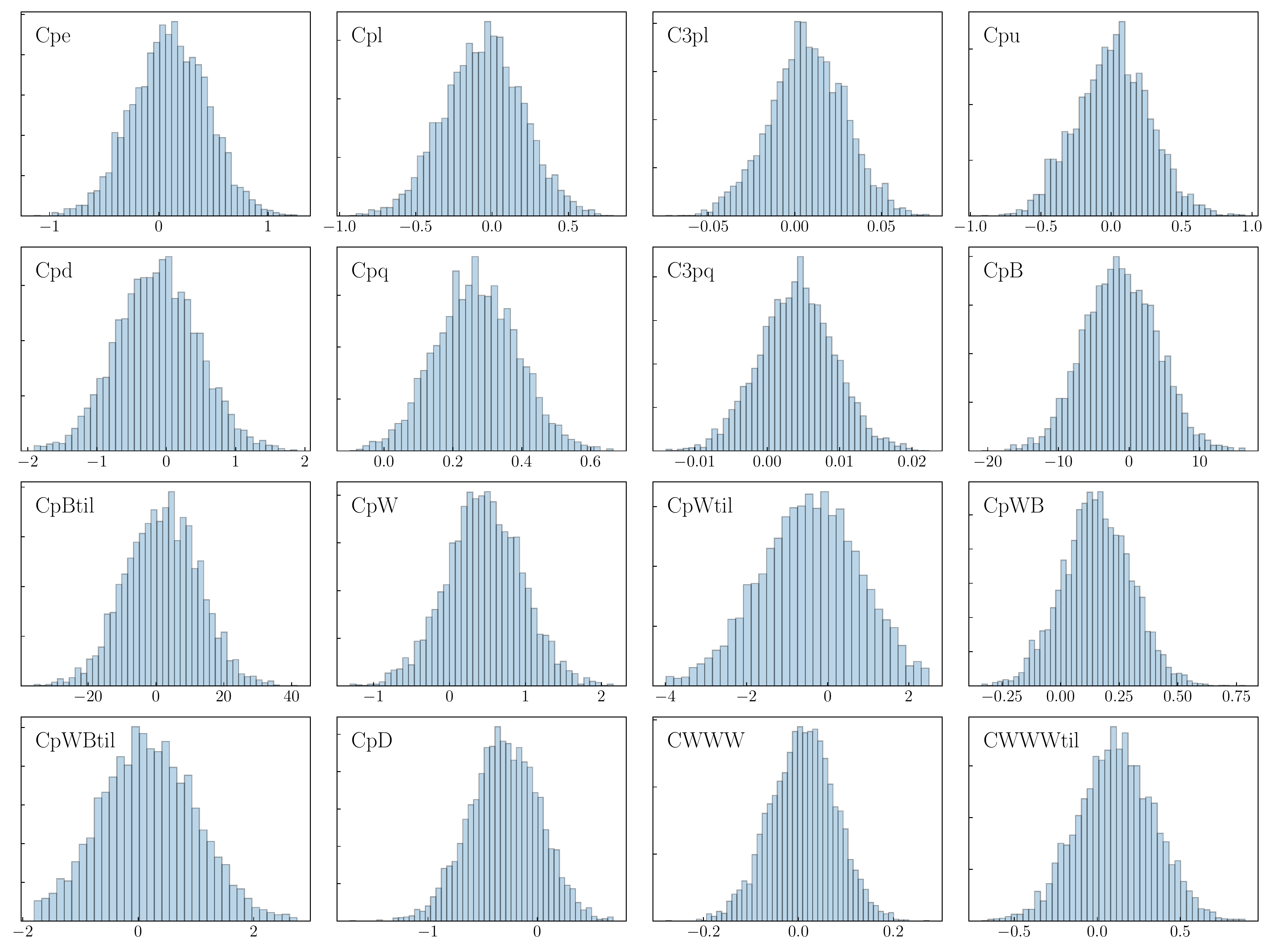}
     \caption{\small The posterior probability distributions associated to each
       of the $n_{\rm op}=16$ coefficients that are constrained in this analysis
       for the baseline dataset. Note that the $x$-axis ranges are different for each coefficient.
     }
    \label{fig:coeff_distributions}
\end{figure}

We now present the constraints on the coefficients
of the dimension-six EFT operators used to interpret the VBS and diboson
cross-sections listed in Table~\ref{tab:chivals}.
In Fig.~\ref{fig:coeff_distributions}, we display
the posterior probability distributions associated to each
of the $16$ coefficients that are constrained in this analysis
for the baseline dataset.
In all cases, we can see that these are approximately Gaussian, as expected for
a linear EFT fit without flat directions.
The latter result is consistent with observations derived from the PCA
in Fig.~\ref{fig:PCA}, and confirm that
the input dataset is sufficient to constrain all 16 independent directions
in the EFT parameter space.

\clearpage

\input{tables/SMEFiT_results_table.tex}

\begin{figure}[htbp]
  \centering
  \includegraphics[width=\textwidth]{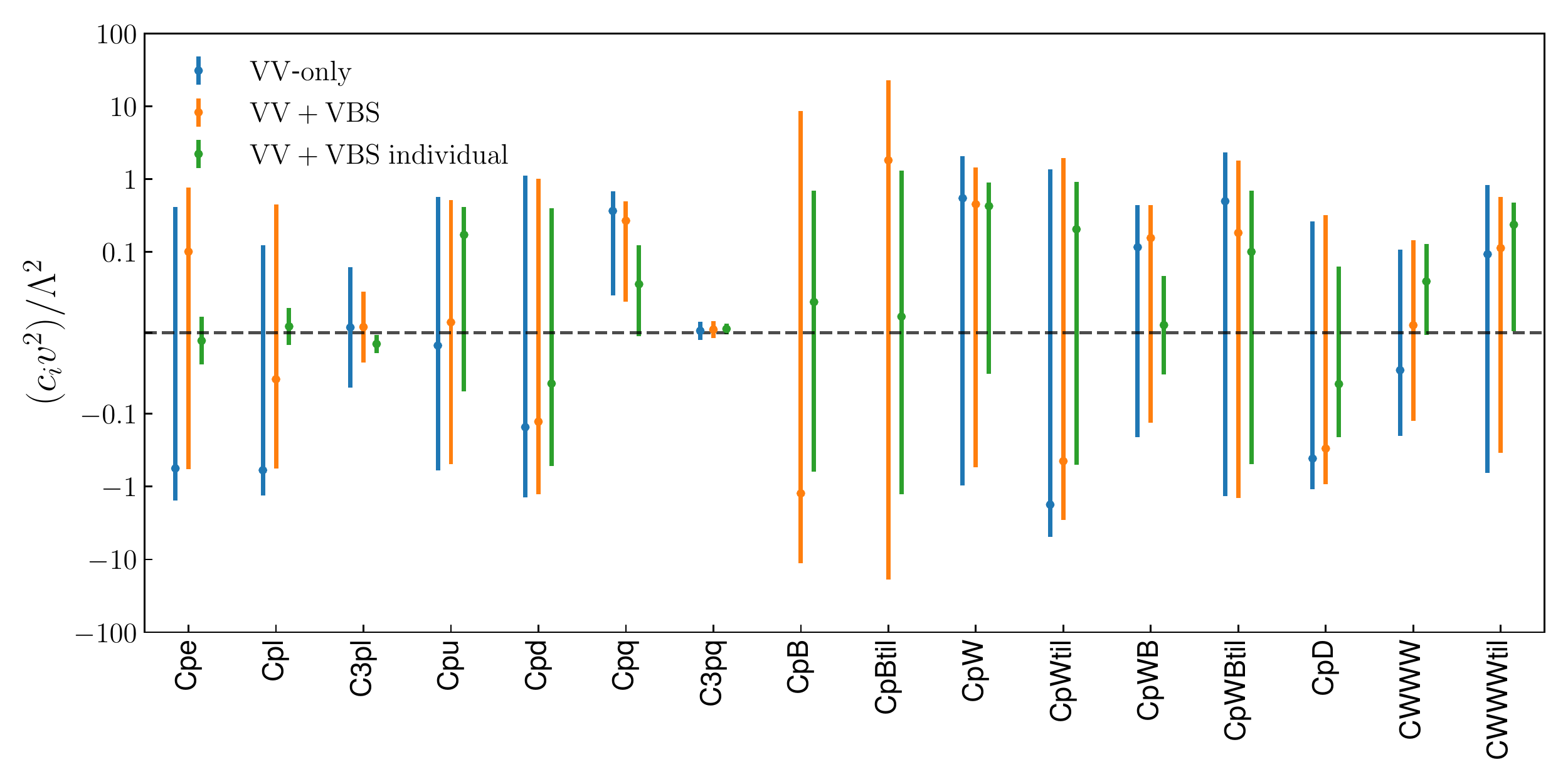}
    \includegraphics[width=\textwidth]{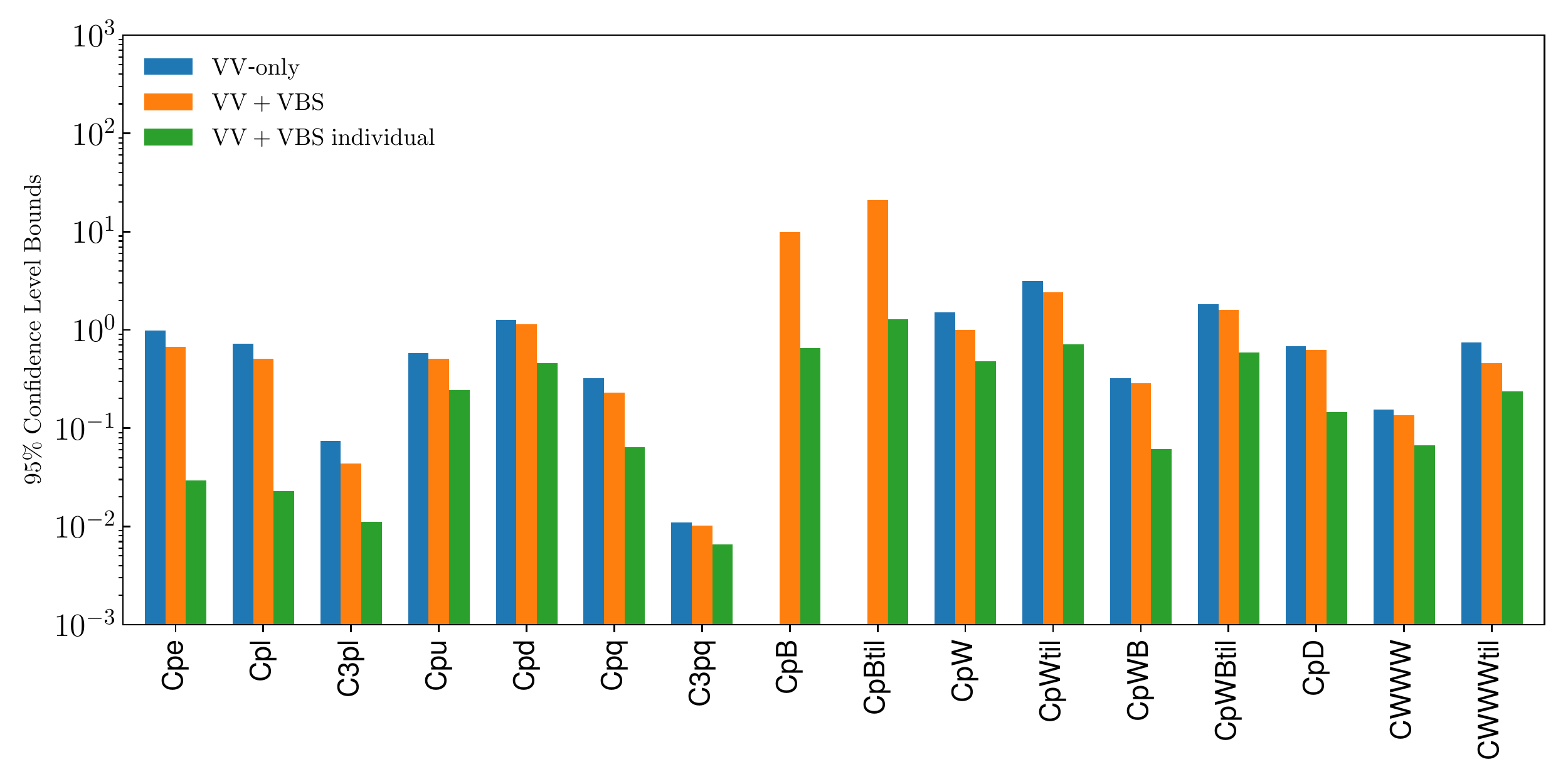}
     \caption{\small Graphical representation
       of the results of  Table~\ref{tab:finalbounds}, displaying the absolute value (upper)
       and the magnitude (bottom panel) of the 95 $\%$ CL intervals associated to each
       of the 16 EFT operators considered here.
       We compare the marginalised results of a diboson-only fit (blue) with the same fit
       once VBS data is added (orange) in both cases when all coefficients are fitted simultaneously.
       For reference, we also show the results of the individual VBS+diboson fits,
       where only one operators is varied at the time and the rest are fixed to their SM value.
     }
    \label{fig:MainBounds}
\end{figure}

From these posterior probability distributions,
the 95\% confidence level intervals associated to each
of the fit coefficients can be evaluated.
Table~\ref{tab:finalbounds} displays these
95\% CL intervals associated
to all 16 degrees of freedom.
Moreover, a comparison is made between the results of the baseline VBS+diboson fit performed at the global (marginalised) and individual levels, as well as with a fit based only on the diboson cross-sections.
In the fourth column (individual fits), only one coefficient is varied at a
time while all others are set to their SM values.
The results of Table~\ref{tab:finalbounds} are also graphically
represented in  Fig.~\ref{fig:MainBounds}, which
displays the absolute value (upper)
and the magnitude (bottom panel) of these 95 $\%$ CL intervals.

From the comparison between the 95\% CL intervals
in Table~\ref{tab:finalbounds} and Fig.~\ref{fig:MainBounds},
several interesting observations can be made.
First, in comparing the results of the combined VBS+diboson fit with the diboson-only analysis, the VBS measurements are seen to improve the bounds provided by the diboson data in a pattern consistent with the Fisher information matrix displayed in Fig.~\ref{fig:FisherMatrix}.
For instance, the bounds on $\bar{c}_{\varphi W}$ improve from $\lc-0.97,+2.1\rc$
to $\lc -0.55,+1.4\rc$, while those on the CP-even (odd) triple gauge operator $\bar{c}_{W}$
($\bar{c}_{\widetilde{W}}$) are reduced
from  $\lc -0.20,+0.11\rc$ ($\lc -0.63,+0.85\rc$) down to  $\lc -0.13,+0.14\rc$ ($\lc -0.35,+0.57\rc$).
In all cases, the VBS data improve the bounds
on the EFT coefficients obtained from the diboson-only fit, highlighting
the consistency and complementarity between the two families of processes.
This result applies both to the CP-even as well as the CP-odd operators.

Another relevant observation from Table~\ref{tab:finalbounds} concerns
the differences between the marginalised and individual fits in the case
of the combined VBS+diboson analysis, which illustrates the role
of the correlations between the operators that modify these two processes.
In the individual fits, one finds more stringent bounds by artificially setting
all other EFT operators to zero, and this distorts the physical
interpretation of the results.
For several operators, the individual bounds underestimate the results
of the 16-dimensional fit by an order of magnitude or more.
This highlights the importance of accounting for all relevant EFT
operators that contribute to a given process rather than just selecting
a subset of them, as has often been the case in the interpretation of VBS measurements.

\begin{figure}[t]
    \centering
    \includegraphics[width=0.7\textwidth]{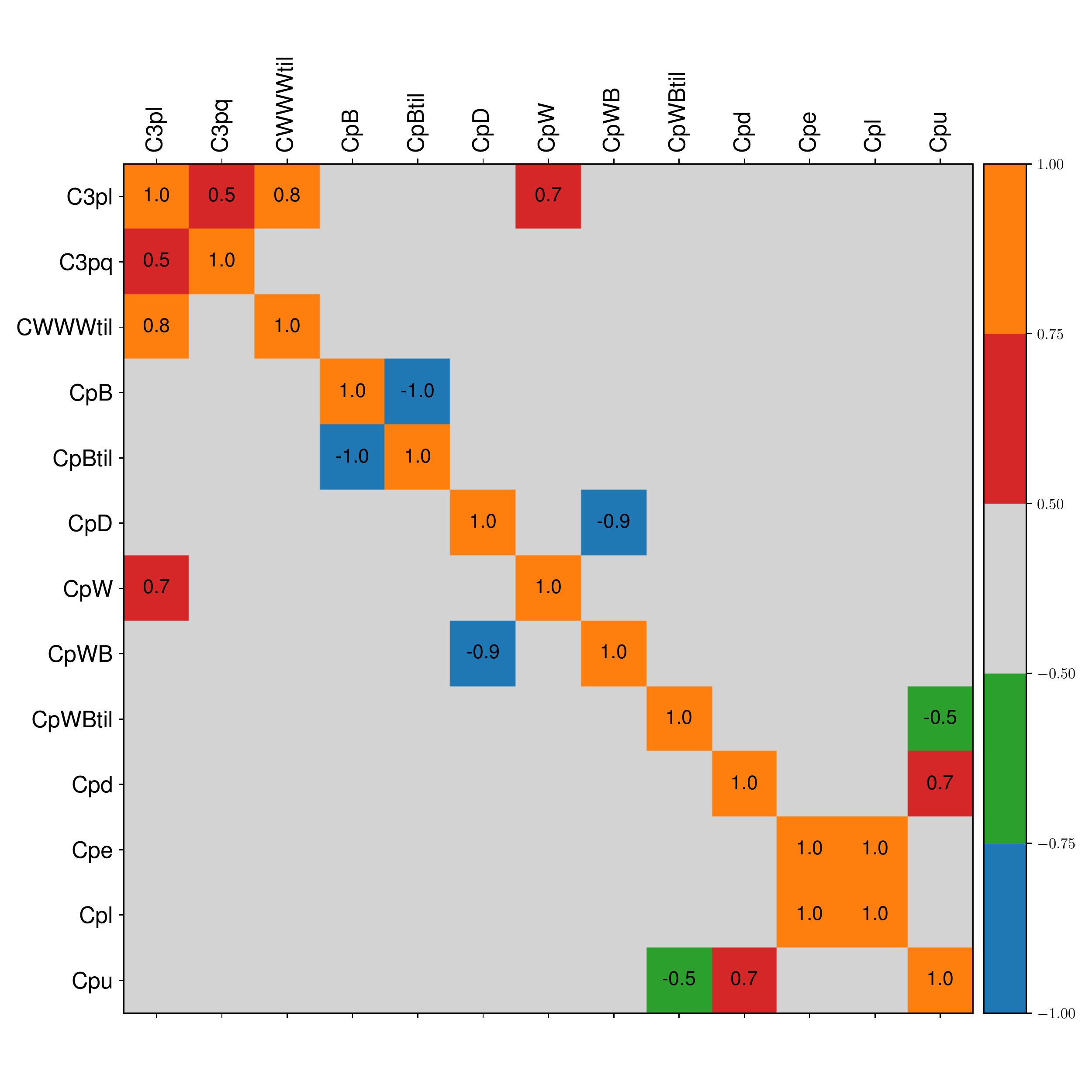}
    \caption{\small The values of the correlation coefficients between the
      operators considered in the fit to the baseline dataset.
      These are categorised
      as positively correlated ($\rho \ge 0.50\ (0.75)$, in red (orange)),
      negatively correlated ($\rho \le 0.50\ (0.75)$, in green (blue)),
      and uncorrelated ($|\rho| \le 0.5$, in grey).
    }
    \label{fig:correcoeff}
\end{figure}

Fig.~\ref{fig:correcoeff} then displays the values of the correlation coefficient between the operators considered in the fit to the baseline dataset.
For some pair-wise combination of operators we observe strong (anti-)correlations
between the fit coefficients,
for example $c_{\varphi B}$ and $c_{\varphi \widetilde{B}}$ are strongly anticorrelated,
and the same holds for $c_{\varphi D}$ and $c_{\varphi W B}$.
However, in most cases, these correlations turn out to be quite small, confirming that our choice
of fitting basis is suitable to describe efficiently the available dataset
in consistency with the PCA results.

\begin{figure}[t]
    \centering
    \includegraphics[width=\textwidth]{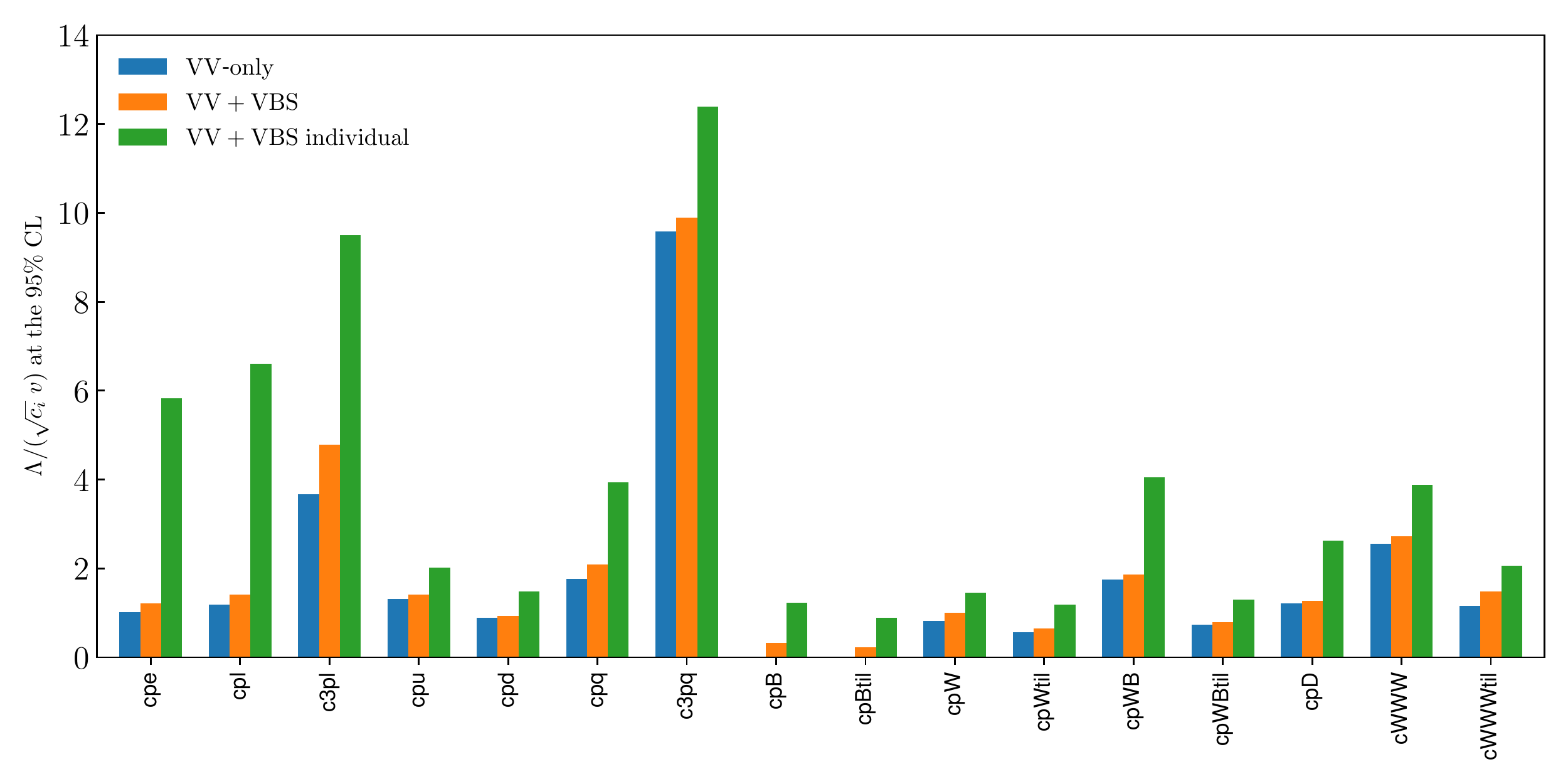}
    \caption{\small Same as the lower panel of Fig.~\ref{fig:MainBounds} now
      representing the 95\% CL bounds on $\Lambda/(v \sqrt{c_i})$. }
    \label{fig:energycoeff}
\end{figure}

Finally, in  Fig.~\ref{fig:energycoeff} we display
the 95\% CL lower bounds on the value of $\Lambda/(v \, \sqrt{c_i})$.
These bounds can be  interpreted
as the lower bounds derived from the EFT fit
on the scale of new physics $\Lambda$ in UV-completions where the corresponding
Wilson coefficients are $c_i=\mathcal{O}\lp 1\rp$. They are again presented as dimensionless quantities, measured in units of vev.
This interpretation can be adjusted to other BSM scenarios, for example
in the case of strongly coupled theories where one expects $c_i=\mathcal{O}\lp 4\pi\rp$.
For several operators, the combined VBS+diboson analysis results in values above 1 TeV for the new physics scale $\Lambda$, for example the triple gauge operator $c_W$ which has $(\Lambda/\sqrt{c_i})\gsim 3 v$ at the 95\% CL.

\begin{figure}[t]
    \centering
    \includegraphics[width=\textwidth]{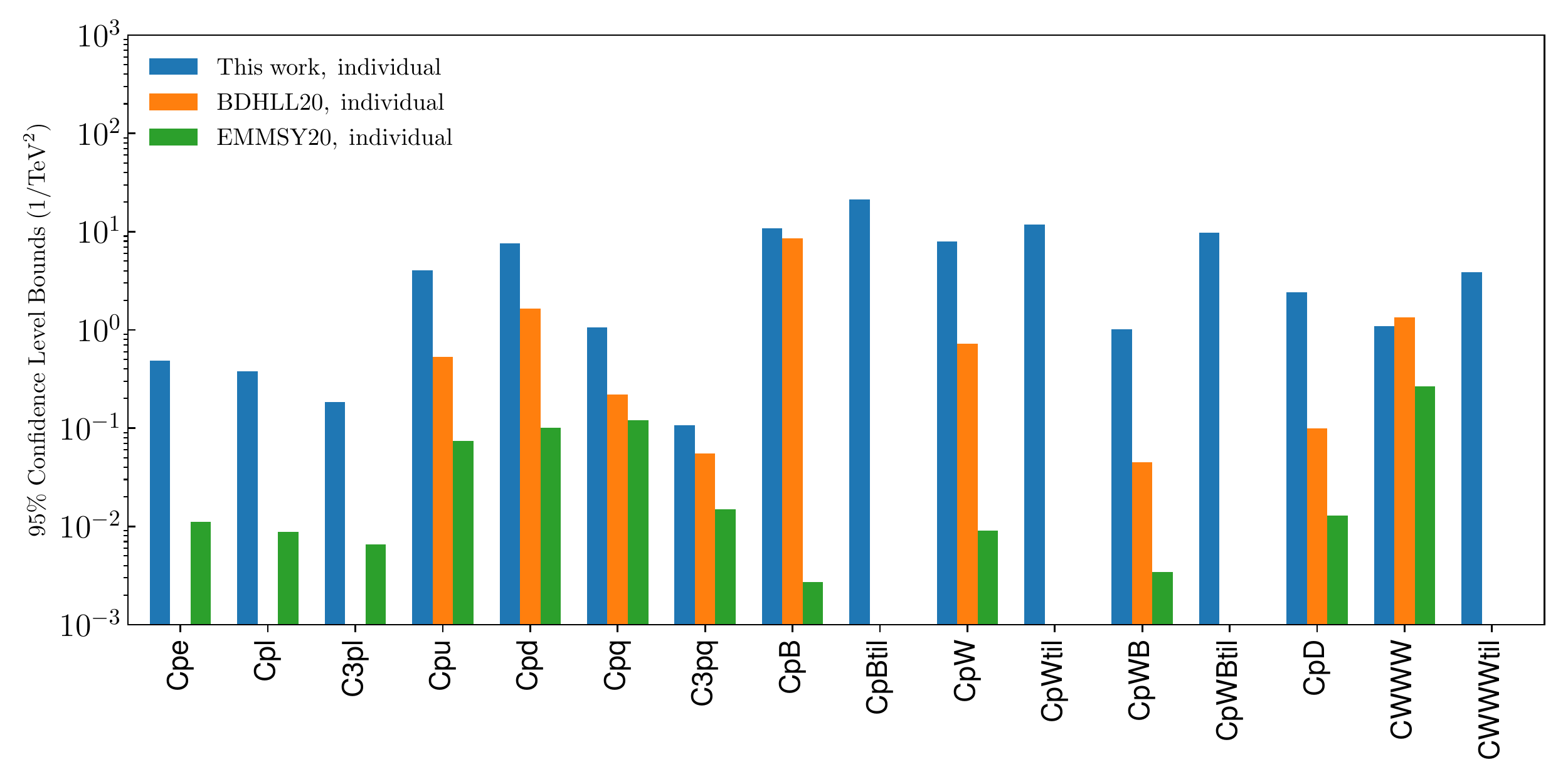}
    \caption{\small Comparison of the individual bounds obtained in this work from the VBS+diboson dataset (shown in Fig.~\ref{fig:MainBounds}) with the corresponding individual bounds obtained in the BDHLL20~\citep{Baglio:2020oqu} and EMMSY20~\citep{Ellis:2020unq} and EFT
      analyses, see text.
      In the three cases, only the linear terms in the EFT expansion are being included and the EFT cross-sections are evaluated
    at leading order. \label{fig:Coeffs_Bar_indivd_comapre}}
\end{figure}

\paragraph{Comparison with other EFT analyses.}
Fig.~\ref{fig:Coeffs_Bar_indivd_comapre} displays a comparison between
the individual bounds obtained in this work, based on the VBS+diboson dataset and shown in Fig.~\ref{fig:MainBounds},
with the corresponding individual bounds obtained in the BDHLL20~\citep{Baglio:2020oqu} and EMMSY20~\citep{Ellis:2020unq} EFT
analyses.
The BDHLL20 fit includes data on diboson cross-sections from the LHC
together with information from the associated production of a Higgs with a vector boson, $hW$ and $hZ$.
EMMSY20 is instead a global EFT interpretation that includes Higgs and top production data together with the EWPOs from LEP
and some diboson cross-sections.
For the three sets of results shown in Fig.~\ref{fig:Coeffs_Bar_indivd_comapre}, only the linear terms in the EFT
expansion are being included and the EFT cross-sections are evaluated at leading order.\footnote{We note that the BDHLL20 analysis has been performed also accounting for NLO QCD corrections in the EFT cross-sections, here we use the LO ones for the sake of comparison.}
Given that these three analyses are based on different subsets of dimension-six operators, a comparison at the level of individual constraints is the most direct way of interpreting similarities or differences.
We also note that CP-odd operators are only considered in this analysis.

For the majority of operators, the global study of EMMSY20 exhibits the superior sensitivity.
Our good determination of $c_{W}$ can be traced back to the inclusion of the $WZ$ differential distributions from ATLAS and CMS, which are also included in BDHLL20, but absent in EMMSY20, where $Zjj$ is included instead. This fact hints that a combined analysis of $WZ$ and $Zjj$ might shed more light on the purely gauge operator.
The results of the global EFT fit lead to more stringent bounds as compared to those from this work and from BDHLL20, especially
for the purely bosonic operators $c_{\varphi B}$, $c_{\varphi W}$ and $c_{\varphi B W}$, which are significantly constrained
both by the EWPOs from LEP as well as Higgs measurements.
For most coefficients, our individual results and those of  BDHLL20 are in good agreement, in particular for bosonic operators $c_{\varphi D}$,  $c_{\varphi B W}$, and $c_{\varphi W}$, $c_{W}$. This is what we would expect, given the datasets chosen.

The comparison of the three works shows that Higgs, LEP and EWPD measurements represent the leading contributions to the parametrisation of BSM effects. There are also enough hints that a global interpretation of the LHC data, independent of older measurements is also a feasible way to go further on the road to the most accurate EFT interpretation.

\subsection{Dataset dependence}

Until now, we have focused only on the analysis of the EFT fit results for the baseline dataset listed in Table~\ref{tab:chivals}.
In the following, we assess the dependence of these results with respect to variations in the input data and theory settings by performing VBS-only fits and studying the impact of the VBS detector-level distributions when added to the VBS-only and to the baseline VBS+diboson fits.
We also present fits where the CP-odd operators are set to zero and only the CP-even ones remain.

\begin{figure}[t]
    \centering
    \includegraphics[width=\textwidth]{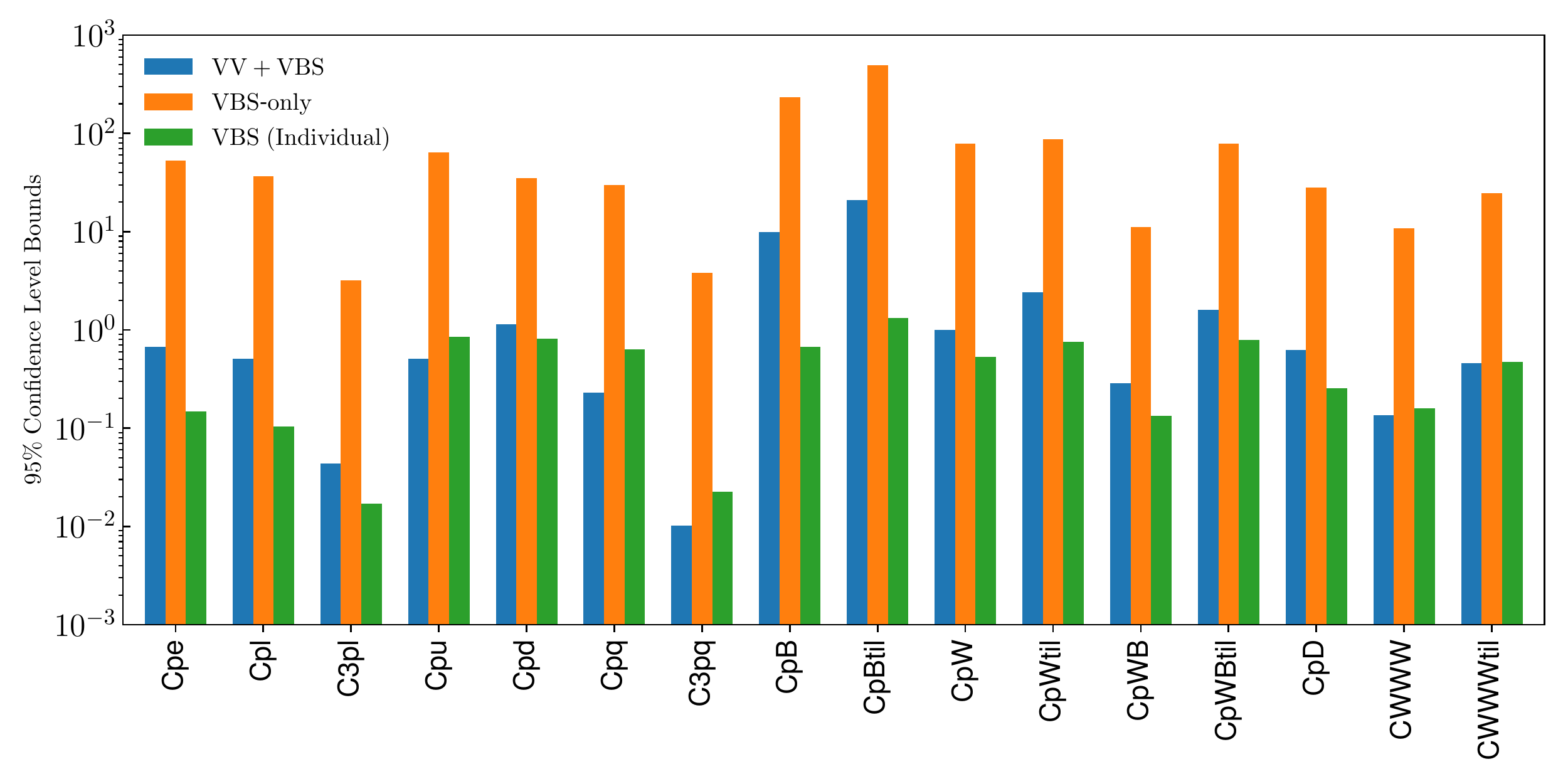}
    \caption{\small Comparison of the  95\% CL intervals in the baseline fit
      with those resulting from marginalised and individual VBS-only fits.
      Only the unfolded VBS cross-section measurements
      listed in Table~\ref{tab:datasettable_VBS} are being included in the fits.
    \label{fig:VBSonlyUnfolded}}
\end{figure}

\paragraph{VBS-only fits.}
First of all, we have verified through a dedicated PCA that flat directions in the EFT parameter space are absent also in the case of a VBS-only fit .
However, the same analysis also reveals that some combinations of coefficients
will be poorly constrained.
The latter result is not unexpected, given that for a VBS-only dataset we have $n_{\rm op}=16$ parameters to fit with only $n_{\rm dat}=18$ data points.
We display in Fig.~\ref{fig:VBSonlyUnfolded} the same 95\% CL intervals
as in the lower panel of Fig.~\ref{fig:MainBounds}, but now comparing the  results of our baseline fit with those obtained from the marginalised and individual VBS-only fits.
By comparing the VBS+diboson with the VBS-only fits, we see that the obtained bounds in the latter case are much looser by a factor between 10 and 100 for most operators.
These findings are consistent with our previous observations that current VBS data provides only a moderate pull when added together with the diboson cross-sections.

However, we would like to emphasize that this result does not imply
that VBS-only fits cannot
provide competitive sensitivity in a EFT analysis, but rather that the available VBS measurements are still scarce and limited by statistics.
In fact, if one compares the results of the marginalised with the individual VBS-only fits, one can see that the individual bounds are notably reduced and become similar, or even better, than in the baseline VBS+diboson analysis.
This implies that VBS processes are endowed with a unique potential to constrain the dimension-six operators of the SMEFT,
but only once sufficient data has been collected to pin down the effects of the individual operators separately.
We will verify this expectation in Sect.~\ref{sec:hllhc} through EFT fits
based on dedicated HL-LHC  projections.

\paragraph{The impact of the VBS detector-level measurements.}
As was discussed in Sect.~\ref{sec:expdata}, one can in principle use detector-level measurements in the EFT fit in addition to the unfolded VBS cross-sections and distributions measured by ATLAS and CMS.
Here we consider the $m_{ZZ}$ and $p_T^{\ell \ell \gamma}$ distributions from CMS and ATLAS in the $ZZjj$ and $\gamma Z jj$ final states respectively, which consist of 15 data points that can be included together with the unfolded VBS cross-section measurements.
Given that our modelling of the detector response is basically
reduced to a flat acceptance correction, we have chosen to remove these
data points from the baseline results presented in the previous section.
We would therefore like to illustrate how these detector-level distributions contain valuable information and are particularly instrumental to realise
a reliable VBS-only EFT dimension-six analysis.

\begin{figure}[t]
    \centering
     \includegraphics[width=0.9\textwidth]{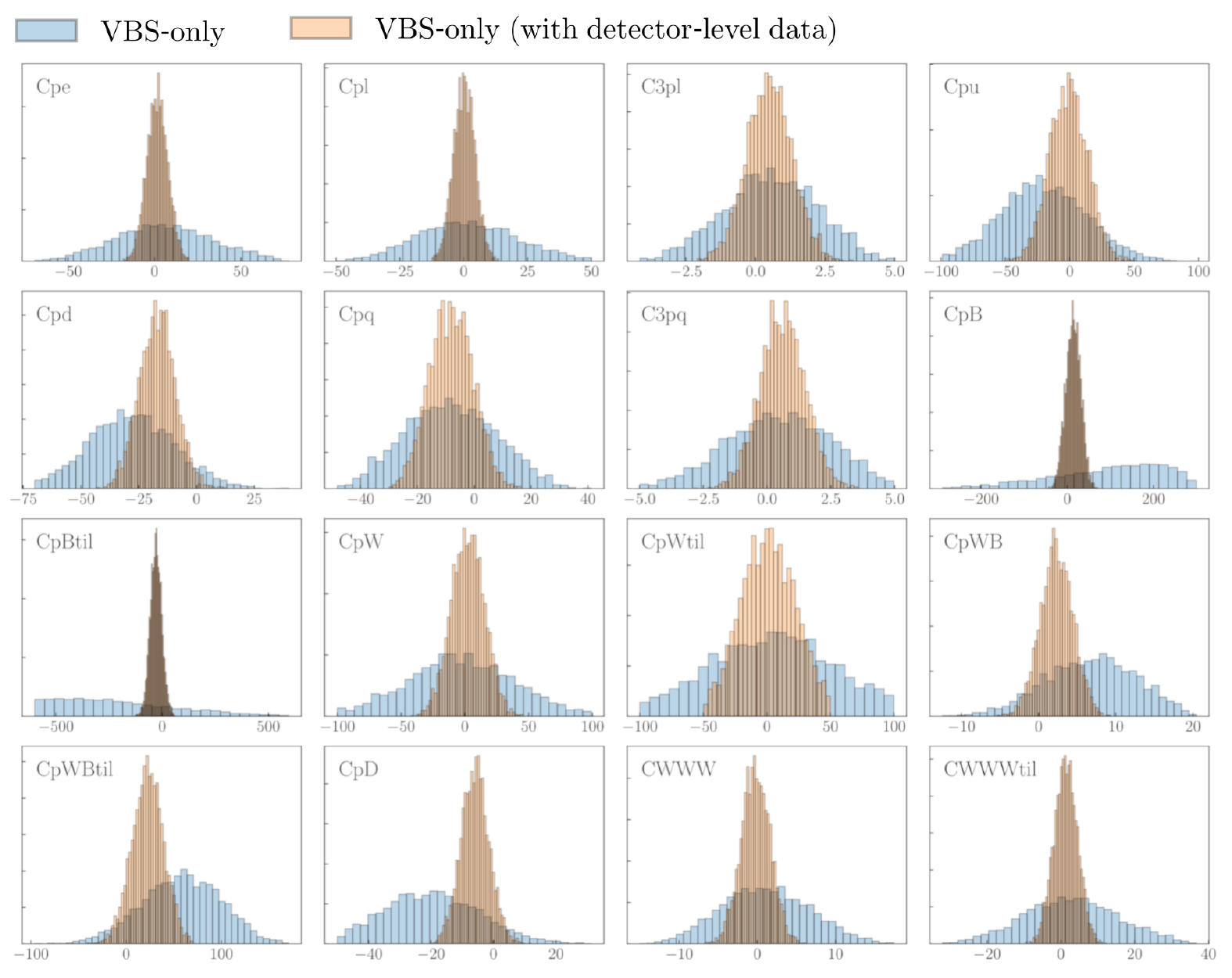}
     \caption{\small Posterior distributions associated to the VBS-only fits
       that include only unfolded cross-sections (blue) and also the
       detector-level distributions (orange).
     }
    \label{fig:coeff_distributions_vbsonly}
\end{figure}

Fig.~\ref{fig:coeff_distributions_vbsonly} displays the
same posterior probability distributions as in Fig.~\ref{fig:coeff_distributions} but
now corresponding to the VBS-only fits.
We compare the results of the analysis based only on unfolded cross-sections with that in which the two detector-level distributions mentioned above are also included.
While the VBS-only fit based on unfolded cross-sections does not exhibit genuine flat directions,
several coefficients end up poorly constrained.
The situation is different once the detector-level distributions
are added to the fit: here the posterior distributions become Gaussian-like, and their width is markedly reduced compared to the previous case.
In particular, the inclusion of
the $m_{ZZ}$ and $p_T^{\ell \ell \gamma}$ detector-level distributions is particularly helpful in strengthening the VBS-only bounds on $c_{\varphi B}$ and its CP-odd counterpart.

\begin{figure}[t]
    \centering
    \includegraphics[width=\textwidth]{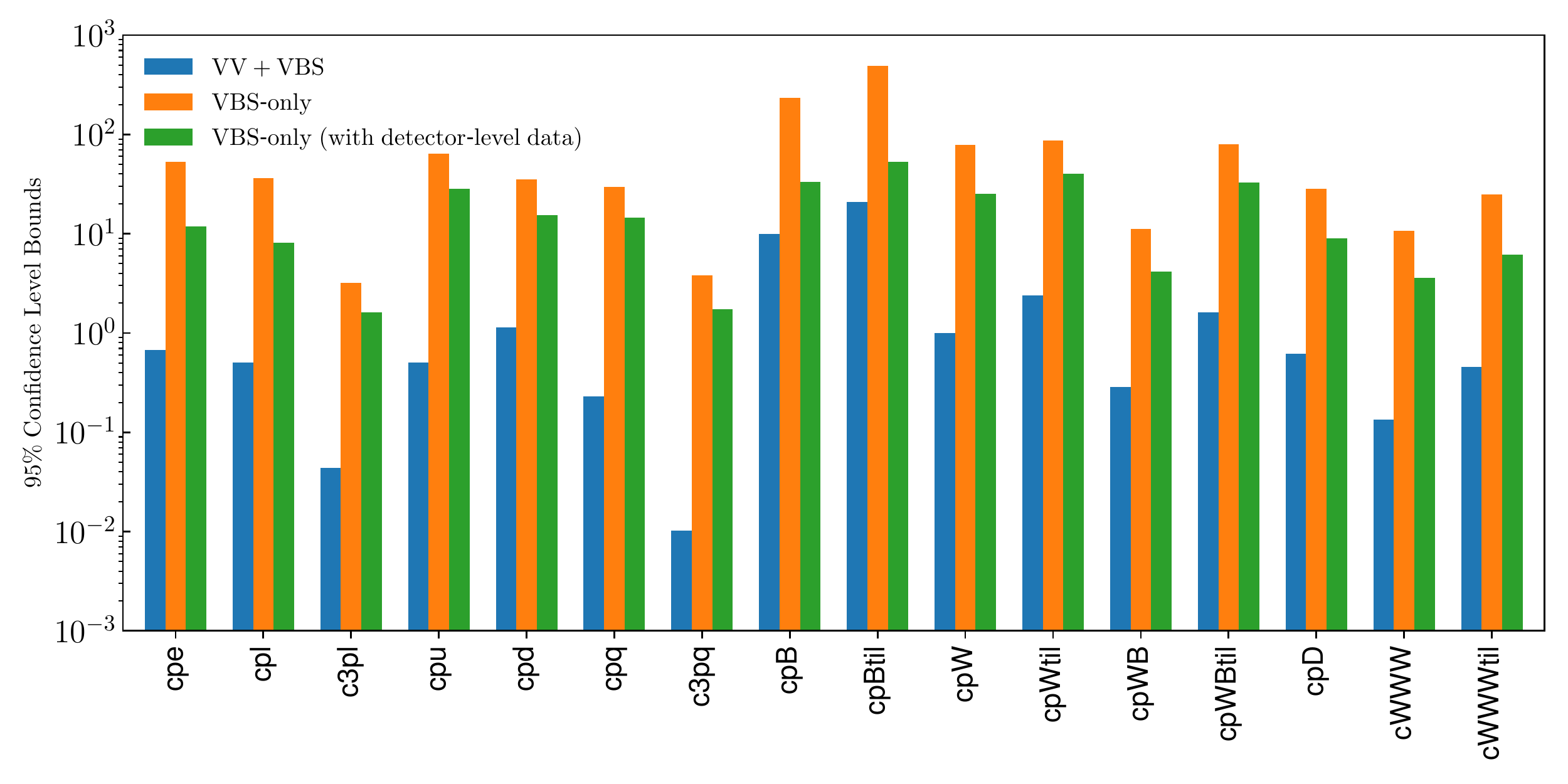}
    \caption{\small Comparison
      of the results of the VBS-only fit based only on unfolded cross-sections with those
      of the same fit where in addition one includes the
detector-level  distributions.
For reference, we also display the results for baseline VBS+diboson
dataset.
    \label{fig:VBSonlyFolded}}
\end{figure}

The 95\% CL intervals associated to the posterior probability distributions of
Fig.~\ref{fig:coeff_distributions_vbsonly} are then represented in
Fig.~\ref{fig:VBSonlyFolded}, where for reference
we also display the results of the baseline VBS+diboson fit.
We find that by adding the detector-level distributions, there is a noticeable improvement in the result of the VBS-only fit, with bounds being reduced by a factor between two and ten depending on the specific operator.
In the case of $c_{\varphi B}$, the resulting bound becomes comparable to that
obtained in the VBS+diboson fit, though in general the VBS-only fit cannot compete with the combined VBS+diboson results even after the addition of the folded data.
These results motivate the release of
all available VBS measurements in terms of unfolded distributions.
We have verified that in the case of the combined VBS+diboson fit, adding the detector-level
measurements leaves the results essentially unaffected, providing a further
justification of our choice of removing them from the baseline dataset.

\begin{figure}[t]
    \centering
    \includegraphics[width=\textwidth]{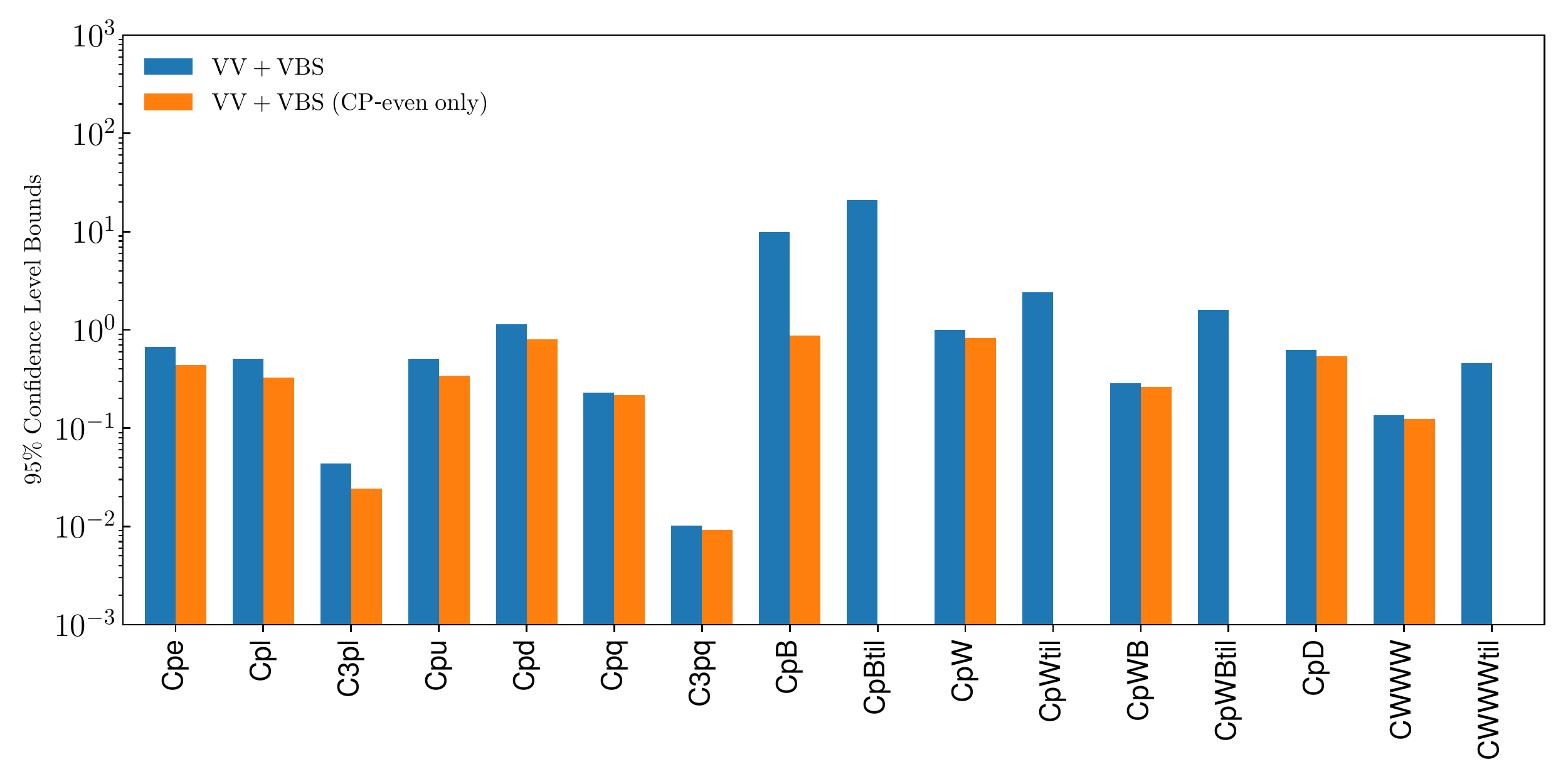}
    \caption{\small Comparing
      the results of the baseline fit with those
      of the same fit where the CP-odd operators
      have been set to zero, such that only the CP-even ones remain.
    \label{fig:bounds_CPeven}}
\end{figure}

\paragraph{The impact of CP-odd operators.}
Finally, we assess how the EFT fit results are modified once only CP-conserving operators are considered.
Fig.~\ref{fig:bounds_CPeven} compares
the results of the baseline VBS+diboson fit with those
of the same fit where the CP-odd operators
have been set to zero, such that only the CP-even ones remain.
In general the differences are quite small, and as expected the fit without CP-violating operators leads to somewhat more stringent bounds.
The only operator for which removing the CP-odd operators has a significant
effect is $c_{\varphi B}$, where a  difference of an order of magnitude
in the 95\% CL bound is observed.
The reason for this behaviour is that, as indicated in the correlation heat map
of  Fig.~\ref{fig:correcoeff}, $c_{\varphi B}$ and $c_{\varphi \widetilde{B}}$
are strongly anti-correlated and thus in general it is rather challenging
to disentangle them.

%% file: tables/chi2_table.tex
\begin{table}[t]
  \footnotesize
    \centering
    \renewcommand{\arraystretch}{1.3}
    \begin{tabular}{l|l|c|c|c}
      Process  & $\quad\quad$ Dataset  $\quad\quad$   & $\quad n_{\rm dat}\quad $  & $\chi^2 /n_{\rm dat}$ (SM)  &
      $\chi^2/n_{\rm dat}$ (EFT) \\
    \toprule
\multirow{6}{*}{Diboson}  
&   \texttt{ATLAS\_WW\_memu}    & 13    &  0.70 &  0.66  \\
&   \texttt{CMS\_WW\_memu}      & 13    &  1.28 &  1.32 \\
&   \texttt{ATLAS\_WZ\_ptz}     & 7     &  1.38 &  0.93 \\
&   \texttt{CMS\_WZ\_ptz}       & 11    &  1.48 &  1.14 \\
&   \texttt{CMS\_ZZ\_mzz}       & 8     &  1.17 &  0.74 \\
&   {\bf Total diboson}               & {\bf 52}    &  {\bf 1.17} & {\bf 0.97}  \\
\midrule
\multirow{11}{*}{VBS}  
&   \texttt{ATLAS\_WWjj\_fid}   & 1     &  0.01 &  0.67 \\
&   \texttt{CMS\_WWjj\_fid}     & 1     &  2.17 &  0.15 \\
&   \texttt{CMS\_WWjj\_mll}     & 3     &  0.31 &  0.45 \\   
&   \texttt{ATLAS\_WZjj\_mwz}   & 5     &  1.60 &  1.52 \\
&   \texttt{CMS\_WZjj\_fid}     & 1     &  0.38 &  0.79 \\
&   \texttt{CMS\_WZjj\_mjj}     & 3     &  1.10 &  0.73 \\   
&   \texttt{ATLAS\_ZZjj\_fid}   & 1     &  0.09 &  0.15 \\
&   \texttt{CMS\_ZZjj\_fid}     & 1     &  0.02 &  0.02 \\
&   \texttt{ATLAS\_AZjj\_fid}   & 1     &  0.00 &  0.25 \\
&   \texttt{CMS\_AZjj\_fid}     & 1     &  0.03 &  0.38 \\
&   {\bf Total VBS }                  & {\bf 18}  & {\bf 0.83} & {\bf 0.75} \\
\midrule
    & \textbf{Total}               &  {\bf 70} & {\bf 1.084}  & {\bf 0.917}  \\
     \bottomrule
    \end{tabular}
    \caption{\small The values of the $\chi^2 /n_{\rm dat}$ for each dataset
      considered in the fit, as well as the totals in each category.
      We indicate the SM (pre-fit) results as well as the best-fit results
      once  EFT effects are accounted for, and separate the diboson (upper)
    from the VBS (bottom) datasets.} 
    \label{tab:chivals}
\end{table}

%% file: tables/SMEFiT_results_table.tex
\begin{table}[htbp]
  \small
  \centering
      \renewcommand{\arraystretch}{1.8}
    \begin{tabular}{c|c|C{2.5cm}|C{2.5cm}|C{2.5cm}}
      \multirow{2}{*}{Class} &  \multirow{2}{*}{ Coefficient} & VBS+diboson  & VBS+diboson &  Diboson-only \\
      &  & (marginalised)  &  (individual)  &  (marginalised)  \\
      \toprule
      \multirow{4}{*}{purely bosonic} &   $\bar{c}_{W}$  &  [-0.13,~0.14] &  [-0.001,~0.13] & [-0.20,~0.11]   \\
      \multirow{4}{*}{(CP-even)}  &  $\bar{c}_{\varphi W}$  & [-0.55,~1.4]  &  [-0.048,~0.91] &  [-0.97,~2.1]  \\
      &$\bar{c}_{\varphi B}$ & [-11,~8.8]  & [-0.62,~0.69]  & ---   \\
      &   $\bar{c}_{\varphi WB}$ & [-0.13,~0.44]  & [-0.050,~0.071]  & [-0.20,~0.44]   \\
      &  $\bar{c}_{\varphi D}$ &  [-0.93,~0.32] & [-0.21,~0.08]  & [-1.09,~0.26]   \\
      \midrule
      \multirow{3}{*}{purely bosonic}    & $\bar{c}_{\widetilde{W}}$ & [-0.35,~0.57] &  [-0.008,~0.46] &  [-0.63,~0.85]  \\
   \multirow{3}{*}{(CP-odd)}   & $\bar{c}_{\varphi \widetilde{W}}$  &  [-2.9,~1.8] &[-0.49,~0.93]   & [-4.9,~1.3]   \\
     & $\bar{c}_{\varphi \widetilde{W}B}$ & [-1.4,~1.8] & [-0.49,~0.69]  &  [-1.3,~2.4]  \\
      &$\bar{c}_{\varphi \widetilde{B}}$&   [-19,~23] & [-1.2,~1.4]  &  ---  \\
      \midrule
      \multirow{7}{*}{two-fermion}    &
      $\bar{c}_{\varphi l}^{(1)}$  & [-0.56,~0.45] & [-0.015,~0.031]  & [-1.3,~0.12]   \\
      &$\bar{c}_{\varphi l}^{(3)}$  & [-0.037,~0.051] & [-0.024,~-0.002]  &  [-0.068,~0.081]  \\
      &$\bar{c}_{\varphi q}^{(1)}$  & [0.043,~0.50] & [-0.007,~0.12]  & [0.038,~0.68]   \\
      &$\bar{c}_{\varphi q}^{(3)}$ & [-0.002,~0.011] & [-0.006,~0.014]  &  [-0.008,~0.013]  \\
      &$\bar{c}_{\varphi e}$ & [-0.58,~0.77] & [-0.038,~0.021]  & [-1.5,~0.41]   \\
      &$\bar{c}_{\varphi u}$ & [-0.49,~0.53] & [-0.073,~0.42]  &  [-0.59,~0.58]  \\
      &$\bar{c}_{\varphi d}$ & [-1.3,~1.0] & [-0.53,~0.39]  &  [-1.4,~1.2]  \\
      \bottomrule
    \end{tabular}
    \caption{\small The 95\% confidence level intervals associated
      to the 16 dimension-six EFT degrees
      of freedom considered in the present analysis.
      We compare the results of fits based on the baseline VBS+diboson dataset
      both at the global (marginalised) and the individual levels,~ as well
      as with those of a fit based only on the diboson cross-sections.
      Results shown here correspond to the dimensionless quantities $\bar{c} = c \cdot v^2/ \Lambda^2 $ and can be rescaled
      for any value of $\Lambda$, or presented as dimensioful quantities by extracting the powers of the vev.
      }
    \label{tab:finalbounds}
\end{table}

%% file: sec-hllhc.tex
\section{Vector boson scattering at the HL-LHC}
\label{sec:hllhc}

While the results presented in the previous section indicate the potential
of VBS measurements for dimension-6 EFT analyses, their
impact is currently limited by statistics.
The ultimate LHC sensitivity required
to constrain the coefficients of these dimension-6 operators
from VBS data will only be achieved
by legacy measurements based on the full HL-LHC luminosity
of $\mathcal{L}\simeq 3$ ab$^{-1}$ per experiment.
With this motivation, we generate HL-LHC pseudo-data for
EW-induced vector boson scattering processes
and quantify their impact on the EFT fit
by comparing the results to those presented in Sect.~\ref{sec:fitresult}.
The strategy adopted here is the same as the one used for the HL-LHC PDF projections
in Refs.~\cite{Khalek:2018mdn,AbdulKhalek:2019mps}, which were subsequently used in the studies presented in the
corresponding Yellow Reports~\cite{Azzi:2019yne,Cepeda:2019klc}.

In order to generate the HL-LHC pseudo-data, we select reference
measurements out of the VBS datasets presented in Sect.~\ref{sec:expdata}.
Table~\ref{tab:datasettable_VBS_HL} presents
the overview of the HL-LHC projections considered in this analysis, which
 include only EW-induced VBS processes
since we assume that the QCD-induced backgrounds can be removed
at the analysis level.
We consider the following differential distributions for each final state:
$m_{\ell\ell}$ for
  $W^{\pm}W^{\pm}jj$, $p_{T}^{\ell \ell \ell}$ and $m_{T}^{WZ}$ in $ZW^{\pm}jj$,
  $m_{ZZ}$    for $ZZjj$, and then
  $ p_{T}^{\gamma \ell \ell}$ and
$m_{\gamma Z}$ in the $\gamma Zjj$ final state,
yielding a total of
  $n_{\rm dat}=61$ datapoints.
The theoretical predictions for these observables
are generated as in Sect.~\ref{sec:expdata} with the same selection and acceptance cuts,
except that they are rescaled to account for the
increase in the center of mass energy from $\sqrt{s}=13$~TeV to $\sqrt{s}=14$~TeV.
We note that the actual
HL-LHC analysis are expected to contain a larger number of bins, as well as a higher reach in energy,
however for simplicity we maintain here the current binning.
The theoretical calculations are generated for the null hypothesis ($\boldsymbol{c}=
\boldsymbol{0}$), with the caveat that better sensitivities would be obtained
in the case of an EFT signal.

The statistical and systematic uncertainties associated to the HL-LHC pseudo-data
are evaluated as follows.
First, we denote $\sigma^{\rm th}_{i}$ as the theoretical prediction for the EW-induced
VBS cross-section in the $i$-th bin of a given differential distribution.
This cross-section includes all relevant selection and acceptance cuts,
as well as the leptonic branching fractions.
The expected number of events in this bin and the associated (relative)
statistical uncertainty $\delta_i^{\rm stat}$ are then given by,
\begin{equation}
  N_{i}^{\rm th}= \sigma^{\rm th}_{i}\times \mathcal{L} \,, \quad \delta_i^{\rm stat}\equiv
  \frac{\delta N_{i}^{\rm stat} }{N_{i}^{\rm th}} = \frac{1}{\sqrt{N_{i}^{\rm th}}} \, .
\end{equation}
Note that the relative statistical uncertainty for 
the number of events and for the cross-sections will be the same, either in the fiducial region or extrapolated to the full phase space.
Here we take the luminosity to be $\mathcal{L}=3$ ab$^{-1}$ and generate two
differential distributions per final state, one from ATLAS and the other from CMS,
as indicated in Table~\ref{tab:datasettable_VBS_HL}.

\input{tables/table_datsaset-VBS-HL}

Concerning the systematic uncertainties, these are also taken from the reference measurements as follows.
If $\delta_{i,j}^{\rm sys}$ denotes the $j$-th relative systematic uncertainty
associated to the $i$-th bin of the reference
measurement,  we assume that the same systematic error 
at the HL-LHC will be given by
$f_{{\rm red},j}\delta_{i,j}^{\rm sys}$, where $f_{{\rm red},j}\simeq 1/2$
is the expected reduction in systematic errors, in agreement with available projections~\cite{CMS:2018mbt,CMS:2018zxa,ATLAS:2018tav,ATLAS:2018ocj}.
Adding in quadrature all systematic uncertainties with the statistical error,
the total relative uncertainty for the $i$-th bin of our HL-LHC projections
will be given by
\begin{equation}
\delta_{{\rm tot},i}^{\rm exp} = \left( \left( \delta_i^{\rm stat}\right)^2 + \sum_{j=1}^{n_{\rm sys}}
\left( f_{{\rm red},j}\delta_{i,j}^{\rm sys} \right)^2\right)^{1/2} \, ,
\end{equation}
where $n_{\rm sys}$ indicates the number of systematic error sources.
Finally, we generate the central values for the HL-LHC pseudo-data projections
by fluctuating the theory prediction by the expected total experimental
uncertainty, namely
\begin{equation}
\sigma^{\rm hllhc}_{i} \equiv \sigma^{\rm th}_{i} \left( 1+ r_i\delta_{{\rm tot},i}^{\rm exp}   \right) \, , \qquad i=1,\ldots,n_{\rm bin} \, ,
\end{equation}
where $r_i$ are univariate Gaussian random numbers.
By construction, one expects that the EFT fit quality to the HL-LHC pseudo-data
to be $\chi^2/n_{\rm bin} \simeq 1$ for a sufficiently large number of bins.

\begin{figure}[htbp]
    \centering
    \subfloat{\includegraphics[width=\textwidth]{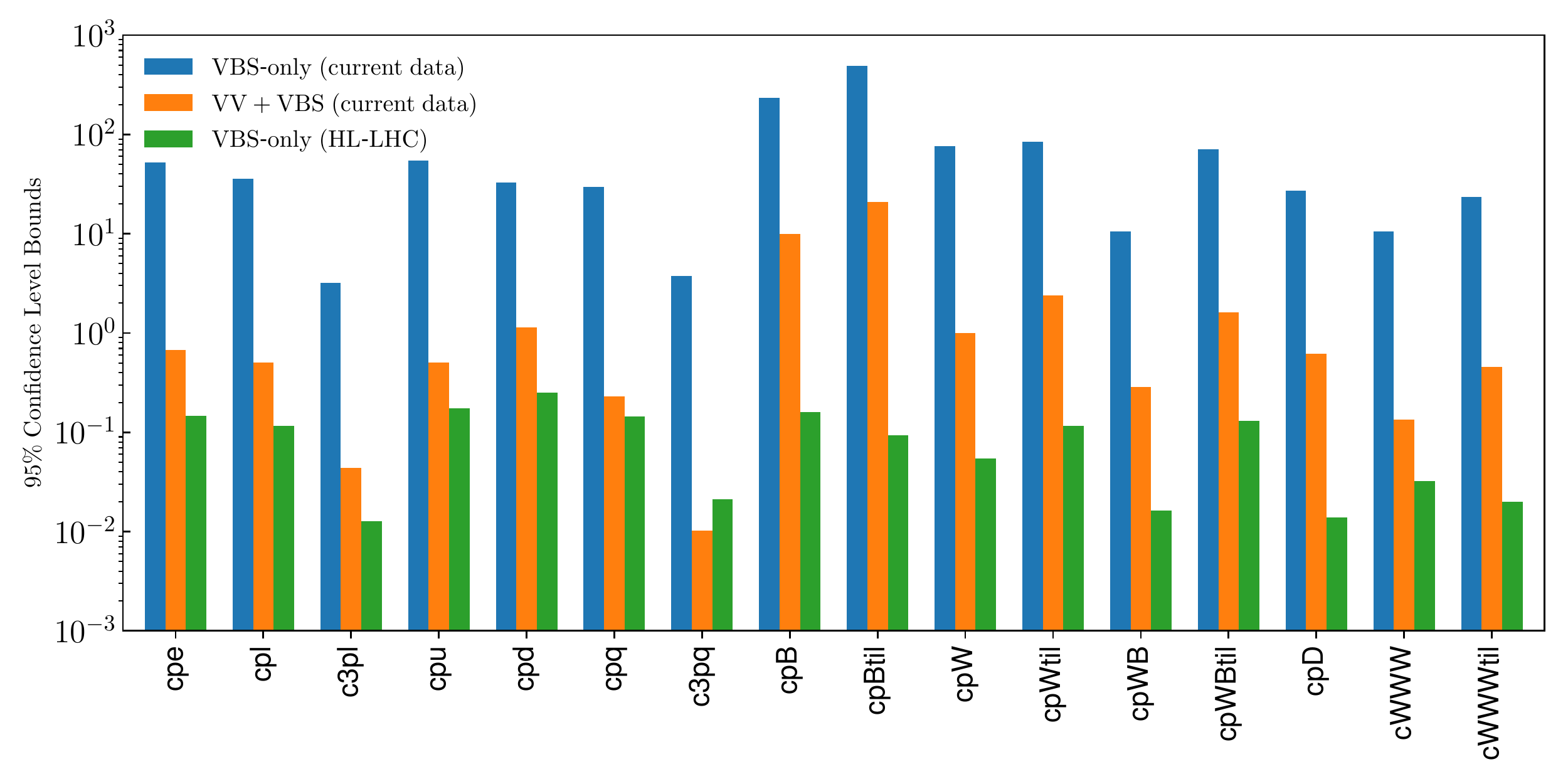}}
    \quad
    \subfloat{\includegraphics[width=\textwidth]{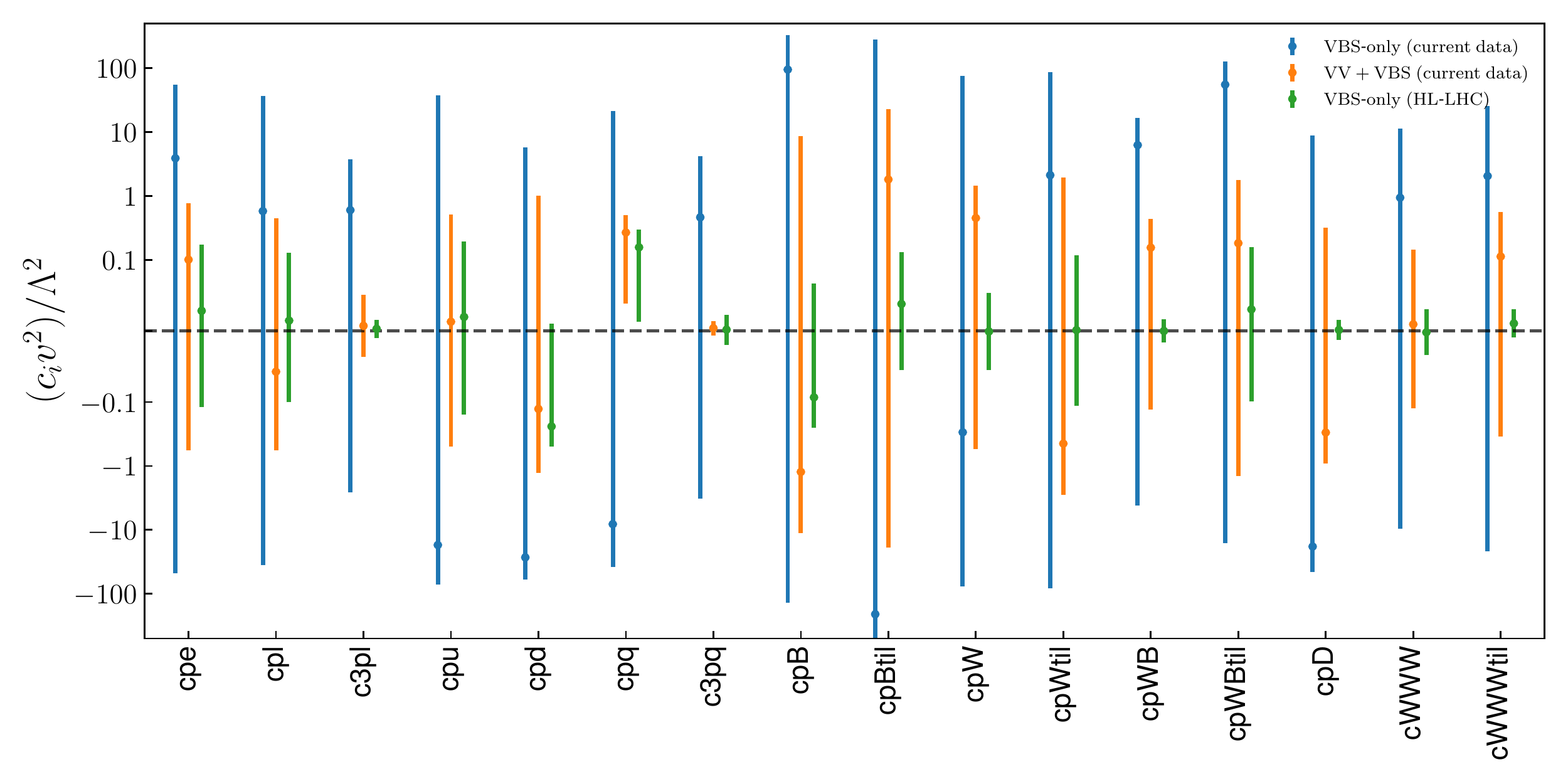}}
    \caption{\small Comparison of the 95\% CL intervals for the EFT coefficients between
      three related analyses: the VBS-only and a the combined diboson+VBS fits based on current data,
      and the VBS-only fit based on the HL-LHC projections listed in
      Table~\ref{tab:datasettable_VBS_HL}.
 }
    \label{fig:HLresults}
\end{figure}

Fig.~\ref{fig:HLresults} displays the comparison of the obtained
95\% CL intervals for the 16 EFT coefficients considered here between
three related analyses.
In particular, EFT fits based on the current measurements, both
for a VBS-only and for a combined diboson+VBS dataset,
are compared with the corresponding results from
the VBS-only fit based on the HL-LHC projections listed in
Table~\ref{tab:datasettable_VBS_HL}.
Here we find that the HL-LHC measurements lead
to a significant impact at the level of the VBS-only fit,
where the current best bounds are improved by up to three orders of magnitude
depending on the specific coefficient.
It is also interesting to note that a VBS-only fit from HL-LHC measurements would even have a superior sensitivity compared to the combined diboson+VBS analysis, especially for the purely bosonic operators where at least a factor of 10 improvement over the current bounds is expected.

The results presented here further highlight
the capability of VBS measurements
for dimension-six EFT studies and the relevance of their integration in the global EFT fit, especially as more luminosity
is accumulated.
While our projections are based on optimistic assumptions such as a clean
separation between the EW- and QCD-induced components of the measurement,
the outstanding performance
of the LHC experiments so far is rather encouraging.

%% file: tables/table_datsaset-VBS-HL.tex
\begin{table}[t]
\begin{center}
\footnotesize
\renewcommand{\arraystretch}{1.8}
\begin{tabular}{c|c|c|c|c|c}
    Final state & Selection & Observable & $n_{\rm dat}$ &  $\mathcal{L}$ (ab$^{-1}$) & Label
    \\
\toprule
    \multirow{2}*{$W^{\pm}W^{\pm}jj$}
    &  EW-induced  & d$\sigma$/d$m_{ll}$
    & 7         & 3  &  \texttt{ATLAS\_WWjj\_mll\_HL} 
    \\
    \cline{2-6}
    & EW-induced   & d$\sigma$/d$m_{ll}$
    & 4         & 3  &  \texttt{CMS\_WWjj\_mll\_HL}   
    \\
    \midrule
    \multirow{2}*{$ZW^{\pm}jj$}
    & EW-induced   & d$\sigma$/d$ p_{T_{\ell \ell \ell}}$
    & 5         &3  &  \texttt{ATLAS\_WZjj\_plll\_HL} 
    \\
    \cline{2-6}
    & EW-induced   & d$\sigma$/d$m_{T}^{WZ}$
    & 5         & 3  & \texttt{CMS\_WZjj\_mwz\_HL}    
    \\
    \midrule
    \multirow{2}*{$ZZjj$}
    & EW-induced  & d$\sigma$/d$m_{ZZ}$
    & 9        & 3  & \texttt{ATLAS\_ZZjj\_mzz\_HL} 
    \\
    \cline{2-6}
    & EW-induced  & d$\sigma$/d$m_{ZZ}$
    & 9        & 3  & \texttt{CMS\_ZZjj\_mzz\_HL}   
    \\
    \midrule
    \multirow{2}*{$\gamma Zjj$}
    & EW-induced  & d$\sigma$/d$ p_{T}^{\gamma \ell \ell}$
    & 13       & 3  & \texttt{ATLAS\_AZjj\_ptlla\_HL} 
    \\
    \cline{2-6}
    & EW-induced  & d$\sigma$/d$m_{\gamma Z}$
    & 9        & 3  & \texttt{CMS\_AZjj\_maz\_HL}     
    \\
    \midrule
    \midrule
    {\bf HL-LHC VBS total  }  &   &  &   {\bf 61}  &   & 
    \\
     \bottomrule
\end{tabular}
\caption{\small Overview of the (EW-induced) VBS HL-LHC projections considered in this analysis.
}
\label{tab:datasettable_VBS_HL}
\end{center}
\end{table}

%% file: sec-summary.tex
\section{Summary and outlook}
\label{sec:summary}

In this work, we have presented an exhaustive investigation of effects from
dimension-six SMEFT operators in the theoretical modelling of vector
boson scattering processes.
By exploiting information provided by the most updated VBS measurements
from ATLAS and CMS, several of which are based on the full Run II data, we have obtained bounds on the relevant SMEFT operators that contribute to this process.
We have demonstrated the overall consistency of the constraints
provided by VBS with those from diboson production, and have highlighted how
VBS measurements provide a useful addition to global EFT interpretations
of LHC data.
Using tailored projections, we have also estimated the improvements in the bounds on these dimension-six operators that can
be expected from the VBS process with the legacy measurements of the HL-LHC,
finding that these measurements will provide a remarkable sensitivity
to several directions in the EFT parameter space.

We emphasize that the goal of this work was not to achieve state-of-the-art bounds on all the dimension-six operators that modify VBS observables.
Such ambition can only be achieved within a dedicated global
EFT fit
that includes all relevant sensitive observables.
These analyses must include, among others,
Higgs production and decay measurements from the LHC and electroweak precision observables
from electron-positron colliders, which by virtue of the electroweak gauge symmetry,
constrain several of the same dimension-six operators that enter
the description of VBS observables, as well as Drell-Yan distributions.
For such an effort, some improvements in the theory
calculations compared to this work
will be required, in particular the use of exact, rather than approximate,
NLO QCD effects in the EFT cross-sections using {\tt SMEFT@NLO} as well as accounting for the quadratic corrections in the EFT expansion.

Most of the previous EFT interpretations of VBS observables from the LHC 
have focused on dimension-eight operators, with the argument that these can modify the quartic gauge couplings while leaving unaffected the triple ones that are purportedly well constrained by other processes.
It would therefore be important to revisit these studies within a consistent EFT analysis that includes the effects of both dimension-six
and dimension-eight operators up to $\mathcal{O}\lp \Lambda^{-4}\rp$.
For instance, it would be important to quantify how the current bounds on dimension-8 operators are modified with the inclusion of the dimension-six ones.
Since there is no cross-talk between the dim-6 and dim-8
operators at this order in the EFT expansion, it would be possible to extend the present analysis by adding the various
sources of quadratic contributions separately.
Such a fully consistent $\mathcal{O}\lp \Lambda^{-4}\rp$ analysis, combined with future measurements from Run III and the HL-LHC, would unlock the ultimate potential of EFT interpretations of VBS data and represent one of the key legacy results from the LHC.

Additional avenues for future research include the EFT interpretation of
novel VBS observables, such as polarised scattering, as well as going beyond the SMEFT by considering other effective theories such as the HEFT or the Electroweak Chiral Lagrangian.
In this respect, we point out that the fitting framework used in this work can be straightforwardly extended to other EFTs, and a fully general dependence of the theory predictions with the EFT coefficients is allowed.

The first measurements of unfolded VBS cross-sections and differential distributions discussed in this work undoubtedly represent a milestone in the LHC program, with profound
implications for our understanding of the gauge sector in the SM and its
extensions.
While current VBS measurements are still statistics-dominated and, for the time
being, provide only a moderate pull in the EFT fit,
we have demonstrated that they provide complementary information
as compared to the more traditional diboson processes.
VBS is therefore poised to play a growing role in global EFT interpretations
in the coming years, especially once high-statistics
measurements become available.\\

\noindent
{\bf \Large Acknowledgments}\\

\noindent
We are grateful to Simone Alioli, Fabio Maltoni, Giampiero Passarino,  Eleni Vryonidou, and Cen Zhang
for discussions about the topics covered in this paper and for feedback on this manuscript.
We thank our colleagues of the  VBScan COST action for many useful
discussions about VBS over the past years.
Special thanks go to the {\tt HEPData} team, in particular Graeme Watt,
and to the ATLAS and CMS analyzers that have provided assistance with the
implementation of the VBS measurements:
Claude Charlot, Roberto Covarelli, Guillelmo G\'omez-Ceballos, and Kristin Lohwasser.
The work of J.~E., G.~M., and J.~R.  is partially
supported by the Netherlands Organization for Scientific
Research (NWO).
R.~G. acknowledges funding  from the ERC Starting Grant REINVENT-714788 and from the Fondazione Cariplo and Regione Lombardia, grant 2017-2070, as well as the UK Science and Technology Facilities Council (STFC) grant ST/P001246/1.